\documentclass[twoside,romanappendices]{IEEEtran}
\usepackage{url}
\usepackage{ifthen}
\usepackage[cmex10]{amsmath} 
\usepackage{longtable}
\usepackage{makecell}
\usepackage{algorithm,algorithmic}
\usepackage{color}
\usepackage[normalem]{ulem}
                             
\usepackage{amssymb,amsthm}
\usepackage{mathtools,bbm}  
\usepackage{xifthen,xparse}
\usepackage{float}
\usepackage{hhline}
\usepackage{caption}

\usepackage{blkarray, bigstrut}
\usepackage{booktabs}

\usepackage[table]{xcolor}
\usepackage{listings}

\definecolor{dkgreen}{rgb}{0,0.6,0}
\definecolor{gray}{rgb}{0.5,0.5,0.5}
\definecolor{mauve}{rgb}{0.58,0,0.82}

\usepackage{tikz,pgfplots,tkz-graph,color}
\pgfplotsset{compat=newest}
\pgfplotsset{plot coordinates/math parser=false}
\usetikzlibrary{decorations.markings,calc,positioning,matrix}

\usepackage{textcomp}

\allowdisplaybreaks

\newtheorem{theorem}{Theorem}
\newtheorem{corollary}[theorem]{Corollary}
\newtheorem{remark}[theorem]{Remark}
\newtheorem{lemma}[theorem]{Lemma}
\newtheorem{definition}[theorem]{Definition}
\newtheorem{example}{Example}

\newenvironment{example*}
  {\addtocounter{example}{-1}\example}
  {\endexample}

\newcommand{\ket}[1]{\left\lvert #1 \right\rangle}
\newcommand{\bra}[1]{\left\langle #1 \right\rvert}
\NewDocumentCommand\ketbra{+m+g}{%
  \IfNoValueTF{#2}
    {\left\lvert #1 \right\rangle \left\langle #1 \right\vert}
  {\left\lvert #1 \right\rangle \left\langle #2 \right\rvert}%
}
\NewDocumentCommand\braket{+m+g}{%
  \IfNoValueTF{#2}
    {\left\langle #1 \vert #1 \right\rangle}
  {\left\langle #1 \vert #2 \right\rangle}%
}

\newcommand{\vecnot}[1]{\underline{#1}}

\newcommand{\MZ}{\mathbb{Z}}
\newcommand{\Fn}{\mathbb{Z}_2^n}

\newcommand{\syminn}[2]{\langle #1, #2 \rangle_{\text{s}}}
\newcommand{\llbr}{[\![}
\newcommand{\rrbr}{]\!]}

\begin{document}

%\title{General Stabilizer Codes that Support\\Integer Symmetric Diagonal (ISD) Gates}
%\title{Synthesizing Stabilizer Codes that Support\\Physical $T$ and $T^{\dagger}$ Gates}
%\title{Stabilizer Codes that Support \red{Transversal} $Z$-Rotations}
\title{On Optimality of CSS Codes for Transversal $T$}

%\author{Narayanan Rengaswamy\\%
%Department of ECE, Duke University, USA}
 \author{%
   \IEEEauthorblockN{Narayanan Rengaswamy,~\IEEEmembership{Graduate Student Member,~IEEE},
                     Robert Calderbank,~\IEEEmembership{Fellow,~IEEE},
                     Michael Newman, and
                     Henry D. Pfister,~\IEEEmembership{Senior Member,~IEEE}}%
   \thanks{N. Rengaswamy, R. Calderbank, and H.D. Pfister are with the
           Department of Electrical and Computer Engineering,
           Duke University,
           Durham, North Carolina 27708, USA.
           M. Newman is with the Departments of Electrical and Computer Engineering, Chemistry and Physics,
           Duke University,
           Durham, North Carolina 27708, USA.
           Email: \{narayanan.rengaswamy, robert.calderbank, henry.pfister\}@duke.edu, mgnewman@umich.edu}%
    \thanks{Part of this work was presented at the 2020 IEEE International Symposium on Information Theory~\cite{Rengaswamy-isit20}.}
%    \thanks{This paper has supplementary downloadable material available at http://ieeexplore.ieee.org, provided by the author. The material includes appendices containing proofs for many results. Contact narayanan.rengaswamy@duke.edu for further questions about this work.}
  }

%\author{Narayanan Rengaswamy}
% \email{narayanan.rengaswamy@duke.edu}
%\author{Robert Calderbank}
% \email{robert.calderbank@duke.edu}
%\author{Henry D. Pfister}
% \email{henry.pfister@duke.edu}
%\affiliation{Department of Electrical and Computer Engineering, Duke University, Durham, North Carolina 27708, USA}

%\date{\today}

{\maketitle}

\begin{abstract}
In order to perform universal fault-tolerant quantum computation, one needs to implement a logical non-Clifford gate. 
Consequently, it is important to understand codes that implement such gates transversally. 
In this paper, we adopt an algebraic approach to characterize all stabilizer codes for which transversal $T$ and $T^{\dagger}$ gates preserve the codespace. 
Our Heisenberg perspective reduces this question to a finite geometry problem that translates to the design of certain classical codes. 
We prove three corollaries of this result: 
(a) For any non-degenerate $[\![ n,k,d ]\!]$ stabilizer code supporting a physical transversal $T$ (which might not be logical $T$), there exists an $[\![ n,k,d ]\!]$ CSS code with the same property; 
(b) Triorthogonal codes form the most general family of CSS codes that realize logical transversal $T$ via physical transversal $T$; 
(c) Triorthogonality is necessary for physical transversal $T$ on a CSS code to realize the logical identity. 
The main tool we use is a recent characterization of a particular family of diagonal gates in the Clifford hierarchy that are efficiently described by symmetric matrices over rings of integers [N. Rengaswamy et al., Phys. Rev. A \textbf{100}, 022304]. 
We refer to these operations as \emph{Quadratic Form Diagonal (QFD)} gates. 
Our framework generalizes all existing stabilizer code constructions that realize logical gates via transversal $T$. 
We provide several examples of codes and briefly discuss connections to decreasing monomial codes, pin codes, generalized triorthogonality and quasitransversality. 
We partially extend these results towards characterizing all stabilizer codes that support transversal $\pi/2^{\ell}$ $Z$-rotations. 
In particular, using Ax's theorem on residue weights of polynomials, we provide an alternate characterization of logical gates induced by transversal $\pi/2^{\ell}$ $Z$-rotations on a family of quantum Reed-Muller codes. 
We also briefly discuss a general approach to analyze QFD gates that might lead to a characterization of all stabilizer codes that support any given physical transversal $1$- or $2$-local diagonal gate.
\end{abstract}


\section{Introduction}
\label{sec:intro}

% {\color{red}
% 1) Motivation: Why are transversal Z-rotations interesting?
% 2) Background: How do people think about transversal Z-rotations?
% 3) What do we do differently?
% 4) Why is this perspective helpful?
% 5) Examples to build intuition + Ax's Theorem --> for nondegenerate, circle back to ket basis.
% 6) Outline of paper
% }

Quantum error correction is vital to build a universal, fault-tolerant quantum computer.
Since such a device must process the information stored in it, we need to devise schemes that fault-tolerantly perform unitary operations on the protected information.
For an $\llbr n,k,d \rrbr$ quantum error-correcting code, any unitary operation on the $k$ logical qubits must be realized via an operation on the $n$ physical qubits that preserves the code subspace.
A \emph{transversal} gate is one in which the physical operation decomposes into a tensor product of individual unitaries on each physical qubit of the code.
Since errors do not propagate within codeblocks during such an operation, these gates are naturally fault-tolerant. 
Hence, transversal implementations of logical gates are highly desirable.
However, the Eastin-Knill theorem shows that there is no QECC that detects at least $1$ error and possesses a universal set of logical gates that can be realized via transversal operations~\cite{Eastin-prl09, Zeng-it07}.
%\red{\sout{In fact, for stabilizer codes, even including those logical gates that can be realized via physical qubit permutations might not yield universality~\cite{Zeng-it07}.}}
Therefore, there is a need to balance the operations that can be implemented transversally and the operations for which other fault-tolerant mechanisms must be devised.

In general, logical Clifford gates are easier to implement than logical $T$ gates.
This is because self-dual CSS codes, i.e., CSS codes where the pure $X$-type and pure $Z$-type stabilizers are constructed from the same classical code, admit a transversal implementation of the logical Clifford group, but not of the logical $T$ gate.
On the other hand, there are some code families such as \emph{triorthogonal codes}~\cite{Bravyi-pra12} and \emph{color codes}~\cite{Kubica-pra15} that realize the logical $T$ gate transversally.
Therefore, a common strategy is to utilize these codes to perform magic state distillation and state injection to apply the logical $T$ gate on the data ~\cite{Bravyi-pra12,Gottesman-nature99,Bravyi-pra05}.
By this approach, circuits on the error-corrected quantum computer will only consist of Clifford operations, augmented by ancillary magic states, and these operations can be realized transversally.
% \red{\sout{However, magic state distillation usually consumes a large proportion of resources for realizing a fault-tolerant quantum computer, although the cost has been decreasing over the years~\cite{Litinski-arxiv19}.
% The above approach is primarily motivated by the fact that physical qubits only have nearest-neighbor connectivity in some hardware systems, such as superconducting qubits.
% Hence, topological codes such as the surface codes are highly suited to such architectures since all operations on these codes are geometrically local.
% The logical Clifford gates are fault-tolerant on these codes and the logical $T$ gate can be realized via magic state distillation using the aforementioned codes.}}

There has also been interest in employing logical smaller angle rotations, compared to the $\pi/8$ rotation of the $T$ gate~\cite{Landahl-arxiv13}.  
This poses heavier requirements on the distillation code, but can also result in shorter gate sequences during compilation.  
In contrast to the difficulty of small-angle logical rotations, the fidelity of physical rotations can increase at finer angles~\cite{Nam-arxiv19}, helping to mitigate the burden of magic state distillation with cumbersome codes.  
Such trapped-ion systems are leading candidates to realize a universal fault-tolerant quantum computer and most experimental efforts focus on stabilizer codes.
Thus, it may be profitable to further understand stabilizer codes supporting smaller-angle transversal $Z$-rotations as well.

% \red{\sout{However, architectures such as trapped-ion systems are less limited by nearest-neighbor connectivity; for the near-future, they are expected to have all-to-all connectivity.
% This provides additional freedom in designing codes and we are not necessarily restricted to topological codes with geometrically local operations. 
% Moreover, arbitrary angle Pauli-rotations are native operations for many hardware architectures. 
% In fact, the fidelity of finer angle rotations has been observed to be superior to coarser angle ones in trapped-ion systems~\cite{Nam-arxiv19}.
% In this paper, we intend to take steps towards systematically understanding how to exploit such native resources by designing codes carefully. 
% This is important because, (logical) circuit decompositions that include such finer angle rotations might produce lower depth circuits when compared to Clifford + $T$ decompositions.
% For near-term quantum computing, the resulting gate complexity gains can be practically significant.}}

In this paper, we take steps towards systematically understanding the construction of general stabilizer codes that support physical transversal $T$ and $T^{\dagger}$ gates as logical operators, and then discuss extensions to transversal finer angle $Z$-rotations.
We also briefly discuss a general method to analyze other diagonal gates that have an efficient representation using symmetric matrices over rings of integers~\cite{Rengaswamy-pra19} (see Section~\ref{sec:general_QFD}), which we refer to as \emph{Quadratic Form Diagonal} (QFD) gates.
% \red{\sout{In principle, we \red{could} analyze other native operations such as finer angle M{\o}lmer-S{\o}rensen gates, which are QFD gates up to \red{local Clifford operators}, \red{but} we leave that for future work.}}
These form a subgroup of all diagonal gates in the Clifford hierarchy which were characterized earlier by Cui et al.~\cite{Cui-physreva17}.
We have shown that all $1$- and $2$-local diagonal gates in the Clifford hierarchy are QFD gates~\cite{Rengaswamy-pra19}.
Fault-tolerance makes it natural to partition the physical qubits into small groups and employ ``generalized'' transversal gates that split into operations on these individual groups.
Indeed, such a scheme has been recently explored by Jochym-O'Connor et al.~\cite{JochymOconnor-prx17}, and can be used to construct a universal set of fault-tolerant gates~\cite{JochymOconnor-prl14}.
%where they argued that fault-tolerance requires that we work with a \emph{fixed} partition of qubits.
In fact, they showed that if we allow the partition to change during computation, then we can obtain a universal set of logical gates through transversal operations alone.
More precisely, on a concatenated code, where the (fixed) partition decides which is the inner code and which is the outer code, a transversal operation on the outer code can be effectively used to realize the operation fault-tolerantly on the overall concatenated code, although the overall operation is not transversal.
Therefore, our general approach to analyze QFD gates allows one to investigate codes that support transversal $1$- and $2$-local diagonal gates, on a partition of qubits into groups of at most two.
This paper is a proof-of-concept for the important case of $Z$-rotations.

Several works have studied the problem of realizing non-trivial logical operators via physical $Z$-rotations~\cite{Bravyi-pra12,Haah-quantum17b,Campbell-pra17,Campbell-prl17,Vuillot-arxiv19}.
These works approach this problem by restricting themselves to \emph{Calderbank-Shor-Steane (CSS)} codes and then examining the action of these gates on the basis states of these codes.
When a $\pi/2^{\ell}$ $Z$-rotation gate 
\begin{align}
\label{eq:Zrotations}
\exp\left( \frac{-\imath\pi}{2^{\ell}} Z \right) = \cos\frac{\pi}{2^{\ell}} \cdot I_2 - \imath \sin\frac{\pi}{2^{\ell}} \cdot Z \equiv \text{diag}\left( 1, e^{2\pi\imath/2^{\ell}} \right)
\end{align}
acts on a qubit in the computational basis $\{ \ket{0}, \ket{1} \}$, it picks up a phase of $\exp\left( \frac{2\pi\imath}{2^{\ell}} \right)$ when acting on $\ket{1}$ and it leaves $\ket{0}$ undisturbed.
Hence, a transversal application of this gate on an $n$-qubit state $\ket{v}, v \in \mathbb{Z}_2^n$, picks up the phase $\exp\left( \frac{2\pi\imath}{2^{\ell}} w_H(v) \right)$, where $w_H(v) \coloneqq \sum_{i=1}^{n} v_i$ denotes the Hamming weight of $v$.
% Therefore, by engineering the vectors in the superpositions forming the CSS basis states to have the desired Hamming weights, these works determined \emph{sufficient} conditions for such transversal $Z$-rotations to realize logical operators on these codes.
Therefore, by engineering the Hamming weights of the binary vectors describing the superposition in the CSS basis states, these works determined \emph{sufficient} conditions for such transversal $Z$-rotations to realize logical operators on these codes.

In contrast to these previous works, we take a Heisenberg approach to this problem by examining the action of the physical operation on the stabilizer group defining the code, naturally generalizing the aforementioned strategy.
Consequently, we are able to derive necessary and sufficient conditions for any stabilizer code to support a physical transversal $T, T^{\dagger}$ gate, without restricting ourselves to CSS codes (see Theorems~\ref{thm:transversal_T},~\ref{thm:transversal_T_Tinv}).
When applied to CSS codes, these conditions translate to constructing a pair of classical codes $C_X$ and $C_Z$ such that $C_Z$ contains a self-dual code supported on each codeword in $C_X$.
% \red{\sout{(We note here that the case of general stabilizer codes is not much different.)}}
Concretely, this result allows us to prove the following corollaries which broadly form ``converses'' to the sufficient conditions derived in the aforementioned works.
\begin{enumerate}
    \item Given an $\llbr n,k,d \rrbr$ \emph{non-degenerate} stabilizer code supporting a physical transversal $T, T^{\dagger}$ gate, there exists an $\llbr n,k,d \rrbr$ CSS code supporting the same operation (see Corollary~\ref{cor:csst_sufficient}).
    An $\llbr n,k,d \rrbr$ stabilizer code is non-degenerate if every stabilizer element has weight at least $d$.
    For degenerate stabilizer codes, this statement holds under an additional assumption on the stabilizer generators.
    Note that the toric and color codes are degenerate codes because the weights of the stabilizer generators are fixed even when the lattice size is increased, i.e., code distance is increased.
However, codes such as quantum Reed-Muller codes are typically non-degenerate since each stabilizer generally has weight at least equal to the code distance.

    \item Triorthogonal codes form the most general family of CSS codes that realize logical transversal $T$ from physical transversal $T$ (see Theorem~\ref{thm:logical_trans_T} and Corollary~\ref{cor:triortho_general}). 
    Here, by ``logical transversal $T$'' we mean that the induced logical operation applies $T$ to each logical qubit.
    
    \item Triorthogonality is necessary for physical transversal $T$ on a CSS code to realize the logical identity (see Theorem~\ref{thm:logical_identity}).
    An additional condition on the logical $X$ operators distinguishes this case from triorthogonal codes where the logical operation is also a transversal $T$.
\end{enumerate}
These results suggest that, for the problem of distilling magic states using physical transversal $T$, CSS codes might indeed be optimal.
We emphasize that we are able to make such conclusions because we focus on the effects of physical operations directly on the stabilizer (and logical Pauli) group(s), rather than just on the basis states of CSS codes, i.e., by taking a ``Heisenberg'' perspective rather than a ``Schr{\"o}dinger'' perspective.

We believe this result opens the way to leverage the rich classical literature on self-dual codes~\cite{Rains-arxiv02,Nebe-2006}, the MacWilliams identities~\cite{Macwilliams-1977} and the McEliece theorems on divisibility of weights~\cite{McEliece-jpl72,McEliece-dm72}, to potentially construct new stabilizer codes with transversal gates. 
Furthermore, this perspective is a new tool for arguing about the best possible scaling achievable for rates and distances of stabilizer codes supporting transversal $T$ gates, or even general $\pi/2^{\ell}$ $Z$-rotations. %(see Section~\ref{sec:stab_codes_Z_rotations}).

Among several examples, we construct a $\llbr 16,3,2 \rrbr$ code where transversal $T$ realizes the logical CC$Z$ (up to Pauli corrections; see Section~\ref{sec:logical_ccz}). 
This code belongs to the compass code family studied in \cite{Li-prx19}.
This is also closely related to Campbell's $\llbr 8,3,2 \rrbr$ color code~\cite{Campbell-blog16} that is defined on a $3$-dimensional cube, and it can be interpreted as three such cubes in a chain.
(The construction can be extended to a chain of arbitrary number of cubes.)
As we show in Example~\ref{eg:Campbell}, the $\llbr 8,3,2 \rrbr$ code belongs to a family of $\llbr 2^m,m,2 \rrbr$ quantum Reed-Muller codes defined on $m$-dimensional cubes.
However, as we discuss in Example~\ref{eg:decreasing_monomial}, the $\llbr 16,3,2 \rrbr$ code can be constructed using the (classical) formalism of \emph{decreasing monomial codes} that was introduced by Bardet et al.~\cite{Bardet-isit16,Bardet-arxiv16}.
This formalism generalizes Reed-Muller and polar codes~\cite{Arikan-it09}, and provides a general framework for synthesizing a large family of codes via evaluations of polynomials.
Recently, Krishna and Tillich~\cite{Krishna-arxiv18} have exploited this framework to construct triorthogonal codes from punctured polar codes for magic state distillation.
% \red{COMMENT: Here, I might expand a bit more.  This code, from the perspective of Earl's [[8,3,2]] code, can be thought of as a color code with only boundaries.  If I recall correctly this can extend indefinitely to a 'chain of cubes'.  However, as you mention, what's important is that it illustrates a connection to decreasing monomial codes that might be exploited to generate new families.  This connection is important because if you don't make it explicitly, the example seems like a simple extension of the [[8,3,2]], but actually moving from the [[8,3,2]] goes from Reed-Muller to another different but famous family.  It is illustrative of a potentially fruitful connection.}
Thus, the $\llbr 16,3,2 \rrbr$ code forms an interesting example because it points towards a general application of the formalism of decreasing monomial codes for transversal $Z$-rotations, where the logical $X$ and $Z$ strings are not necessarily identical as in the standard presentation of triorthogonal codes.
Such asymmetry in logical operators and hence the $X$- and $Z$-distances of the codes, which can also exist in triorthogonal codes, might be useful in scenarios of biased noise as well~\cite{Tuckett-prl18}. 
Hence, this formalism provides more flexibility in designing codes as well as analyzing them.

Finally, we extend this approach beyond $T$ gates and establish conditions for a stabilizer code to support a transversal $\pi/2^{\ell}$ $Z$-rotation (see Theorem~\ref{thm:transversal_Z_rot}).
However, the conditions we derive involve trigonometric quantities on the weights of vectors describing the stabilizer, and we are unable to distill finite geometric conditions without making a simplifying assumption.
Therefore, we have yet to establish a full generalization of Theorems~\ref{thm:transversal_T} and~\ref{thm:transversal_T_Tinv} to general $\pi/2^{\ell}$ $Z$-rotations.
Note that we only discuss $Z$-rotations of the form in~\eqref{eq:Zrotations} because non-trivial error-detecting stabilizer codes only support rotations belonging to the Clifford hierarchy~\cite{JochymOconnor-prx17,Cui-physreva17}.
However, we are able to study a family of $\llbr 2^m, \binom{m}{r}, 2^r \rrbr$ quantum Reed-Muller codes, where $1 \leq r \leq m/2$ and $r$ divides $m$, and provide an alternative perspective that highlights the logical operation realized by a transversal $\exp\left( \frac{-\imath\pi}{2^{m/r}} Z \right)$ gate.
This recovers the well-known $\llbr 8,3,2 \rrbr$ code of Campbell~\cite{Campbell-blog16}, and also provides new information about the family of codes discussed in~\cite{Haah-quantum17b,Campbell-pra17}.
By the ``CSS sufficiency'' intuition above, this ties back to the ``Schr{\"o}dinger'' perspective of past works.

The paper is organized as follows.
Section~\ref{sec:prelim} establishes the necessary background and notation, which includes the Pauli group, QFD gates, and stabilizer codes. % and classical Reed-Muller codes.
Section~\ref{sec:general_QFD} outlines the general approach to analyze stabilizer codes that support a given physical QFD gate.
Section~\ref{sec:stab_codes_T_Tinv} establishes the necessary and sufficient conditions for a stabilizer code to support a given pattern of $T$ and $T^{\dagger}$ gates on the physical qubits.
Section~\ref{sec:logical_T} derives conditions for physical transversal $T$ to realize logical transversal $T$ on CSS codes, and proves that triorthogonal codes are the most general CSS codes that satisfy this property.
Sections~\ref{sec:logical_ccz} and~\ref{sec:product_of_CCZs} discuss examples of codes where physical transversal $T$ realizes logical $\text{CC}Z$ gates, and explain how these codes satisfy the conditions derived in Theorem~\ref{thm:transversal_T}.
The discussion on related work in~\cite{Haah-quantum17b} is provided at the end of Section~\ref{sec:product_of_CCZs}.
Section~\ref{sec:stab_codes_Z_rotations} partially extends the results in Section~\ref{sec:stab_codes_T_Tinv} and explicitly derives the logical operation realized by transversal $Z$-rotations on the aforementioned family of quantum Reed-Muller codes.
The relation to quantum pin codes and quasitransversality is discussed at the end of Section~\ref{sec:stab_codes_Z_rotations}.
Finally, Section~\ref{sec:conclusion} concludes the paper and discusses future directions.

\section{Preliminaries and Notation}
\label{sec:prelim}

\subsection{The Pauli or Heisenberg-Weyl Group}
\label{sec:pauli_gp}

The single qubit Pauli operators are the unitaries
\begin{align}
I_2 \coloneqq 
\begin{bmatrix}
1 & 0 \\
0 & 1
\end{bmatrix},\ 
X \coloneqq 
\begin{bmatrix}
0 & 1 \\
1 & 0
\end{bmatrix}, \ 
Z \coloneqq 
\begin{bmatrix}
1 & 0 \\
0 & -1
\end{bmatrix}, \ 
Y \coloneqq \imath X Z, 
%= 
%\begin{bmatrix}
%0 & -\imath \\
%\imath & 0
%\end{bmatrix},
\end{align}
where $\imath \coloneqq \sqrt{-1}$.
They satisfy $X^2 = Z^2 = Y^2 = I_2$.
Let $A \otimes B$ denote the Kronecker product between matrices $A$ and $B$.
For $n$ qubits, given $a = [\alpha_1,\alpha_2,\ldots,\alpha_n], b = [\beta_1,\beta_2,\ldots, \beta_n] \in \mathbb{Z}^n$, where $\mathbb{Z}$ denotes the ring of integers, we define the operators
\begin{align}
D(a,b) & \coloneqq X^{\alpha_1} Z^{\beta_1} \otimes X^{\alpha_2} Z^{\beta_2} \otimes \cdots \otimes X^{\alpha_n} Z^{\beta_n}, \\
E(a,b) & \coloneqq \left( \imath^{\alpha_1 \beta_1} X^{\alpha_1} Z^{\beta_1} \right) \otimes \cdots \otimes \left( \imath^{\alpha_n \beta_n} X^{\alpha_n} Z^{\beta_n} \right) \nonumber \\
  & = \imath^{ab^T \bmod 4} D(a,b).
\end{align}
The unitaries $E(a,b)$ are Hermitian and satisfy $E(a,b)^2 = I_N$, where $N \coloneqq 2^n$, but the unitaries $D(a,b)$ can have order $1, 2$ or $4$.
Although Pauli operators are usually represented as binary vectors, we use the generalized notation of integer vectors to essentially keep track of phases and signs more carefully, as we discussed in~\cite{Rengaswamy-pra19}.
%for some technical reasons discussed in~\cite{Rengaswamy-pra19} that will affect the computations in this paper.
The $n$-qubit \emph{Heisenberg-Weyl group} (or \emph{Pauli group}) is defined as $HW_N \coloneqq \{ \imath^{\kappa} D(a,b)\ \text{for\ all}\ a,b \in \mathbb{Z}_2^n, \kappa \in \mathbb{Z}_4 \}$. 
We use the notation $\mathbb{Z}_{2^{\ell}} = \{0,1,2,\ldots,2^{\ell} - 1\}$ to denote the set of integers modulo $2^{\ell}$ for some integer $\ell > 0$.

For $v = [v_1, v_2, \ldots, v_n] \in \mathbb{Z}_2^n$, let $\ket{v} = e_v = \ket{v_1} \otimes \ket{v_2} \otimes \cdots \otimes \ket{v_n}$ denote the standard basis vector with entry $1$ in the position indexed by $v$ and $0$ elsewhere, and let $\bra{v} = \ket{v}^{\dagger}$, the Hermitian transpose of $\ket{v}$.
An arbitrary $n$-qubit quantum state can be written as $\ket{\psi} = \sum_{v \in \mathbb{Z}_2^n} \alpha_v \ket{v} \in \mathbb{C}^N$, where $\alpha_v \in \mathbb{C}$ satisfy $\sum_{v \in \mathbb{Z}_2^n} | \alpha_v |^2 = 1$ as per the Born rule~\cite{Wilde-2013}, and $\mathbb{C}$ denotes the field of complex numbers.
It is easy to check that $X \ket{0} = \ket{1}, X \ket{1} = \ket{0}, Z \ket{0} = \ket{0}, Z \ket{1} = - \ket{1}$.
Hence, we can write $E(a,0) = \sum_{v \in \mathbb{Z}_2^n} \ketbra{v \oplus a}{v}, E(0,b) = \sum_{v \in \mathbb{Z}_2^n} (-1)^{vb^T} \ketbra{v}$.
Throughout the paper, $\oplus$ denotes modulo $2$ addition and $+$ denotes the usual addition over integers.
Also, all binary and integer-valued vectors will be row vectors while complex-valued vectors will be column vectors.
For $x = [x_1,x_2,\ldots,x_n] ,y = [y_1,y_2,\ldots,y_n] \in \mathbb{Z}_2^n$, $x \ast y = [x_1 y_1, x_2 y_2, \ldots, x_n y_n]$.

Using the fact that $XZ = -ZX$, we can prove the following identities for Pauli matrices (e.g., see~\cite{Rengaswamy-arxiv18}).
\begin{align}
\label{eq:Eab_multiply}
E(a,b) E(c,d) & = \imath^{bc^T - ad^T \bmod 4} E(a+c, b+d) \nonumber \\
  & = (-1)^{\syminn{[a,b]}{[c,d]}} E(c,d) E(a,b), \\
\syminn{[a,b]}{[c,d]} & \coloneqq [a,b] \, \Omega\, [c,d]^T, \ \Omega \coloneqq 
\begin{bmatrix}
0 & I_n \\
I_n & 0
\end{bmatrix}, \\
\label{eq:Eab_2x}
E(a, b + 2x) & = \imath^{a(b+2x)^T} D(a,b + 2x) = %(-1)^{ax^T} \imath^{ab^T} D(a,b) =
 (-1)^{ax^T} E(a,b).
\end{align}
Therefore, two Pauli matrices $E(a,b), E(c,d)$ commute if and only if $\syminn{[a,b]}{[c,d]} = 0$, and they anti-commute otherwise.

The Pauli operators $\{ \frac{1}{\sqrt{N}} E(a,b),\, a,b \in \mathbb{Z}_2^n \}$ form an orthonormal basis for all unitary matrices under the trace inner product $\langle A,B \rangle_{\text{Tr}} \coloneqq \text{Tr}(A^{\dagger} B)$, where $A^{\dagger}$ represents the Hermitian transpose of $A$.
Therefore, given any matrix $U \in \mathbb{U}_N$, where $\mathbb{U}_N$ denotes the group of $N \times N$ unitary matrices, we can express it as
\begin{align}
U = \sum_{a,b \in \mathbb{Z}_2^n} \text{Tr}\left( \frac{1}{\sqrt{N}} E(a,b) \cdot U \right) \cdot \frac{1}{\sqrt{N}} E(a,b). 
% = \frac{1}{N} \sum_{a,b \in \mathbb{Z}_2^n} \text{Tr}(E(a,b) U) \ E(a,b).
\end{align}
Note that $\text{Tr}(E(a,b)) = 0$ unless $E(a,b) = E(0,0) = I_N$, in which case $\text{Tr}(E(0,0)) = N$.

If $U = \sum_{v \in \mathbb{Z}_2^n} \phi_v \ketbra{v}$ is diagonal (in the standard coordinate basis), then for $a \neq 0$ we observe that
\begin{align}
& \text{Tr}(E(a,b) U) \nonumber \\
  & \coloneqq \text{Tr}\left[ \imath^{ab^T} \sum_{v \in \mathbb{Z}_2^n} \ketbra{v \oplus a}{v} \cdot \sum_{v' \in \mathbb{Z}_2^n} (-1)^{v'b^T} \ketbra{v'} \cdot U \right] \\
  & = \imath^{ab^T} \sum_{v \in \mathbb{Z}_2^n} (-1)^{vb^T} \bra{v} U \ket{v \oplus a} = 0.
%
%  & = \imath^{ab^T} \sum_{v \in \mathbb{Z}_2^n} (-1)^{vb^T} \text{Tr}\left[ \ketbra{v \oplus a}{v} \cdot U \right] \\
%  & = \imath^{ab^T} \sum_{v \in \mathbb{Z}_2^n} (-1)^{vb^T} \bra{v} U \ket{v \oplus a} \\
%  & = \imath^{ab^T} \sum_{v \in \mathbb{Z}_2^n} (-1)^{vb^T} \phi_{v \oplus a} \braket{v}{v \oplus a} \\
%  & = 0.
\end{align}
Hence, $\text{Tr}(E(a,b) U) \neq 0$ if and only if $a = 0$, and $\text{Tr}(E(0,b) U) = \sum_{v \in \mathbb{Z}_2^n} (-1)^{vb^T} \phi_{v}$.

\subsection{Quadratic Form Diagonal (QFD) Gates}
\label{sec:QFD_gates}

The \emph{Clifford hierarchy} of unitary operators was defined by Gottesman and Chuang~\cite{Gottesman-nature99} in order to demonstrate that universal quantum computation can be realized via quantum teleportation if one has access to Bell-state preparation, Bell-basis measurements and arbitrary single-qubit operations.
The first level $\mathcal{C}^{(1)}$ of the hierarchy is defined to be the Pauli group, i.e., $\mathcal{C}^{(1)} = HW_N$.
For $\ell \geq 2$, the levels $\mathcal{C}^{(\ell)}$ are defined recursively as
\begin{align}
\mathcal{C}^{(\ell)} \coloneqq \{ U \in \mathbb{U}_N \colon U E(a,b) U^{\dagger} \in \mathcal{C}^{(\ell - 1)}\ \forall\ E(a,b) \in HW_N \}.
\end{align}
By this definition, the second level is the Clifford group, $\mathcal{C}^{(2)} = \text{Cliff}_N$, that is fundamental to quantum computation.
Equivalently, the Clifford group can also be defined as the automorphism group of $HW_N$.
Any given $g \in \text{Cliff}_N$ satisfies
\begin{align}
\label{eq:cliff_action}
g E(a,b) g^{\dagger} = \pm E([a,b] F_g),
\end{align}
where $F_g \in \mathbb{Z}_2^{2n \times 2n}$ satisfies $F_g\, \Omega\, F_g^T = \Omega$ and hence is called a \emph{binary symplectic matrix}~\cite{Rengaswamy-arxiv18}.
$\text{Cliff}_N$ can be generated using the unitaries \emph{Hadamard, Phase}\footnote{We use the notation ``$P$'' for the phase gate and reserve the commonly used notation of ``$S$'' for stabilizer groups.}, \emph{Controlled-$Z$ (C$Z$)} and \emph{Controlled-NOT (C$X$)} defined respectively as
\begin{align}
H & \coloneqq \frac{1}{\sqrt{2}}
\begin{bmatrix}
1 & 1 \\
1 & -1
\end{bmatrix}, \ 
P \coloneqq 
\begin{bmatrix}
1 & 0 \\
0 & \imath
\end{bmatrix}, \nonumber \\
\text{C}Z_{ab} & \coloneqq \ketbra{0}_a \otimes (I_2)_b + \ketbra{1}_a \otimes Z_b, \nonumber \\ 
\text{C}X_{a \rightarrow b} & \coloneqq \ketbra{0}_a \otimes (I_2)_b + \ketbra{1}_a \otimes X_b.
\end{align}
The subscripts for C$Z$ and C$X$ denote the indices of the two qubits involved, and C$Z$ is symmetric with respect to both qubits. 
However, for C$X$, $a \rightarrow b$ indicates that when the control qubit $a$ is in state $\ket{1}$ the target qubit $b$ is flipped by applying the $X$ gate, and when $a$ is in state $\ket{0}$ the target $b$ is left undisturbed.
Note that $H^{\dagger} = H$ and $HZH = X$ implies $\text{C}Z_{ab} = ((I_2)_a \otimes H_b)\, \text{C}X_{a \rightarrow b}\, ((I_2)_a \otimes H_b)$.
It is well-known that Clifford unitaries along with any unitary from a higher level, say from $\mathcal{C}^{(3)}$, can be used to approximate any unitary operator arbitrarily well, and hence form a universal set for quantum computation~\cite{Boykin-arxiv99}.
The widely used choice for the non-Clifford unitary is the ``$\pi/8$'' gate or the $T$ gate, defined as 
\begin{align}
T \coloneqq 
\begin{bmatrix}
1 & 0 \\
0 & e^{\imath\pi/4}
\end{bmatrix} = \sqrt{P} = Z^{1/4} \equiv 
\begin{bmatrix}
e^{-\imath\pi/8} & 0 \\
0 & e^{\imath\pi/8}
\end{bmatrix} = e^{-\frac{\imath\pi}{8} Z}.
\end{align}

The work in~\cite{Rengaswamy-pra19} considered diagonal unitaries of the form $\tau_R^{(\ell)} = \sum_{v \in \Fn} \xi^{v R v^T \bmod 2^{\ell}} \ketbra{v}$, where $\ell \in \mathbb{Z}_{>1}$, $\xi \coloneqq \exp(\frac{2\pi\imath}{2^{\ell}})$, and $R$ is a symmetric matrix over $\MZ_{2^{\ell}}$.
It explicitly calculates their action on an $n$-qubit (Hermitian) Pauli matrix to be
\begin{align}
\label{eq:tau_Eab}
\tau_R^{(\ell)} E(a,b) ( \tau_R^{(\ell)} )^{\dagger} & = \xi^{\phi(R,a,b,\ell)} E(a_0, b_0 + a_0 R) \, \tau_{\tilde{R}(R,a,\ell)}^{(\ell-1)}, \\
\phi(R,a,b,\ell) & \coloneqq (1 - 2^{\ell-2}) a_0 R a_0^T \nonumber \\
  & \hspace*{2cm} + 2^{\ell-1} (a_0 b_1^T + b_0 a_1^T), \\
\tilde{R}(R,a,\ell) & \coloneqq (1 + 2^{\ell-2}) D_{a_0 R} - (D_{\bar{a}_0} R D_{a_0} \nonumber \\
  & \hspace*{1cm} + D_{a_0} R D_{\bar{a}_0} + 2 D_{a_0 R D_{a_0}}).
\end{align}
Here, $D_x$ represents a diagonal matrix with the diagonal set to the vector $x$, and $\bar{x} = \vecnot{1} - x$ with $\vecnot{1}$ representing the vector whose entries are $1$.
We write $a = a_0 + 2a_1 + 4a_2 + \ldots, b = b_0 + 2b_1 + 4b_2 + \ldots \in \mathbb{Z}^n$ with $a_i,b_i \in \mathbb{Z}_2^n$.
With this notation, $b_0 + a_0 R$ is an integer sum and the definition of $E(a,b)$ has been suitably generalized to integer vectors $a,b$ as we did in~\cite{Rengaswamy-pra19}.
Note that $D(a,b)$ is unaffected by this generalization since it does not have an overall phase factor $\imath^{ab^T}$ and $X^2 = Z^2 = I_2$, i.e., $D(a,b) = D(a_0,b_0)$.
Whenever we only consider binary vectors $a,b$ we will replace $a_0$ in the above expressions with $a$, so then $\phi(R,a,b,\ell) = (1 - 2^{\ell-2}) a R a^T$.

Equation~\eqref{eq:tau_Eab} naturally extends the action of the Clifford group given in~\eqref{eq:cliff_action} to a large class of diagonal unitaries in the Clifford hierarchy.
In this case the symplectic matrix
$\Gamma_R = \begin{bmatrix}
I_n & R \\
0 & I_n
\end{bmatrix}$ is defined over $\MZ_{2^{\ell}}$ and still satisfies $\Gamma_R\, \Omega\, \Gamma_R^T = \Omega\ (\text{mod}\ 2)$.
The results of~\cite{Rengaswamy-pra19} also show that $\tau_R^{(\ell)} \in \mathcal{C}^{(\ell)}$, and that all $1$- and $2$-local diagonal unitaries in the Clifford hierarchy can be represented using an integer symmetric matrix $R$.
There are also many higher-locality diagonal gates that can be represented in this manner, and some examples are provided in~\cite{Rengaswamy-pra19}.
Henceforth, we will refer to these type of diagonal unitaries as \emph{Quadratic Form Diagonal} (QFD) gates.
%Here, $D_x$ represents a diagonal matrix with the diagonal set to the vector $x$, and $\bar{x} = \vecnot{1} - x$ with $\vecnot{1}$ representing the vector whose entries are $1$.
%We write $a = a_0 + 2a_1 + 4a_2 + \ldots, b = b_0 + 2b_1 + 4b_2 + \ldots \in \mathbb{Z}^n$ with $a_i,b_i \in \mathbb{Z}_2^n$.
%With this notation, $b_0 + a_0 R$ is an integer sum and the definition of $E(a,b)$ has been suitably generalized to integer vectors $a,b$ as we did in~\cite{Rengaswamy-pra19}.
%Note that $D(a,b)$ is unaffected by this generalization since it does not have an overall phase factor $\imath^{ab^T}$ and $X^2 = Z^2 = I_2$, i.e., $D(a,b) = D(a_0,b_0)$.
%Whenever we only consider binary vectors $a,b$ we will replace $a_0$ in the above expressions with $a$, so then $\phi(R,a,b,\ell) = (1 - 2^{\ell-2}) a R a^T$.

\begin{example}
\label{eg:T_gate}
\normalfont
Consider $n = 1, \ell = 3, \xi = e^{\imath\pi/4}, R = [\, 1 \, ]$ and hence $\tau_R^{(\ell)} = T$.
Since this is diagonal in the standard basis, it commutes with $Z = E(0,1)$.
For $E(1,0)$ with $a = 1$, $\phi(R,a,b,\ell) = (1 - 2) aRa^T = -1$ and $\tilde{R}(R,a,\ell) = (1 + 2) D_1 - (D_0 R D_1 + D_1 R D_0 + 2 D_1) = [\, 1 \, ]$.
Hence,~\eqref{eq:tau_Eab} implies $TXT^{\dagger} = \tau_R^{(3)} E(1,0) (\tau_R^{(3)})^{\dagger} = e^{-\imath\pi/4} E(1,1) \tau_{\tilde{R}(R,1,3)}^{(2)} = e^{-\imath\pi/4} Y P$.
It is easy to see that the phase gate can be expressed as $P = \frac{(I_2 + Z)}{2} + \imath \frac{(I_2 - Z)}{2}$.
Therefore,
\begin{align}
TXT^{\dagger} & = e^{-\imath\pi/4} \left[ \frac{Y + \imath X}{2} + \frac{\imath(Y - \imath X)}{2} \right] \nonumber \\
  & = \frac{e^{-\imath\pi/4}}{2} \left[ (1 + \imath) (Y + X) \right] \nonumber \\
  & = \frac{1}{2} \frac{(1 - \imath)}{\sqrt{2}} (1 + \imath) (Y + X) = \frac{X + Y}{\sqrt{2}},
\end{align}
which is a well-known identity cast in our framework.
We will use the identity $TXT^{\dagger} = e^{-\imath\pi/4} Y P$ extensively in this paper.
\end{example}

It will be convenient to expand $\tau_R^{(\ell)}$ in the Pauli basis. 
Using the observation we made earlier for a general diagonal unitary $U$, for $x \in \mathbb{Z}_2^n$ we define 
\begin{align}
\label{eq:tau_coeff}
c_{R,x}^{(\ell)} & \coloneqq \frac{1}{\sqrt{2^n}} \text{Tr}\left[ E(0,x) \tau_R^{(\ell)} \right] = \frac{1}{\sqrt{2^n}} \sum_{v \in \mathbb{Z}_2^n} (-1)^{vx^T} \xi^{vRv^T} \nonumber \\
\Rightarrow \tau_R^{(\ell)} & = \frac{1}{\sqrt{2^n}} \sum_{x \in \mathbb{Z}_2^n} c_{R,x}^{(\ell)} E(0,x).
\end{align}
Using this Pauli expansion for $\tau_{\tilde{R}(R,a,\ell)}^{(\ell-1)}$ in~\eqref{eq:tau_Eab}, and assuming $a,b \in \mathbb{Z}_2^n$, we get
\begin{align}
& \tau_R^{(\ell)} E(a,b) ( \tau_R^{(\ell)} )^{\dagger} \nonumber \\
  & = \xi^{\phi(R,a,b,\ell)} E(a, b + a R) \, \tau_{\tilde{R}(R,a,\ell)}^{(\ell-1)} \\
  & = \xi^{\phi(R,a,b,\ell)} E(a, b + a R) \cdot \frac{1}{\sqrt{2^n}} \sum_{x \in \mathbb{Z}_2^n} c_{\tilde{R}(R,a,\ell),x}^{(\ell-1)} E(0,x) \\
\label{eq:tau_Eab_expand}
  & = \frac{1}{\sqrt{2^n}} \xi^{\phi(R,a,b,\ell)} \sum_{x \in \mathbb{Z}_2^n} c_{\tilde{R}(R,a,\ell),x}^{(\ell-1)} \imath^{-ax^T} E(a, b + aR + x).
\end{align}
The primary challenge here is to determine which coefficients are non-zero for given $R,a,\ell$, and also their values.

\subsection{Stabilizer Codes}
\label{sec:stabilizer_codes}

A \emph{stabilizer group} $S$ is a commutative subgroup of the Pauli group $HW_N$ with Hermitian elements that does not include $-I_N$.
If $S$ has dimension $r$, then it can be generated as $S = \langle \nu_i E(c_i,d_i) ; i = 1,\ldots,r \rangle$, where $\nu_i \in \{ \pm 1 \}, c_i, d_i \in \mathbb{Z}_2^n$ and $E(c_i,d_i), E(c_j,d_j)$ commute for all $i \neq j, i,j \in \{1,\ldots,r\}$, i.e., $\syminn{[c_i,d_i]}{[c_j,d_j]} = c_i d_j^T + d_i c_j^T = 0$ (mod $2$).
Recollect that commuting $N \times N$ Hermitian matrices can be simultaneously diagonalized and hence have a common basis of eigenvectors that span $\mathbb{C}^N$.
Given a stabilizer group $S$, the corresponding \emph{stabilizer code}~\cite{Nielsen-2010} is the subspace $V(S)$ spanned by all eigenvectors in the common eigenbasis of $S$ that have eigenvalue $+1$, i.e., $V(S) \coloneqq \{ \ket{\psi} \in \mathbb{C}^N \colon g \ket{\psi} = \ket{\psi}\ \text{for\ all}\ g \in S \}$.
The subspace $V(S)$ is called an $\llbr n,k,d \rrbr$ stabilizer code because it encodes $k \coloneqq n-r$ \emph{logical} qubits into $n$ \emph{physical} qubits. 
The minimum distance $d$ is defined to be the minimum weight of any operator\footnote{Weight of a Pauli operator refers to the number of qubits on which it acts non-trivially, i.e., as $X, Y$ or $Z$.} in $\mathcal{N}_{HW_N}(S) \setminus S$.
Here, $\mathcal{N}_{HW_N}(S)$ denotes the normalizer of $S$ inside $HW_N$, %i.e., 
\begin{align}
\label{eq:normalizer_S}
\mathcal{N}_{HW_N}(S) & \coloneqq \{ \imath^{\kappa} E(a,b) \in HW_N \colon E(a,b) E(c,d) E(a,b) = \nonumber \\
  & \qquad E(c,d)\ \text{for\ all}\ \nu E(c,d) \in S,\ \kappa \in \mathbb{Z}_4 \}.
\end{align}

Given a Hermitian Pauli matrix $E(c,d)$ and $\nu \in \{ \pm \}$, it is easy to show that $\frac{I_N + \nu E(c,d)}{2}$ is the projector on to the $\nu$-eigenspace of $E(c,d)$.
Therefore, the projector on to the code subspace $V(S)$ of the stabilizer code defined by $S$ is given by
\begin{align}
\Pi_S \coloneqq \prod_{i=1}^{r} \frac{\left( I_N + \nu_i E(c_i,d_i) \right)}{2} = \frac{1}{2^r} \sum_{j = 1}^{2^r} \epsilon_j E(a_j, b_j),
\end{align}
where $\epsilon_j \in \{ \pm 1 \}$ in the last equality is a character of the group $S$, and hence is determined by the product of signs of the generators of $S$ that produce $E(a_j,b_j)$, i.e., $\epsilon_j E(a_j,b_j) = \prod_{t \in J \subseteq \{1,\ldots,r\}} \nu_t E(c_t,d_t)$ for a unique subset $J$.

A \emph{CSS (Calderbank-Shor-Steane) code} is a special type of stabilizer code defined by a stabilizer $S$ whose generators split into strictly $X$-type and strictly $Z$-type operators.
Consider two classical binary codes $C_1,C_2$ such that $C_2 \subset C_1$, and let $C_1^{\perp}, C_2^{\perp}$ represent their respective dual codes ($C_1^{\perp} \subset C_2^{\perp}$).
Define the stabilizer $S \coloneqq \langle \nu_c E(c,0), \nu_d E(0,d), c \in C_2, d \in C_1^{\perp} \rangle$ for some suitable $\nu_c, \nu_d \in \{ \pm 1 \}$.
Let $C_1$ be an $[n,k_1]$ code and $C_2$ be an $[n,k_2]$ code such that $C_1$ and $C_2^{\perp}$ can correct up to $t$ errors.
Then $S$ defines an $\llbr n, k_1-k_2, \geq 2t+1 \rrbr$ CSS code\footnote{We say the distance is at least $2t+1$ because distance of the code is the minimum weight of any vector in $(C_1 \setminus C_2) \cup (C_2^{\perp} \setminus C_1^{\perp})$, and not just $C_1 \cup C_2^{\perp}$.} that we will represent as CSS($X,C_2 ; Z, C_1^{\perp}$).
If $G_2 \in \mathbb{Z}_2^{k_2 \times n}$ and $G_1^{\perp} \in \mathbb{Z}_2^{(n-k_1) \times n}$ represent generator matrices for the codes $C_2$ and $C_1^{\perp}$, respectively, then a binary generator matrix for $S$ can be written as
\begin{align}
\setlength\aboverulesep{0pt}\setlength\belowrulesep{0pt}
    \setlength\cmidrulewidth{0.5pt}
G_S = 
\begin{blockarray}{ccc}
 n & n &   \\
\begin{block}{[c|c]c}
\hspace*{1cm} & G_1^{\perp} & n - k_1 \\
\cmidrule(lr){1-2}
G_2 & \hspace*{1cm} & k_2 \\
\end{block}
\end{blockarray}.
\end{align}

\section{Main Results and Discussion}
\label{sec:main_results}

Given a stabilizer code, we need to develop a scheme to perform fault-tolerant universal quantum computation on the logical qubits protected by the code, because otherwise the code can only be used as a quantum memory.
In~\cite{Rengaswamy-arxiv18}, we produced a systematic algorithm that can synthesize all (equivalence classes of) Clifford circuits on the physical qubits that realize a given logical Clifford operator (on the logical qubits).
We will refer to this as the Logical Clifford Synthesis (LCS) algorithm%
\footnote{Our implementation of the LCS algorithm, along with several general purpose subroutines, is available at: \url{https://github.com/nrenga/symplectic-arxiv18a}.}.
As mentioned before, for universal quantum computation, we also need to determine a way to synthesize circuits that realize at least one non-Clifford logical operator.
Moreover, the simplest example of a fault-tolerant circuit is a \emph{transversal} operator, that splits as a tensor product of individual single-qubit operators on the physical qubits of the code, since errors on individual qubit lines do not spread to other qubits.
Since fault-tolerant realizations of logical non-Clifford operators are harder to produce, we begin by asking a question that is motivated by the easiest non-Clifford operator to engineer.
%we begin by asking the following question that is practically motivated:

\begin{center}
What kind of stabilizer codes support a transversal operator composed of $T$ and $T^{\dagger}$ gates on the physical qubits?
\end{center}

In other words, what structure is needed in the stabilizer $S$ so that the code subspace $V(S)$ is preserved under the application of a given pattern of $T$ and $T^{\dagger}$ (and identity) gates on the physical qubits?
This reverses the strategy employed in the LCS algorithm, where we translated a logical operator to a physical operator. 
The above question is more practically motivated because single-qubit $Z$-rotations are some of the easiest examples of non-Clifford gates that can be performed in the lab, e.g., for trapped ion systems these gates are actually \emph{native} operations~\cite{Linke-nas17}, and we want to make the maximum use of them.
% Indeed, we also need to determine what logical operator the given pattern of $T$ and $T^{\dagger}$ gates realize, assuming the stabilizer has the necessary structure.
% We address this question as well for some codes and logical operators, especially the case when a transversal application of just the $T$ gate realizes the logical transversal $T$ on all the $k$ logical qubits encoded by a CSS($X,C_2 ; Z, C_1^{\perp}$) code.
Assuming the stabilizer has the necessary structure, we also need to determine what logical operator is realized by the given pattern of $T$ and $T^{\dagger}$ gates.
We also address this question for some codes and logical operators, e.g., we consider the case when a transversal application of the $T$ gate realizes the logical transversal $T$ on all the $k$ logical qubits encoded by a CSS($X,C_2 ; Z, C_1^{\perp}$) code.
This establishes a tight connection with \emph{triorthogonal codes} defined by Bravyi and Haah~\cite{Bravyi-pra12}. 
Subsequently, we provide a partial answer to the extension of the above question to $Z$-rotations above level $3$ of the Clifford hierarchy~\cite{Landahl-arxiv13,Haah-quantum17b,Campbell-pra17,Campbell-prl17,Vuillot-arxiv19}.
Finally, we produce a family of $\llbr 2^m, \binom{m}{r}, 2^r \rrbr$ quantum Reed-Muller codes, where $1 \leq r \leq m/2$ and $r$ divides $m$, and show that the transversal $\pi/2^{m/r}$ $Z$-rotation is a logical operator on these codes.
Furthermore, we also derive the exact logical operation realized by this transversal gate on these codes.

We begin by outlining the general strategy for understanding when a physical QFD gate preserves a stabilizer codespace.

\subsection{General Approach for QFD Gates}
\label{sec:general_QFD}

The key idea in addressing the above question is the following: a physical operator $U \in \mathbb{U}_N$ preserves the code subspace of a stabilizer code defined by a stabilizer group $S$ if and only if $U \Pi_S U^{\dagger} = \Pi_S$.
This is because two operators preserve each others' eigenspaces if and only if they commute.
Here, we say an operator $A$ preserves the eigenspace of another operator $B$ if it holds that for any eigenvector $v$ of $B$ with eigenvalue $b$, $Av$ is also an eigenvector of $B$ with eigenvalue $b$.
% This is because two operators possess the same eigenspace if and only if they commute.
% This can be seen as follows.
% Any such unitary $U$ must indeed satisfy $\Pi_S U \Pi_S = U \Pi_S$, which states that when $U$ acts on a code state the result should also be a code state, i.e., the left most $\Pi_S$ on the left hand side of the equation acts trivially as the identity.
% Rewriting this equality, we observe that
% \begin{align}
% U \Pi_S = \Pi_S U \Pi_S \Leftrightarrow \Pi_S = (U^{\dagger} \Pi_S U) \Pi_S \Leftrightarrow (I_N - U^{\dagger} \Pi_S U) \Pi_S = 0 \Leftrightarrow U \Pi_S = \Pi_S U.
% \end{align}
% The last implication is true because rank($U^{\dagger} \Pi_S U$) = rank($\Pi_S$) and $(I_N - U^{\dagger} \Pi_S U)$ is a projector on to the subspace orthogonal to the code subspace $V(S)$ defined by $S$.
Thus, $U$ is a valid logical operator for $S$ if and only if
\begin{align}
U \Pi_S U^{\dagger} = \frac{1}{2^r} \sum_{j = 1}^{2^r} \epsilon_j U E(a_j,b_j) U^{\dagger} & = \frac{1}{2^r} \sum_{j = 1}^{2^r} \epsilon_j E(a_j,b_j) \nonumber \\
  & = \Pi_S.
\end{align}
If $U = \tau_R^{(\ell)}$ for some $\ell \geq 2$ and $R$ symmetric over $\MZ_{2^\ell}$, then by~\eqref{eq:tau_Eab} we need
\begin{align}
\tau_R^{(\ell)} \Pi_S (\tau_R^{(\ell)})^{\dagger} & = \frac{1}{2^r} \sum_{j = 1}^{2^r} \epsilon_j \tau_R^{(\ell)} E(a_j,b_j) (\tau_R^{(\ell)})^{\dagger} \\
  & = \frac{1}{2^r} \sum_{j = 1}^{2^r} \epsilon_j \frac{1}{\sqrt{2^n}} \xi^{\phi(R,a_j,b_j,\ell)} \nonumber \\
  & \quad \sum_{x \in \Fn} c_{\tilde{R}(R,a_j,\ell),x}^{(\ell-1)} \imath^{-a_j x^T} E(a_j, b_j + a_j R + x) \\
  & = \frac{1}{2^r} \sum_{j = 1}^{2^r} \epsilon_j E(a_j,b_j).
\end{align}
This shows one important use of the formula~\eqref{eq:tau_Eab} derived in~\cite{Rengaswamy-pra19}.
As mentioned at the end of Section~\ref{sec:QFD_gates}, the primary problem here is to determine which coefficients are non-zero for given $R,a,\ell$, and also their values.
In principle, we can solve for all the conditions on $S$ that are necessary (and sufficient) for this equality. 
So, if we want to take advantage of an operation that we can do ``easily'' in the lab, then we can use the above approach to derive codes accordingly; if the operation is also a QFD gate, then we can exactly use the above equations. 
However, solving the above equality for arbitrary $R,\ell$ might be hard.
In this paper, we solve the equality completely when $\tau_R^{(3)}$ is a transversal combination of $T$ and $T^{\dagger}$ gates. 
We also provide a nearly complete solution for the transversal application of higher level $Z$-rotations.

\subsection{Stabilizer Codes that Support $T$ and $T^{\dagger}$ Gates}
\label{sec:stab_codes_T_Tinv}

We begin with a formula for the physical transversal $T$ gate which, given several applications, is of independent interest. %could be of interest on its own for several applications.

\begin{lemma}
\label{lem:conj_by_trans_T}
Let $E(a,b) \in HW_N, N = 2^n$, for some $a,b \in \mathbb{Z}_2^n$.
Then the transversal $T$ gate acts on $E(a,b)$ as
\begin{align}
T^{\otimes n} E(a,b) \left( T^{\otimes n} \right)^{\dagger} = \frac{1}{2^{w_H(a)/2}} \sum_{y \preceq a} (-1)^{b y^T} E(a, b \oplus y),
\end{align}
where $w_H(a) = aa^T$ is the Hamming weight of $a$, and $y \preceq a$ denotes that $y$ is contained in the support of $a$.
\begin{proof}
%See Appendix~\ref{sec:proof_conj_by_trans_T}.
This result is a special case of Lemma~\ref{lem:conj_by_trans_T_Tinv}, which we prove in Appendix~\ref{sec:proof_conj_by_trans_T_Tinv}.
\end{proof}
\end{lemma}

Using this lemma, we state our first result which partially answers the above question.

\begin{theorem}[Transversal $T$]
\label{thm:transversal_T}
% Let $S = \langle \nu_i E(c_i,d_i) ; i = 1,\ldots,r \rangle$ define an $\llbr n,n-r \rrbr$ stabilizer code, with arbitrary $\nu_i \in \{ \pm 1 \}$. % and $c_i, d_i \in \mathbb{Z}_2^n$ satisfying $\syminn{[c_i,d_i]}{[c_j,d_j]} = 0$ for any $i,j \in \{ 1,\ldots, r \}$.
% Then the transversal application of the $T$ gate realizes a logical operation on $V(S)$ if and only if the following are true.
% \begin{enumerate}

% \item For any $\epsilon_j E(a_j,b_j) \in S$ with non-zero $a_j$, $w_H(a_j)$ is even, where $w_H(a_j)$ represents the Hamming weight of $a_j \in \mathbb{Z}_2^n$.

% \item Let $Z_S \coloneqq \{ z \in \mathbb{Z}_2^n \colon \epsilon_z E(0,z) \in S\ \text{for\ some}\ \epsilon_z \in \{ \pm 1 \} \}$. 
%       For any $\epsilon_j E(a_j,b_j) \in S$ with non-zero $a_j$, $Z_S$ contains a dimension $w_H(a_j)/2$ self-dual code $A_j$ that is supported on $a_j$, i.e., there exists a subspace $A_j \subseteq \{ y \in Z_S \colon y \preceq a_j \}$ such that $y z^T = 0$ (mod 2) for any $y,z \in A_j$ (including $y = z$) and $\text{dim}(A_j) = w_H(a_j)/2$.
      
% \item For each $z \in \mathbb{Z}_2^n$ such that $z \in A_j$ for some $j \in \{ 1,\ldots,2^r \}$, $\imath^{w_H(z)} E(0,z) \in S$.      

% \end{enumerate} 
% 
Let $S = \langle \nu_i E(c_i,d_i) ; i = 1,\ldots,r \rangle$ define an $\llbr n,n-r,d \rrbr$ stabilizer code, with arbitrary $\nu_i \in \{ \pm 1 \}$, and denote the elements of $S$ by $\epsilon_j E(a_j, b_j), j = 1,2,\ldots,2^r$. % and $c_i, d_i \in \mathbb{Z}_2^n$ satisfying $\syminn{[c_i,d_i]}{[c_j,d_j]} = 0$ for any $i,j \in \{ 1,\ldots, r \}$.
If the transversal application of the $T$ gate  preserves the code space $V(S)$ and hence realizes a logical operation on $V(S)$, then the following are true.
\begin{enumerate}

\item For any $\epsilon_j E(a_j,b_j) \in S$ with non-zero $a_j$, $w_H(a_j)$ is even, where $w_H(a_j)$ represents the Hamming weight of $a_j \in \mathbb{Z}_2^n$.

\item  For any $\epsilon_j E(a_j,b_j) \in S$ with non-zero $a_j$, define $Z_j \coloneqq \{ z \preceq a_j \colon \epsilon_z E(0,z) \in S\ \text{for\ some}\ \epsilon_z \in \{ \pm 1 \} \}$% 
	  \footnote{Although using the above notation it is true that $\epsilon_z E(0,z) = \epsilon_{j'} E(a_{j'},b_{j'})$ for some $j' \in \{ 1,2,\ldots,2^r \}$ with $a_{j'} = 0$, we use the notation with $z$ for convenience and also because it actually refers to pure $Z$-type stabilizers.}.
      Then $Z_j$ contains its dual computed only on the support of $a_j$, i.e., on the ambient dimension $w_H(a_j)$.
      Equivalently, $Z_j$ contains a dimension $w_H(a_j)/2$ self-dual code $A_j$ that is supported on $a_j$, i.e., there exists a subspace $A_j \subseteq Z_j$ such that $y z^T = 0$ (mod 2) for any $y,z \in A_j$ (including $y = z$) and $\text{dim}(A_j) = w_H(a_j)/2$.
      
\item Let $\tilde{Z}_j \subseteq \mathbb{Z}_2^{w_H(a_j)}$ denote the subspace $Z_j$ where all positions outside the support of $a_j$ are punctured (dropped).
      Then, for each $z \in \mathbb{Z}_2^n$ such that $\tilde{z} \in (\tilde{Z}_j)^{\perp}$ for some $j \in \{ 1,\ldots,2^r \}$, we have $\epsilon_z = \imath^{zz^T}$, i.e., $\imath^{zz^T} E(0,z) \in S$.
      Here, $(\tilde{Z}_j)^{\perp}$ denotes the dual of $Z_j$ taken over this punctured space with ambient dimension $w_H(a_j)$.
      (Also, $Z_j \supseteq (\tilde{Z}_j)^{\perp}$ with zeros added outside the support of $a_j$.)

\end{enumerate} 
Conversely, if the first two conditions above are satisfied, and if the third condition holds for all $z \in A_j$ instead of just the dual of (the punctured) $Z_j$, then transversal $T$ preserves the code space $V(S)$ and hence induces a logical operation.
\begin{proof}
See Appendix~\ref{sec:proof_transversal_T}.
\end{proof}
\end{theorem}

\begin{remark}
\label{rem:transversal_T_Pauli_correction}
\normalfont
Note that, since $A_j$ is supported on $a_j$, the ambient dimension of vectors in $A_j$ is essentially $w_H(a_j)$.
So $A_j$ is a $[w_H(a_j), w_H(a_j)/2]$ self-dual code embedded in $\mathbb{Z}_2^n$.
The last point is requiring that the $Z$-stabilizers arising from vectors in the subspaces $A_j$ have the correct sign, given by $\imath^{w_H(z)} = \imath^{zz^T} \in \{ \pm 1 \}$.
If this is not taken care of, then an appropriate Pauli operator has to be applied before and after transversal $T$ in order to make a valid logical operator.
Indeed, this Pauli operator is essentially fixing the signs of the $Z$-stabilizers as required.
Hence, although the necessary and sufficient directions of the theorem differ in the last sign condition, for all practical purposes one can take the signs to be imposed on all of $A_j$ instead of just its subspace that is identified with $(\tilde{Z}_j)^{\perp}$.
These Pauli corrections preserve the code parameters $\llbr n,n-r,d \rrbr$.
\end{remark}

% \begin{remark}
% \label{rem:transversal_T_Pauli_correction}
% Owing to a subtlety in the proof of Theorem~\ref{thm:transversal_T}, it appears that there might sometimes be an overall Pauli application required before transversal $T$, even when a given code satisfies Theorem~\ref{thm:transversal_T} exactly.
% However, we believe this might not be necessary, and that this is a minor issue in the proof that can be fixed.
% Also, we are yet to observe any examples where this phenomenon occurs.
% \end{remark}

Let us now look at a simple example constructed using this theorem that will clarify the requirements above.

\begin{example}
\label{eg:cssT_622}
\normalfont
Define a $\llbr 6,2,2 \rrbr$ CSS code by the following stabilizer generator matrix.
\begin{align}
\setlength\aboverulesep{0pt}\setlength\belowrulesep{0pt}
    \setlength\cmidrulewidth{0.5pt}
G_S = 
\left[
\begin{array}{cccccc|cccccc}
1 & 1 & 1 & 1 & 1 & 1 & 0 & 0 & 0 & 0 & 0 & 0 \\
\hline 
0 & 0 & 0 & 0 & 0 & 0 & 1 & 1 & 0 & 0 & 0 & 0 \\
0 & 0 & 0 & 0 & 0 & 0 & 0 & 0 & 1 & 1 & 0 & 0 \\
0 & 0 & 0 & 0 & 0 & 0 & 0 & 0 & 0 & 0 & 1 & 1 \\
\end{array}
\right].
\end{align}
The right half of the last $3$ rows form the generators of $Z_S$ for this code.
Since there is only one non-trivial $a_j$ in this case, we see that $Z_S = A_1$ with $a_1 = [1,1,1,1,1,1]$.
Hence, the stabilizer generators are $X^{\otimes 6} = X_1 X_2 \cdots X_6, - Z_1 Z_2, - Z_3 Z_4, - Z_5 Z_6$, since the generators of $Z_S$ have weight $2$.
Multiplying $X^{\otimes 6}$ and the product of these three $Z$-stabilizers, we see that $Y^{\otimes 6} \in S$.
We can define the logical $X$ operators for this code to be $\bar{X}_1 = X_1 X_2, \bar{X}_2 = X_3 X_4$, since these are linearly independent and commute with all stabilizers.
Using the identity we observed in Example~\ref{eg:T_gate}, we see that 
\begin{align}
T^{\otimes 6} X_1 X_2 (T^{\otimes 6})^{\dagger} & = e^{-\imath \cdot 2\pi/4} (Y_1 P_1) (Y_2 P_2) \nonumber \\ % = -\imath \cdot (\imath X_1 Z_1 P_1) (\imath X_2 Z_2 P_2)
  & \equiv -\imath (X_1 X_2) (P_1 P_2),
\end{align}
since $-Z_1 Z_2 \in S$.
We observe that $(P_1 P_2) X^{\otimes 6} (P_1 P_2)^{\dagger} = Y_1 Y_2 X_3 X_4 X_5 X_6 \equiv X^{\otimes 6}$ up to the stabilizer $-Z_1 Z_2$, so $P_1 P_2$ indeed preserves $V(S)$.
But $(P_1 P_2) (X_1 X_2) (P_1 P_2)^{\dagger} = Y_1 Y_2 = (X_1 X_2) (-Z_1 Z_2) \equiv X_1 X_2$, and $P_1 P_2$ obviously commutes with $\bar{X}_2$, so $P_1 P_2$ is essentially the logical identity gate.
A similar reasoning holds for $P_3 P_4$.
Therefore, up to a global phase, the transversal $T$ preserves the logical operators $\bar{X}_1$ and $\bar{X}_2$, so in this case the transversal $T$ gate realizes just the logical identity (up to a global phase).
This can also be checked explicitly by writing the logical basis states $\ket{x_1 x_2}_L$ for $x_i \in \mathbb{Z}_2$:
\begin{align}
\ket{x_1 x_2}_L = \bar{X}_1^{x_1} \bar{X}_2^{x_2} \cdot \frac{1}{\sqrt{2}} \left( \ket{010101} + \ket{101010} \right).
\end{align}
If the $Z$-stabilizer generators were instead taken to be $Z_1 Z_2, Z_3 Z_4, Z_5 Z_6$, then the superposition above in $\ket{00}_L$ will be $(\ket{000000} + \ket{111111})$.
Therefore, $T^{\otimes 6} X_1 X_3 X_5$ will be a valid logical operator (that still implements the logical identity).

Given that $S$ has the necessary structure given by Theorem~\ref{thm:transversal_T}, note that we can freely add another $Z$-stabilizer generator that commutes with $X^{\otimes 6}$, e.g., $Z_1 Z_3 Z_4 Z_6 \leftrightarrow [1,0,1,1,0,1] \notin Z_S$.
This does not affect the transversal $T$ property: once $T^{\otimes n} \Pi_S (T^{\otimes n})^{\dagger} = \Pi_S$, the mapping $\Pi_S \mapsto \Pi_S \cdot \frac{(I_N + E(0,z))}{2}$ still preserves the equality since $(I_N + E(0,z))$ is diagonal.
\end{example}

Now we generalize Lemma~\ref{lem:conj_by_trans_T} and Theorem~\ref{thm:transversal_T} to $T$ and $T^{\dagger}$ gates, which addresses the initial question completely.

\begin{lemma}
\label{lem:conj_by_trans_T_Tinv}
For $N = 2^n$ and $a,b \in \mathbb{Z}_2^n$, let $E(a,b) \in HW_N$ and choose $t_1, t_7 \in \mathbb{Z}_2^n$ such that $t_1 \ast t_7 = 0$ (i.e., $\text{supp}(t_1)\, \cap\, \text{supp}(t_7) = \emptyset$).
Define $t = t_1 + 7 t_7 \in \{0,1,7\}^n , t' = t_1 + t_7 \in \mathbb{Z}_2^n$.
Then the physical operation $T^{\otimes t}$ acts on $E(a,b)$ as
\begin{align}
& T^{\otimes t} E(a,b) \left( T^{\otimes t} \right)^{\dagger} \nonumber \\
  & = \frac{1}{2^{w_H(a \ast t')/2}} \sum_{y \preceq (a \ast t')} (-1)^{(b + t_7) y^T} E(a, b \oplus y),
\end{align}
where $T^{\otimes t}$ denotes that $T$ (resp. $T^{\dagger} = T^7$) is applied to the qubits in the support of $t_1$ (resp. $t_7$).
\begin{proof}
See Appendix~\ref{sec:proof_conj_by_trans_T_Tinv}.
\end{proof}
\end{lemma}

\begin{corollary}
\label{cor:conj_by_trans_T_Tinv}
% Let $E(a,b) \in HW_N, N = 2^n$, for some $a,b \in \mathbb{Z}_2^n$, and let $t_j \in \mathbb{Z}_2^n, j = 1,2,\ldots,7$, such that $t_j \ast t_{j'} = 0$ (i.e., $\text{supp}(t_j) \cap \text{supp}(t_{j'}) = \emptyset$) for $j \neq j'$.
Let $E(a,b) \in HW_N$ for $N = 2^n$ and $a,b \in \mathbb{Z}_2^n$.
For $j,j' \in \{ 1,2,\ldots,7 \}$, let $t_j \in \mathbb{Z}_2^n$ and assume that $t_j \ast t_{j'} = 0$ (i.e., $\text{supp}(t_j) \cap \text{supp}(t_{j'}) = \emptyset$) for $j \neq j'$.
Define $t \coloneqq \sum_{j=1}^{7} j t_j \in \mathbb{Z}_8^n , \tilde{t}_1 \coloneqq t_1 + t_5, \tilde{t}_2 \coloneqq t_2 + t_6, \tilde{t}_3 \coloneqq t_3 + t_7 \in \mathbb{Z}_2^n$.
Then the physical operation $T^{\otimes t}$ acts on $E(a,b)$ as
\begin{align}
& T^{\otimes t} E(a,b) \left( T^{\otimes t} \right)^{\dagger} \nonumber \\
  & = \frac{ (-1)^{a (t_3 + t_4 + t_5 + t_6)^T} }{2^{w_H(a \ast (\tilde{t}_1 + \tilde{t}_3))/2}} \nonumber \\
  & \quad \sum_{(a \ast \tilde{t}_2) \preceq z \preceq (a \ast (\tilde{t}_1 + \tilde{t}_2 + \tilde{t}_3))}  (-1)^{(b + \tilde{t}_3) z^T} E\left( a, b \oplus z \right),
\end{align}
where $T^{\otimes t}$ denotes that $T^j$ is applied to the qubits in the support of $t_j$.
\begin{proof}
See Appendix~\ref{sec:proof_cor_conj_by_trans_T_Tinv}.
\end{proof}
\end{corollary}

% \begin{corollary}
% \label{cor:conj_by_trans_T_Tinv}
% Let $E(a,b) \in HW_N, N = 2^n$, for some $a,b \in \mathbb{Z}_2^n$, and let $\tilde{t}_j \in \mathbb{Z}_2^n, j = 1,2,\ldots,7$, such that $\tilde{t}_j \ast \tilde{t}_{j'} = 0$ (i.e., $\text{supp}(\tilde{t}_j) \cap \text{supp}(\tilde{t}_{j'}) = \emptyset$) for $j \neq j'$.
% Define $t \coloneqq \sum_{j=1}^{7} j \tilde{t}_j \in \mathbb{Z}_8^n , t_1 \coloneqq \tilde{t}_1 + \tilde{t}_5, t_2 \coloneqq \tilde{t}_2 + \tilde{t}_6, t_3 \coloneqq \tilde{t}_3 + \tilde{t}_7 \in \mathbb{Z}_2^n$.
% Then the physical operation $T^{\otimes t}$ acts on $E(a,b)$ as
% \begin{align}
% T^{\otimes t} E(a,b) \left( T^{\otimes t} \right)^{\dagger} = \frac{ (-1)^{a (\tilde{t}_3 + \tilde{t}_4 + \tilde{t}_5 + \tilde{t}_6)^T} }{2^{w_H(a \ast (t_1 + t_3))/2}} \sum_{(a \ast t_2) \preceq z \preceq (a \ast (t_1 + t_2 + t_3))}  (-1)^{(b + t_3) z^T} E\left( a, b \oplus z \right),
% \end{align}
% where $T^{\otimes t}$ denotes that $T^j$ is applied to the qubits in the support of $\tilde{t}_j$.
% \begin{proof}
% See Appendix~\ref{sec:proof_conj_by_trans_T_Tinv}.
% \end{proof}
% \end{corollary}

\begin{theorem}[Transversal $T(t)$]
\label{thm:transversal_T_Tinv}
Let $S = \langle \nu_i E(c_i,d_i) ; i = 1,\ldots,r \rangle$ define an $\llbr n,n-r,d \rrbr$ stabilizer code as in Theorem~\ref{thm:transversal_T}.
Let $t = t_1 + 7 t_7, t_1 \ast t_7 = 0,$ with supports of $t_1, t_7 \in \mathbb{Z}_2^n$ indicating the qubits on which $T$ and $T^{\dagger} = T^7$ are applied, respectively.
Define $t' = t_1 + t_7 \in \mathbb{Z}_2^n$.
% Then the application of the $T^{\otimes t}$ gate realizes a logical operation on $V(S)$ if and only if the following are true.
If the application of the $T^{\otimes t}$ gate realizes a logical operation on $V(S)$, then the following are true.
% \begin{enumerate}

% \item For any $\epsilon_j E(a_j,b_j) \in S$ with non-zero $a_j$, $w_H(a_j \ast t')$ is even.

% \item Let $Z_S \coloneqq \{ z \in \mathbb{Z}_2^n \colon \epsilon_z E(0,z) \in S \}$ for some $\{ \epsilon_z \}$. 
%       For any $\epsilon_j E(a_j,b_j) \in S$ with non-zero $a_j$, $Z_S$ contains a dimension $w_H(a_j \ast t')/2$ self-dual code $A_{j,t'}$ that is supported on $a_j \ast t'$, i.e., there exists a subspace $A_{j,t'} \subseteq \{ y \in Z_S \colon y \preceq (a_j \ast t') \}$ such that $y z^T = 0$ (mod 2) for any $y,z \in A_{j,t'}$ (including $y = z$) and $\text{dim}(A_{j,t'}) = w_H(a_j \ast t')/2$.
      
% \item For each $z \in \mathbb{Z}_2^n$ such that $z \in A_{j,t'}$ for some $j \in \{ 1,\ldots,2^r \}$, $\imath^{w_H(z) + 2  t_7 z^T} E(0,z) \in S$.      

% \end{enumerate} 
\begin{enumerate}

\item For any $\epsilon_j E(a_j,b_j) \in S$ with non-zero $a_j$, $w_H(a_j \ast t')$ is even, where $w_H(a_j \ast t')$ represents the Hamming weight of $(a_j \ast t') \in \mathbb{Z}_2^n$.

\item  For any $\epsilon_j E(a_j,b_j) \in S$ with non-zero $a_j$, define $Z_{j,t'} \coloneqq \{ z \preceq (a_j \ast t') \colon \epsilon_z E(0,z) \in S\ \text{for\ some}\ \epsilon_z \in \{ \pm 1 \} \}$. 
      Then $Z_{j,t'}$ contains its dual computed only on the support of $(a_j \ast t')$, i.e., on the ambient dimension $w_H(a_j \ast t')$.
      Equivalently, $Z_j$ contains a dimension $w_H(a_j \ast t')/2$ self-dual code $A_{j,t'}$ that is supported on $(a_j \ast t')$, i.e., there exists a subspace $A_{j,t'} \subseteq Z_{j,t'}$ such that $y z^T = 0$ (mod 2) for any $y,z \in A_{j,t'}$ (including $y = z$) and $\text{dim}(A_{j,t'}) = w_H(a_j \ast t')/2$.
      
\item Let $\tilde{Z}_{j,t'} \subseteq \mathbb{Z}_2^{w_H(a_j \ast t')}$ denote the subspace $Z_{j,t'}$ where all positions outside the support of $(a_j \ast t')$ are punctured (dropped).
      Then, for each $z \in \mathbb{Z}_2^n$ such that $\tilde{z} \in (\tilde{Z}_{j,t'})^{\perp}$ for some $j \in \{ 1,\ldots,2^r \}$, we have $\epsilon_z = \imath^{zz^T + 2t_7 z^T}$, i.e., $\imath^{zz^T + 2t_7 z^T} E(0,z) \in S$.
      Here, $(\tilde{Z}_{j,t'})^{\perp}$ denotes the dual of $Z_{j,t'}$ taken over this punctured space with ambient dimension $w_H(a_j \ast t')$.
      ($Z_{j,t'} \supseteq (\tilde{Z}_{j,t'})^{\perp}$ with zeros added outside the support of $(a_j \ast t')$.)

\end{enumerate} 
Conversely, if the first two conditions above are satisfied, and if the third condition holds for all $z \in A_{j,t'}$ instead of just the dual of (the punctured) $Z_{j,t'}$, then transversal $T$ preserves the code space $V(S)$ and hence induces a logical operation.
\begin{proof}
The proof is along the same lines as for Theorem~\ref{thm:transversal_T}, but adapted suitably to the general case in Lemma~\ref{lem:conj_by_trans_T_Tinv}.
\end{proof}
\end{theorem}

Notice that the above two results reduce to Lemma~\ref{lem:conj_by_trans_T} and Theorem~\ref{thm:transversal_T}, respectively, when $t_1 = [1,1,\ldots,1]$ and $t_j = [0,0,\ldots,0]$ for $j = 2,3,\ldots,7$.
The main difference is that in this general scenario, the conditions in Theorem~\ref{thm:transversal_T} are applied to the intersection of the support of $a_j$ and $(t_1 + t_7)$.
% For example, Theorem~\ref{thm:transversal_T_Tinv} implies that $T$ or $T^{\dagger}$ applied to a single qubit will never preserve the code subspace of any stabilizer code whose stabilizer is not composed of only $Z$-type operators, since then $w_H(a_j \ast (t_1 + t_7)) \in \{ 0,1 \}$ and there cannot be a dimension $1/2$ self-dual code supported on $1$ bit.

\begin{example*}[contd.]
\normalfont
Assume that now we want to apply $T$ and $T^{\dagger}$ according to $t_1 = [1,0,1,0,1,0]$ and $t_7 = [0,1,0,1,0,1]$, respectively.
Since $t' = t_1 + t_7 = [1,1,\ldots,1]$, $a_j \ast (t_1 + t_7) = a_j$ always and so the first two conditions of Theorem~\ref{thm:transversal_T_Tinv} reduce to the transversal $T$ case.
However, the last condition needs the sign for the $Z$-stabilizer generators to be $\imath^{2 + 2} = 1$, so we need to change the stabilizer to be $S = \langle X^{\otimes 6}, Z_1 Z_2, Z_3 Z_4, Z_5 Z_6 \rangle$.
Then the superposition for $\ket{00}_L$ will indeed be $(\ket{000000} + \ket{111111})$, and it is easy to verify that $T^{\otimes t}$ fixes the logical basis states $\ket{00}_L, \ket{01}_L, \ket{10}_L, \ket{11}_L$, so that it also realizes the logical identity.
\end{example*}

In principle, we can generalize Theorem~\ref{thm:transversal_T_Tinv} to the case of arbitrary powers of $T$ by using Corollary~\ref{cor:conj_by_trans_T_Tinv}.
However, the derivation is more complicated and the final conditions are not fully clear because the summation in Corollary~\ref{cor:conj_by_trans_T_Tinv} is over a coset and not a subspace as in Lemma~\ref{lem:conj_by_trans_T_Tinv}.
Hence, this generalization still remains open.

Using these results, we can refine the CSS construction to produce codes that support a desired pattern of $T$ and $T^{\dagger}$ gates.
Note that the first two conditions in Theorem~\ref{thm:transversal_T_Tinv} only depend on $(t_1 + t_7)$ and not individually on $t_1$ and $t_7$, i.e., on the union of their supports. 
Hence, any pattern of $T$ and $T^{\dagger}$ on the support of $(t_1 + t_7)$ will preserve the code subspace, up to an initial Pauli application that produces the right signs for the $Z$-stabilizers as prescribed by the last condition in Theorem~\ref{thm:transversal_T_Tinv}.

\begin{corollary}[CSS-T Codes]
\label{cor:css-t_codes}
Let $t_1, t_7 \in \mathbb{Z}_2^n$ be such that $t_1 \ast t_7 = 0$, and define $t = t_1 + 7 t_7, t' = t_1 + t_7$.
Consider a code CSS($X, C_2 ; Z, C_1^{\perp}$) with stabilizer $S$, such that $w_H(x \ast t')$ is even for all $x \in C_2$.
For each $x \in C_2$, let $C_1^{\perp}$ contain a dimension $w_H(x \ast t')/2$ self-dual code $A_{x,t'}$ supported on $(x \ast t')$. 
Moreover, for all $x \in C_2$ and for each $z \in A_{x,t'}$, let $\imath^{w_H(z) + 2 t_7 z^T} E(0,z) \in S$.
Notice that this means $(C_2 \ast t') \subset C_1^{\perp} \subset C_2^{\perp}$, since $(x \ast t') \in A_{x,t'}$.
Then $T^{\otimes t}$ is a valid logical operator for CSS($X, C_2 ; Z, C_1^{\perp}$).
If for all $x \in C_2$ and for all $z \in A_{x,t'}$ we have $t_7 z^T \equiv 0$ (mod $2$), then $T^{\otimes t'}$ (which is composed of only $T$ gates) is also a valid logical operator, as the sign constraints for $E(0,z)$ are now independent of $t_7$.
\end{corollary}

\begin{remark}
\normalfont
Intuitively, a CSS-T code (for transversal $T$) is determined by two classical codes $C_2 \subset C_1$ such that for every codeword $x \in C_2$, there exists a dimension $w_H(x)/2$ self-dual code in $C_1^{\perp}$ supported on $x$.
This also means that $C_1 \ast C_2 \subseteq C_1^{\perp}$ for the following reason.
For $a \in C_1, x \in C_2$, it is clear that $a$ is orthogonal to every vector in $C_1^{\perp}$.
In particular, $a$ is orthogonal to the self-dual code $C_x \subset C_1^{\perp}$ supported on $x$.
But, for any $z \in C_x$, we have $az^T = (a \ast x)z^T = 0$.
This means $a \ast x \in C_x \subset C_1^{\perp}$ since $C_x$ is self-dual.
We think that this observation might make it more convenient to derive some properties of CSS-T codes, since there is a good literature on the star product~\cite{Randriambololona-aagct15}.
\end{remark}

Corollary~\ref{cor:css-t_codes} also suggests that there might not be a significant advantage in working with general stabilizer codes, rather than just CSS codes, as far as $T$ and $T^{\dagger}$ gates are concerned.
This is because, by Theorem~\ref{thm:transversal_T_Tinv}, there is always a large asymmetry required between the number of stabilizer elements that have at least one $X$ (or $Y$) in them, and the number of purely $Z$-type stabilizer elements. 
Hence, altering the pure $X$-type stabilizers into $X,Y$-type stabilizers might not provide much gain, say, in terms of the distance of the code.
The next corollary confirms this intuition for non-degenerate stabilizer codes.

\begin{definition}
An $\llbr n,k,d \rrbr$ stabilizer code is non-degenerate if every stabilizer element has weight at least $d$.
\end{definition}

The general notion of degeneracy is that, for a distance $d$ non-degenerate code, there cannot be two weight at most $t = \lfloor \frac{d-1}{2} \rfloor$ errors that have the same action on the code space, i.e., their product produces a stabilizer (whose weight is at most $2t < d$).
This is what makes degeneracy a purely quantum phenomenon since on a classical linear code two weight $t$ errors cannot combine to produce a codeword on a $d \geq 2t + 1$ code.
Therefore, our definition directly states that all stabilizers have weight at least $d$, which can be seen to be equivalent to the above (cf.~\cite{Gottesman-arxiv98}).

Note that the toric and color codes are degenerate (CSS) codes because the weights of the stabilizer generators are fixed even when the lattice size is increased, i.e., code distance is increased.
However, codes such as quantum Reed-Muller codes are typically non-degenerate since each stabilizer generally has weight at least equal to the code distance.

\begin{corollary}[Sufficiency of CSS-T Codes]
\label{cor:csst_sufficient}
Consider an $\llbr n, k, d \rrbr$ non-degenerate stabilizer code generated by the matrix 
%\begin{align*}
$G_S = 
\begin{bmatrix}
 A & B \\ 
 C & 0 \\
 0 & D 
\end{bmatrix}$
%\end{align*}
that satisfies the transversal $T(t)$ property (Theorem~\ref{thm:transversal_T_Tinv}). 
Then the CSS code generated by 
$G_S = 
\begin{bmatrix}
 A & 0 \\ 
 C & 0 \\
 0 & D 
\end{bmatrix}$
has parameters $\llbr n, \geq k, \geq d \rrbr$ and also satisfies the transversal $T(t)$ property for the same $t \in \{0,1,7\}^n$.
\begin{proof}
See Appendix~\ref{sec:proof_csst_sufficient}.
\end{proof}
\end{corollary}

\begin{remark}[Degenerate Codes]
\normalfont
Observe that the arguments above can be extended to the case when the given stabilizer code is degenerate, but now the distance of the new CSS code constructed above is lower bounded only by the minimum weight of $D$, which can be strictly less than $d$.
More explicitly, the minimum weight of $\langle A,C \rangle^{\perp} \setminus D$ is still $d$ since this space is strictly outside the stabilizer but in the normalizer of the given stabilizer code.
So the distance of the CSS code mainly depends on the minimum weight of $D^{\perp} \setminus \langle A,C \rangle$ and the vector $z$ at the end can be assumed to be taken from this subspace.
As a result, such a $z$ with weight less than $d$ cannot also belong to $B^{\perp}$ since otherwise this would contradict the assumption that the given stabilizer code has distance $d$.
Therefore, under the assumption that for the given stabilizer code any vector $z \in D^{\perp} \setminus (\langle A,C \rangle \cup B^{\perp})$ has weight at least $d$, the above corollary can be extended to the degenerate case.
We leave the more general problem of addressing the full extension of the above corollary to the degenerate case for future work.
\end{remark}

Motivated by the above corollary, all our examples in this paper are CSS-T codes (including the $\llbr 6,2,2 \rrbr$ code in Example~\ref{eg:cssT_622}).
The $\llbr 6,2,2 \rrbr$ code is not just a corner case where the transversal $T$ gate realizes the logical identity.
The following result provides necessary and sufficient conditions for this to happen.

\begin{theorem}[Logical Identity]
\label{thm:logical_identity}
Let $S$ be the stabilizer for an $\llbr n,k,d \rrbr$ CSS-T code CSS($X, C_2 ; Z, C_1^{\perp}$).
Let the logical Pauli $X$ group be $\bar{X} = \langle E(x_i,0) ; i = 1,\ldots,k \rangle$. 
Then the transversal $T$ gate on the $n$ physical qubits realizes the logical identity operation if and only if the following are true.
\begin{enumerate}

\item For each $E(x,0) \in \bar{X}$, $\imath^{w_H(x)} E(0,x)$ must be a stabilizer.

\item For each $E(x,0) \in \bar{X}$ and $\epsilon_a E(a,0) \in S$, $\imath^{w_H(x \ast a)} E(0,x \ast a)$ must be a stabilizer.

\item For any two logical Paulis $E(x,0), E(y,0) \in \bar{X}$, $\imath^{w_H(x \ast y)} E(0, x \ast y)$ must be a stabilizer.

\item For any two $X$-type stabilizers $E(a,0), E(b,0) \in S$, $\imath^{w_H(a \ast b)} E(0, a \ast b)$ must be a stabilizer.

\end{enumerate}
\begin{proof}
See Appendix~\ref{sec:proof_logical_identity}.
We will see shortly that the last three conditions essentially constitute the property of \emph{triorthogonality} for the generator matrix $G_1$ for the classical binary code $C_1$.
See the proof for a more detailed argument.
\end{proof}
\end{theorem}

\subsection{Realizing Logical $T$ Gates with Transversal $T$}
\label{sec:logical_T}

\begin{figure*}
\begin{center}

\scalebox{0.93}{%
\begin{tikzpicture}

\node (Z) at (0,0) {$\{ 0 \}$};
\node (C2) at (0,1.5) {$C_2$};
\node (C1) at (0,3) {$C_1$};
\node (F2m) at (0,4.5) {$\mathbb{F}_2^{15}$};

\path[draw] (Z) -- (C2) node[midway,left] {{$4$}} -- (C1) node[midway,left] {{$1$}} -- (F2m) node[midway,left] {{$10$}};
%\path[draw,<->,black] (-0.4,0) -- (-0.4,3) node [midway,left] {$k_1$}; 
%\path[draw,<->,black] (0.4,0) -- (0.4,1.5) node [midway,right] {$k$}; 

\node[text width=3.5cm] at (2,2.75) {$=$ Punctured\\ \hspace*{0.35cm} RM($1,4$)};

\node[text width=3.5cm] at (2,1.3) {$=$ Simplex\\ \hspace*{0.4cm} code};

\node (Zp) at (3,0) {$\{ 0 \}$};
\node (C1p) at (3,1.5) {$C_1^{\perp}$};
\node (C2p) at (3,3) {$C_2^{\perp}$};
\node (F2m) at (3,4.5) {$\mathbb{F}_2^{15}$};

\path[draw] (Zp) -- (C1p) node[midway,left] {{$10$}} -- (C2p) node[midway,left] {{$1$}} -- (F2m) node[midway,left] {{$4$}};
%\path[draw,<->,black] (2.9,0) -- (2.9,3) node [midway,right] {$n - k_2$}; 
%\path[draw,<->,black] (0.4,0) -- (0.4,1.5) node [midway,right] {$k$}; 

\node[text width=3.5cm] at (5.1,2.8) {$=$ Hamming\\ \hspace*{0.5cm} code};

\node[text width=3.5cm] at (5.1,1.3) {$=$ Even weight\\ \hspace*{0.4cm} subcode};

\node at (11.5,2.5) {\small $\setlength\aboverulesep{0pt}\setlength\belowrulesep{0pt}
    \setlength\cmidrulewidth{0.5pt}
G_1^{\perp} \coloneqq 
\begin{blockarray}{cccccccccccccccc}
 \\
\begin{block}{[ccccccccccccccc]c}
% \\
1 & 0 & 1 & 0 & 1 & 0 & 1 & 0 & 1 & 0 & 1 & 0 & 1 & 0 & 1 & x_1 \\
0 & 1 & 1 & 0 & 0 & 1 & 1 & 0 & 0 & 1 & 1 & 0 & 0 & 1 & 1 & x_2 \\
0 & 0 & 0 & 1 & 1 & 1 & 1 & 0 & 0 & 0 & 0 & 1 & 1 & 1 & 1 & x_3 \\
0 & 0 & 0 & 0 & 0 & 0 & 0 & 1 & 1 & 1 & 1 & 1 & 1 & 1 & 1 & x_4 \\
\cmidrule(lr){1-15}
0 & 0 & 1 & 0 & 0 & 0 & 1 & 0 & 0 & 0 & 1 & 0 & 0 & 0 & 1 & x_1 x_2 \\
0 & 0 & 0 & 0 & 1 & 0 & 1 & 0 & 0 & 0 & 0 & 0 & 1 & 0 & 1 & x_1 x_3 \\
0 & 0 & 0 & 0 & 0 & 0 & 0 & 0 & 1 & 0 & 1 & 0 & 1 & 0 & 1 & x_1 x_4 \\
0 & 0 & 0 & 0 & 0 & 1 & 1 & 0 & 0 & 0 & 0 & 0 & 0 & 1 & 1 & x_2 x_3 \\
0 & 0 & 0 & 0 & 0 & 0 & 0 & 0 & 0 & 1 & 1 & 0 & 0 & 1 & 1 & x_2 x_4 \\
0 & 0 & 0 & 0 & 0 & 0 & 0 & 0 & 0 & 0 & 0 & 1 & 1 & 1 & 1 & x_3 x_4 \\
% \\
\end{block}
\end{blockarray}$};

\node[align=center] at (11.5,0) {The first $4$ rows of $G_1^{\perp}$ form $G_2$, so $C_2 \subset C_1^{\perp}$};

\end{tikzpicture}
}
\caption{\label{fig:RM15}The CSS($X, C_2 ; Z, C_1^{\perp}$) construction for the $\llbr 15,1,3 \rrbr$ quantum Reed-Muller code.}

\end{center}
\end{figure*}
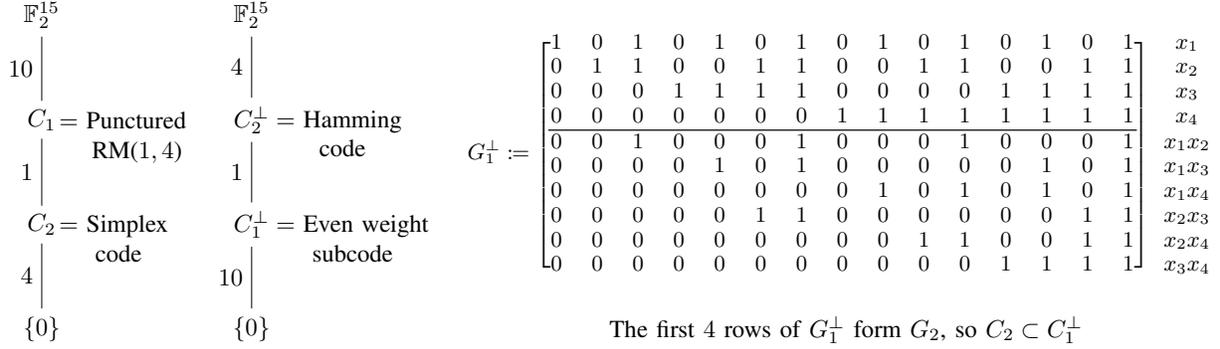

Let us begin by constructing the well-known $\llbr 15,1,3 \rrbr$ (punctured) quantum Reed-Muller code~\cite{Anderson-prl14,Quan-jpmt18} that supports a transversal $T$, using the conditions in Theorem~\ref{thm:transversal_T_Tinv}.
The construction is shown in Fig.~\ref{fig:RM15}.

The generator matrix $G_2$ for the simplex code $C_2$ that produces all $X$-type stabilizers (which are all weight $8$) is formed by the first $4$ rows of $G_1^{\perp}$, as shown in Fig.~\ref{fig:RM15}.
\setcounter{MaxMatrixCols}{20}
%\begin{align}
%G_2 \coloneqq
%\begin{bmatrix}
%1 & 0 & 1 & 0 & 1 & 0 & 1 & 0 & 1 & 0 & 1 & 0 & 1 & 0 & 1 \\
%0 & 1 & 1 & 0 & 0 & 1 & 1 & 0 & 0 & 1 & 1 & 0 & 0 & 1 & 1 \\
%0 & 0 & 0 & 1 & 1 & 1 & 1 & 0 & 0 & 0 & 0 & 1 & 1 & 1 & 1 \\
%0 & 0 & 0 & 0 & 0 & 0 & 0 & 1 & 1 & 1 & 1 & 1 & 1 & 1 & 1   
%\end{bmatrix}.
%\end{align}
Notice that this is obtained by \emph{shortening} RM($1,4$): take the generator matrix for the Reed-Muller code RM($1,4$), remove the first row of all $1$s, and then remove the first column which is all $0$s in the remaining matrix.
In other words, let $x_1,x_2,x_3,x_4$ be binary variables that also represent degree-$1$ monomials, with $x_1$ being the least significant bit and $x_4$ being the most significant bit.
Then, the rows of $G_2$ from top to bottom are $x_1,x_2,x_3,x_4$ respectively, with the first coordinate removed.
Similarly, since the dual of RM($1,4$) is the $[16,11,4]$ extended Hamming code RM($2,4$), the dual $C_1^{\perp}$ of the \emph{punctured} RM($1,4$) code $C_1$ is obtained by shortening RM($2,4$).
Therefore, the rows of $G_1^{\perp}$ must be the degree-$1$ monomials $x_1,x_2,x_3,x_4$ and the degree-$2$ monomials $x_i x_j$ for $i < j$, with the first coordinate removed.
%\begin{align}
%\setlength\aboverulesep{0pt}\setlength\belowrulesep{0pt}
%    \setlength\cmidrulewidth{0.5pt}
%G_1^{\perp} \coloneqq 
%\begin{blockarray}{cccccccccccccccc}
% \\
%\begin{block}{[ccccccccccccccc]c}
%1 & 0 & 1 & 0 & 1 & 0 & 1 & 0 & 1 & 0 & 1 & 0 & 1 & 0 & 1 & x_1 \\
%0 & 1 & 1 & 0 & 0 & 1 & 1 & 0 & 0 & 1 & 1 & 0 & 0 & 1 & 1 & x_2 \\
%0 & 0 & 0 & 1 & 1 & 1 & 1 & 0 & 0 & 0 & 0 & 1 & 1 & 1 & 1 & x_3 \\
%0 & 0 & 0 & 0 & 0 & 0 & 0 & 1 & 1 & 1 & 1 & 1 & 1 & 1 & 1 & x_4 \\
%\cmidrule(lr){1-15}
%0 & 0 & 1 & 0 & 0 & 0 & 1 & 0 & 0 & 0 & 1 & 0 & 0 & 0 & 1 & x_1 x_2 \\
%0 & 0 & 0 & 0 & 1 & 0 & 1 & 0 & 0 & 0 & 0 & 0 & 1 & 0 & 1 & x_1 x_3 \\
%0 & 0 & 0 & 0 & 0 & 0 & 0 & 0 & 1 & 0 & 1 & 0 & 1 & 0 & 1 & x_1 x_4 \\
%0 & 0 & 0 & 0 & 0 & 1 & 1 & 0 & 0 & 0 & 0 & 0 & 0 & 1 & 1 & x_2 x_3 \\
%0 & 0 & 0 & 0 & 0 & 0 & 0 & 0 & 0 & 1 & 1 & 0 & 0 & 1 & 1 & x_2 x_4 \\
%0 & 0 & 0 & 0 & 0 & 0 & 0 & 0 & 0 & 0 & 0 & 1 & 1 & 1 & 1 & x_3 x_4 \\
%\end{block}
%\end{blockarray}.
%\end{align}

Since all vectors in $C_2$ have weight $8$, the first condition of Theorem~\ref{thm:transversal_T} is satisfied.
Now consider, say, the $X$-stabilizer arising from the monomial $x_1$ belonging to $C_2$.
By direct observation of the rows of $G_1^{\perp}$, we see that the monomials $x_1, x_1 x_2, x_1 x_3$, and $x_1 x_4$ are linearly independent vectors contained in the support of $x_1$.
If we project these vectors onto just the support of $x_1$, i.e., drop the $x_1$ in the description of the monomials, then these $4$ vectors form the monomials $1, \tilde{x}_1, \tilde{x}_2, \tilde{x}_3$ in the space of $3$ binary variables.
By definition, these generate the Reed-Muller code RM($1,3$), which is also the $[8,4,4]$ extended Hamming code that is self-dual.
Since all other codewords in $C_2$ are also degree-$1$ polynomials, the same argument as above can be applied to them.
Therefore, the second condition of Theorem~\ref{thm:transversal_T} is satisfied as well.
Finally, since all generating codewords of $C_1^{\perp}$ have Hamming weight $4$ or $8$, the last condition of Theorem~\ref{thm:transversal_T} produces no negative signs for the $Z$-stabilizers.
Hence, the $\llbr 15,1,3 \rrbr$ quantum Reed-Muller code supports the transversal $T$ gate, and it can be checked that this realizes the logical $T^{\dagger}$ gate on the single encoded qubit.

We can also construct CSS codes where the physical transversal $T$ realizes logical transversal $T$.
In fact, \emph{triorthogonal codes} introduced by Bravyi and Haah~\cite{Bravyi-pra12} serve exactly this purpose, although they allow for an additional Clifford correction beyond the strict (physical) transversal $T$ operation.
As our next result, using our methods we show a ``converse'' that triorthogonality is not only sufficient but also necessary if we desire to realize logical transversal $T$ via (strict) physical transversal $T$ (using a CSS-T code).
We first repeat the definition of a triorthogonal matrix for clarity.

\begin{definition}[Triorthogonality~\cite{Bravyi-pra12}]
\label{def:triorthogonality}
A $p \times q$ binary matrix $G$ is said to be triorthogonal if and only if the support of any pair and triple of its rows has even overlap, i.e., $w_H(G_a \ast G_b) \equiv 0$ (mod $2$) for any two rows $G_a$ and $G_b$ for $1 \leq a < b \leq p$, and $w_H(G_a \ast G_b \ast G_c) \equiv 0$ (mod $2$) for all triples of rows $G_a, G_b, G_c$ for $1 \leq a < b < c \leq p$.
\end{definition}

Before we state our next result, let us clarify some unfortunate ambiguity in the terminology of triorthogonal codes.
First, a \emph{binary triorthogonal code} is one that has a generator matrix that satisfies the above triorthogonality property.
Second, a \emph{quantum CSS triorthogonal code} as defined by Bravyi and Haah starts with a binary triorthogonal code and adds constraints to some of its rows that are used to describe generators for the logical $X$ operators of the resulting CSS code.
This family of codes satisfy the property that when transversal $T$ is applied to the physical qubits of the triorthogonal code, along with possibly an additional Clifford operation (correction), then it induces a transversal $T$ on the logical qubits.
But, in the literature, sometimes the terminology is more casual where triorthogonal codes are described as codes that realize logical transversal $T$ via physical transversal $T$.
Finally, we note that the notion of triorthogonality is closely related to \emph{triply-even} codes, but they are distinct objects~\cite{Haah-pra18}.

\begin{theorem}[Logical Transversal $T$]
\label{thm:logical_trans_T}
Let $S$ be the stabilizer for an $\llbr n,k,d \rrbr$ CSS-T code CSS($X, C_2 ; Z, C_1^{\perp}$).
Let $G_1 = 
\begin{bmatrix}
G_{C_1/C_2} \\
G_2
\end{bmatrix}$ be a generator matrix for the classical code $C_1 \supset C_2$ such that the rows $x_i, i = 1,\ldots,k$, of $G_{C_1/C_2}$ form a generating set for the coset space $C_1/C_2$ that produces the logical $X$ group of the CSS-T code, i.e., $\bar{X} = \langle E(x_i,0) ; i = 1,\ldots,k \rangle$.
Then the physical transversal $T$ gate realizes the logical transversal $T$ gate, without any Clifford correction as in~\cite{Bravyi-pra12}, if and only if the matrix $G_1$ is triorthogonal and the following condition holds true:
\begin{align}
x & = \bigoplus_{i=1}^{k} c_i x_i,\ c_i \in \{0,1\} \nonumber \\
\Rightarrow w_H(x \oplus a) & \equiv w_H(c) \ (\bmod\ 8)\ \text{for\ all}\ a \in C_2.
\end{align}
\begin{proof}
See Appendix~\ref{sec:proof_logical_trans_T}.
\end{proof}
\end{theorem}

\begin{corollary}
\label{cor:triortho_general}
The triorthogonal construction introduced by Bravyi and Haah in~\cite{Bravyi-pra12} is the most general CSS-T family that realizes logical transversal $T$ from physical transversal $T$.
\begin{proof}
See Appendix~\ref{sec:proof_triortho_general}.
\end{proof}
\end{corollary}

%Therefore, the triorthogonal construction introduced by Bravyi and Haah for realizing logical transversal $T$ from physical transversal $T$ is essentially the most general CSS-T family that achieves this purpose.
While the Hamming weight condition above can be hard to check in practice, using the Bravyi-Haah recipe still implies that one has to calculate a final Clifford correction.
We suspect that CSS-T codes constructed using classical monomial codes, such as Reed-Muller codes or more general decreasing monomial codes~\cite{Bardet-isit16,Krishna-arxiv18}, might possess simple ways to check the Hamming weight condition above, since the weight distribution of some of these codes are known.

Observe that triorthogonality is a common condition for realizing either logical transversal $T$ or logical identity from physical transversal $T$, since the last three conditions in Theorem~\ref{thm:logical_identity} constitute the property of triorthogonality.
Indeed, if $\imath^{xy^T} E(0,x \ast y) \in S$ for any $x,y \in C_1/C_2$, then we need $xy^T \equiv 0$ (mod $2$) for $x \neq y$ and $w_H(a \ast (x \ast y)) \equiv 0, w_H(z \ast (x \ast y)) \equiv 0$ (mod $2$) for any $a \in C_2, z \in C_1/C_2, z \notin \{ x,y \}$, all because $E(0,x \ast y)$ needs to commute with $X$-type stabilizers and logical $X$ operators.
Similarly, the other conditions of triorthogonality can be derived from Theorem~\ref{thm:logical_identity}.

Therefore, the essential difference between transversal $T$ realizing logical transversal $T$ or logical identity is the following: for the former we need the Hamming weight condition above which in part implies $w_H(x_i) \equiv 1$ (mod $8$), while for the latter we need $\imath^{w_H(x)} E(0,x) \in S$ which implies $w_H(x) \equiv 0$ (mod $2$), and these are mutually contradictory.
Note that even if we permit a Clifford correction and omit the Hamming weight condition above, the proof of Theorem~\ref{thm:logical_trans_T} implies that the constraint $w_H(x_i) \equiv 1$ (mod $8$) is still necessary, so the contradiction remains.
Even in the Bravyi-Haah recipe, they impose that $w_H(x_i) \equiv 1$ (mod $2$).
We will construct a Reed-Muller family of CSS-T codes shortly, where we explicitly state a condition that differentiates between when the physical transversal $T$ realizes the logical identity and when it realizes some non-trivial logical operator.

Finally, with respect to the earlier discussion of degenerate codes, there appears to be limited examples of degenerate triorthogonal codes. 
Prominently, the 3D color codes are degenerate and realize logical $T$ via physical combinations of $T$ and $T^{-1}$ on different qubits.
Thus, these codes are triorthogonal (which permits diagonal Clifford corrections that can change $T$ to $T^{-1}$)~\cite{Watson-pra15}.

% Next we provide the conditions on an $\llbr n,k,d \rrbr$ CSS-T code for realizing a $T$ gate on, say, the $i$-th logical qubit via physical transversal $T$, when $k > 1$.
% If the automorphism group of the CSS-T code is \emph{transitive on the logical qubits}, then one can use the physical transversal $T$ to apply the $T$ gate on any other logical qubit as follows: first apply a permutation from the automorphism group on the physical qubits that brings the desired logical qubit to the position $i$, and then apply $T^{\otimes n}$.

% \begin{theorem}[Logical $T$]
% Let $S$ be the stabilizer for an $\llbr n,k,d \rrbr$ CSS-T code CSS($X, C_2 ; Z, C_1^{\perp}$).
% \end{theorem}

\subsection{Realizing Logical CC$Z$ via Transversal $T$}
\label{sec:logical_ccz}

The controlled-controlled-$Z$ (CC$Z$) gate is defined as the unitary $\text{CC}Z \coloneqq \text{diag}(1,1,1,1,1,1,1,-1)$, which is a $3$-qubit gate that applies the Pauli $Z$ operator on the third qubit if and only if the first two qubits are in state $\ket{1}$.
Similar to the C$Z$ gate, this unitary is symmetric with respect to all the three qubits involved in the operation.

\begin{example}[$\llbr 8,3,2 \rrbr$ Color Code]
\label{eg:Campbell}
\normalfont
First we revisit the construction of the $\llbr 8,3,2 \rrbr$ color code of Campbell~\cite{Campbell-blog16}, since it is now well-known, and show how it satisfies Theorem~\ref{thm:transversal_T}.
The code can be defined by considering the $8$ physical qubits to be the vertices of a cube.
There is a single $X$-type stabilizer generator that is defined by $X$ on all the vertices.
There are $4$ independent $Z$-type generators that are defined by $Z$ on the vertices of ($4$ independent) faces of the cube.
So the $X$-type stabilizers come from the $[8,1,8]$ classical repetition code, which can be written as the Reed-Muller code RM($0,3$).
It is easy to verify that the $Z$-type stabilizers come from the $[8,4,4]$ extended Hamming code, which is also the self-dual Reed-Muller code RM($1,3$).
So, there are $14$ elements of weight $4$, one element of weight $8$, and the identity.
By appropriately defining the logical $X$ strings from faces of the cube, it can be shown that transversal $T$ realizes logical CC$Z$ on this code.
This code is also a special case of Theorem~\ref{thm:QRM_family} for $m = 3, r = 1$, which generalizes to any $m$ (and $r = 1$) by the conditions of the theorem.
% Thus, this is a family of $\llbr 2^m, m, 2 \rrbr$ codes defined on $m$-dimensional cubes similar to the 3D code above.
Thus, this is a family of $\llbr 2^m, m, 2 \rrbr$ CSS($X, C_2 ; Z, C_1^{\perp}$) codes defined on $m$-dimensional cubes as $C_2 = \text{RM}(0,m), C_1 = \text{RM}(1,m)$, similar to the 3D code above.
\end{example}

\begin{example}[$\llbr 16,3,2 \rrbr$ Bacon-Shor-like Code]
\normalfont
Now we construct a $\llbr 16,3,2 \rrbr$ Bacon-Shor-like code using the conditions of Theorem~\ref{thm:transversal_T} and show that the transversal $T$ realizes the logical CC$Z$ gate (up to Paulis).  
In particular, this code belongs to the compass code family studied in \cite{Li-prx19}.
Although the $\llbr 8,3,2 \rrbr$ code is smaller while having essentially the same properties, we will demonstrate shortly that the $\llbr 16,3,2 \rrbr$ code can be constructed from \emph{decreasing monomial codes}~\cite{Bardet-isit16,Bardet-arxiv16}.
While this framework has been recently used by Krishna and Tillich~\cite{Krishna-arxiv18} to construct triorthogonal codes from punctured polar codes, this example has non-identical logical $X$ and $Z$ generators unlike the standard presentation of triorthogonal codes~\cite{Bravyi-pra12}.
%Therefore, this points to a more general use of decreasing monomial codes to construct CSS codes with transversal $Z$-rotations.

\begin{figure}
% \begin{center}

\centering

\includegraphics[scale=0.1,keepaspectratio]{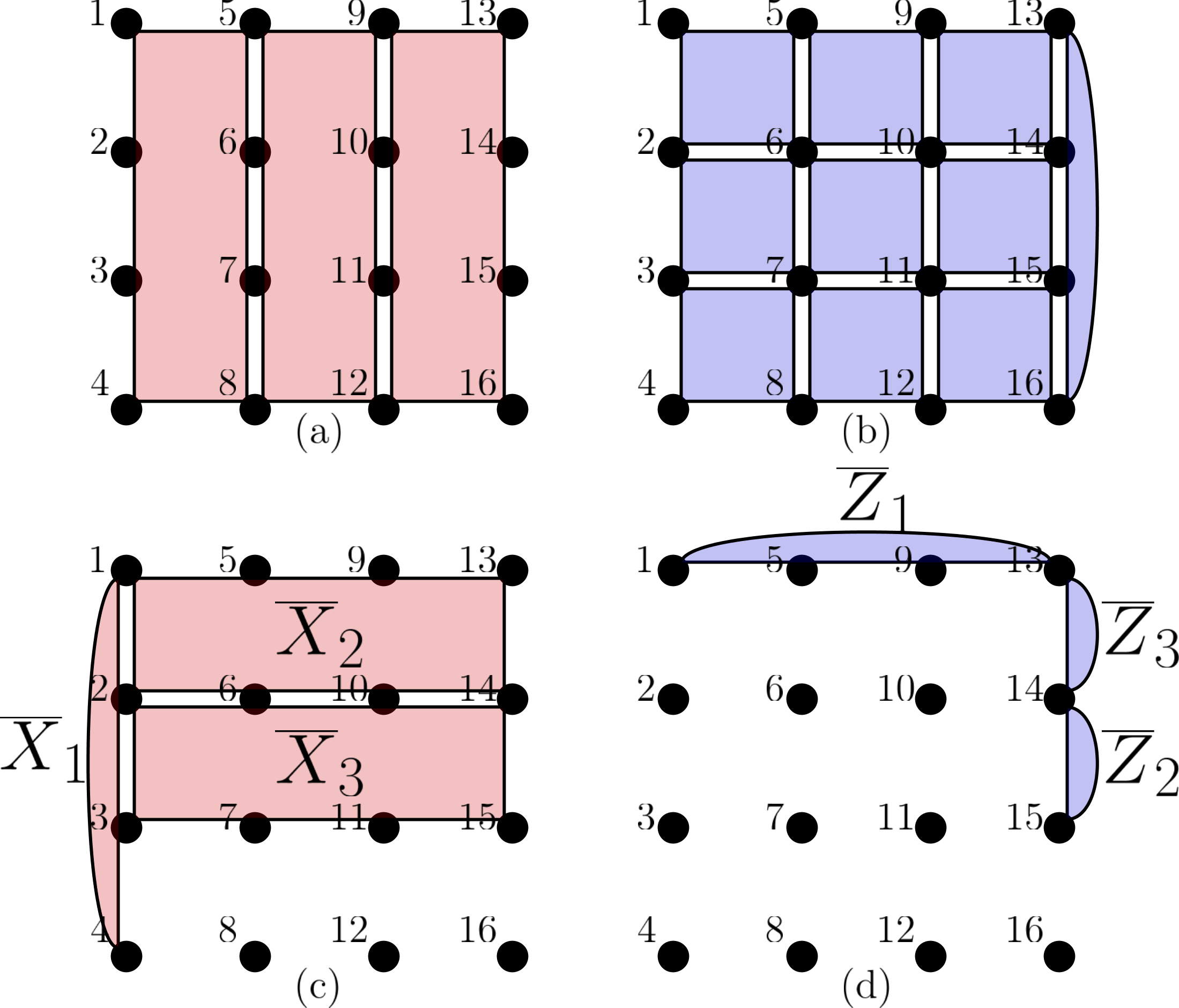}

\caption{\label{fig:CSST_16_3_2}A $\llbr 16,3,2 \rrbr$ CSS-T code where transversal $T$ realizes logical CC$Z$ (up to logical Paulis). (a) The $3$ weight-$8$ $X$-type stabilizer generators. (b) The $10$ weight-$4$ $Z$-type stabilizer generators. (c) The $3$ $X$-type logical Pauli generators. (d) The corresponding $3$ $Z$-type logical Pauli generators. The red and blue colors are included to differentiate between $X$-type and $Z$-type operators, respectively, for visual clarity.}

% \end{center}
\end{figure}

The construction of the $\llbr 16,3,2 \rrbr$ CSS-T code CSS($X, C_2 ; Z, C_1^{\perp}$) is shown in Fig.~\ref{fig:CSST_16_3_2}.
The code $C_2$ is generated by three weight-$8$ vectors that are represented as the vertical rectangles in Fig.~\ref{fig:CSST_16_3_2}(a).
It is easy to check that all but one non-zero vector in $C_2$ are weight-$8$ and there is one vector which is all $1$s.
The code $C_1^{\perp}$ is generated by $10$ weight-$4$ vectors, $9$ of which are represented as plaquette operators and the last one corresponds to the vertical string $Z_5 Z_6 Z_7 Z_8$ (or $Z_{13} Z_{14} Z_{15} Z_{16}$ in Fig.~\ref{fig:CSST_16_3_2}(b)).
Consider the first $X$-type generator $X_1 X_2 \cdots X_8$.
In its support, there are $3$ plaquette weight-$4$ strings and $1$ vertical weight-$4$ string, all of which are linearly independent and have mutually even overlap,
Hence, these clearly form a self-dual code which is in fact the $[8,4,4]$ extended Hamming code.
This can be checked for all the other vectors in $C_2$, so the first two conditions of Theorem~\ref{thm:transversal_T} are satisfied.
The last condition imposes no negative signs to the $Z$-type stabilizers since all of them have weight at least $4$.
Therefore, transversal $T$ preserves the code subspace.

To see that the realized logical operator is CC$Z$, consider the action of $T^{\otimes 16}$ on $\bar{X}_1 = X_1 X_2 X_3 X_4$.
Recollect that CC$Z$ on qubits $a,b,c$ maps $X_a \mapsto X_a\, \text{CZ}_{bc}, X_b \mapsto X_b\, \text{CZ}_{ac}, X_c \mapsto X_c\, \text{CZ}_{ab}$, and C$Z$ on qubits $e,f$ maps $X_e \mapsto X_e Z_f, X_f \mapsto X_f Z_e$.
%We observe that
\begin{align}
T^{\otimes 16} \bar{X}_1 \left( T^{\otimes 16} \right)^{\dagger} & = e^{-\frac{\imath\pi}{4} \cdot 4} (Y_1 Y_2 Y_3 Y_4) (P_1 P_2 P_3 P_4) \nonumber \\
  & = - \bar{X}_1 (P_1^{\dagger} P_2^{\dagger} P_3^{\dagger} P_4^{\dagger}).
\end{align}
We need to show that $U \coloneqq P_1^{\dagger} P_2^{\dagger} P_3^{\dagger} P_4^{\dagger} \equiv \bar{\text{CZ}}_{23}$.
Recollecting that $P^{\dagger} X P = -Y$, we notice that 
\begin{align}
U \bar{X}_2 U^{\dagger} & = Y_1 Y_2 X_5 X_6 X_9 X_{10} X_{13} X_{14} % = - \bar{X}_2 (Z_1 Z_2) 
\equiv - \bar{X}_2 \bar{Z}_3, \\
U \bar{X}_3 U^{\dagger} & = Y_2 Y_3 X_6 X_7 X_{10} X_{11} X_{14} X_{15} % = - \bar{X}_3 (Z_2 Z_3) 
\equiv - \bar{X}_3 \bar{Z}_2,
\end{align}
since $Z_1 Z_2 Z_{13} Z_{14}, Z_2 Z_3 Z_{14} Z_{15} \in S$.
Thus, up to signs, we have verified that $P_1^{\dagger} P_2^{\dagger} P_3^{\dagger} P_4^{\dagger} \equiv \bar{\text{CZ}}_{23}$.
Hence, $T^{\otimes 16}$ acts like logical CC$Z$ on $\bar{X}_1$, and similar calculations can be done to verify the other relations for CC$Z$.
In this case, the signs can be fixed by checking the relations for $(\bar{X}_2 \bar{X}_3)\, \bar{\text{CC}Z}\, (\bar{X}_2 \bar{X}_3)$, since this is the precise logical operator realized by $T^{\otimes 16}$.

% We checked this by explicitly calculating the superposition states for the $8$ logical computational basis states (on a computer), and observing the action of $T^{\otimes 16}$ on them.
% The effective logical diagonal gate is $\text{diag}(1,1,1,1,-1,1,1,1) = (X_2 X_3)\, \text{CC}Z\, (X_2 X_3)$, where we take the computational basis states to be $\ket{x_1 x_2 x_3}_L$ with $\ket{x_i}_L \in \{ \ket{0}_L, \ket{1}_L \}$ having flip operator $\bar{X}_i$ in Fig.~\ref{fig:CSST_16_3_2}.
% Another way to see this directly is as follows.
% The state $\ket{000}_L = \frac{1}{\sqrt{8}} \sum_{c \in C_2} \ket{c}$ and we know that there are $6$ weight-$8$ vectors, one all $0$s vector and one all $1$s vector in $C_2$.
% Therefore, the superposition for $\ket{100}_L = \bar{X}_1 \ket{000}_L$ can be directly observed to have vectors of weight $4$ modulo $8$, and hence transversal $T$ picks up a $(-1)$ when acting on $\ket{100}_L$.
\end{example}

\begin{example}[$\llbr 16,3,2 \rrbr$ Decreasing Monomial Code]
\label{eg:decreasing_monomial}
\normalfont
An equivalent $\llbr 16,3,2 \rrbr$ code can be constructed as a decreasing monomial code as follows, using the monomial description of Reed-Muller codes we discuss in Appendix~\ref{sec:rm_codes}.
Define the code $C_2$ as the space generated by the monomials $G_2 = \{1, x_1, x_2\}$, and the code $C_1$ as the space generated by $G_1 = G_2 \cup \{ x_3, x_4, x_1 x_2 \}$.
Hence, the logical $X$ group is generated by $G_X = \{ x_3, x_4, x_1 x_2 \}$.
While Reed-Muller codes always include all monomials up to some degree as the generators, decreasing monomial codes with maximum degree $r$ might include only some of the degree $r$ monomials among the generators.
However, the code must include all monomials of degree up to $r-1$, and the degree $r$ terms must be chosen according to a partial order as described in~\cite{Bardet-arxiv16}.
In the construction above, both $C_2$ and $C_1$ are decreasing monomial codes.
Using the formalism in~\cite{Bardet-arxiv16}, it is easy to see that the dual codes $C_1^{\perp}$ and $C_2^{\perp}$ are generated respectively by $G_1^{\perp} = \{ 1,x_1,x_2,x_3,x_4,x_1 x_2,x_1 x_3,x_1 x_4,x_2 x_3,x_2 x_4 \}$ and $G_2^{\perp} = G_1^{\perp} \cup \{ x_3 x_4,x_1 x_2 x_3,x_1 x_2 x_4 \}$.
So the logical $Z$ group is generated by $G_Z = \{ x_1 x_2 x_4, x_1 x_2 x_3, x_3 x_4 \}$, where we have rewritten the generators in an order such that they form corresponding pairs with logical $X$ generators in $G_X$.
In other words, we see that the corresponding entries in $G_X$ and $G_Z$ multiply to the full monomial $x_1 x_2 x_3 x_4$, which is the only monomial whose evaluation has odd weight, and hence the pairs anti-commute as required.
Similarly, multiplying terms from $G_X$ and $G_Z$ that are not pairs does not yield the full monomial, thereby ensuring they have even overlap.
Finally, we see that the product of the three logical $X$ generators produces the full monomial, which means their triple product has odd weight.
This is precisely one of the requirements in the \emph{generalized triorthogonality} conditions established by Haah and Hastings~\cite{Haah-quantum17b}, which is a special case of \emph{quasitransversality} established earlier by Campbell and Howard~\cite{Campbell-pra17}, in order to ensure that transversal $T$ performs a logical CC$Z$ on a CSS code.
The other requirements in their conditions can also be quickly verified simply using the fact that the only monomial of odd weight is the full monomial.

To see that this also satisfies Theorem~\ref{thm:transversal_T}, consider for example the $X$-stabilizer corresponding to the monomial $x_1 \in G_2$. 
We observe that the elements $x_1, x_1 x_2, x_1 x_3, x_1 x_4 \in G_1^{\perp}$ are supported on $x_1$.
When we project down to $x_1$, i.e., consider only the support of $x_1$, we get the monomials $1, \tilde{x}_1 = x_2, \tilde{x}_2 = x_3, \tilde{x}_3 = x_4$ that precisely generate the self-dual code RM($1,3$).
A similar analysis can be made for other elements in $C_2$.
Moreover, since the elements in $G_1^{\perp}$ have weights $4,8$, or $16$, the last condition of Theorem~\ref{thm:transversal_T} does not introduce any negative signs for the $Z$-stabilizers.
Therefore, we have used the decreasing monomial codes formalism to produce an equivalent $\llbr 16,3,2 \rrbr$ code, where only the logical $X$ and $Z$ generators have changed in comparison to the construction above.
We believe this is not just one special case but points to a general construction of CSS codes using this formalism that support transversal $Z$-rotations.
\end{example}

% Next we generalize this example by providing conditions for a CSS-T code such that transversal $T$ realizes the logical CC$Z$ gate on the desired $3$ logical qubits.

% \begin{theorem}[Logical CC$Z$]
% Let $S$ be the stabilizer for an $\llbr n,k,d \rrbr$ CSS-T code CSS($X, C_2 ; Z, C_1^{\perp}$).
% \end{theorem}

\subsection{Realizing Products of $\mathcal{C}^{(3)}$ gates with Transversal $T$}
\label{sec:product_of_CCZs}

We demonstrate two examples where transversal $T$ realizes a logical diagonal gate at the $3$rd level that is not a single elementary gate but a product of elementary gates.
These codes have been partially discussed in recent works~\cite{Haah-quantum17b,Campbell-pra17} but we describe the general family and later derive the exact logical operation realized by transversal $T$ on these codes.

\begin{example}[$\llbr 64,15,4 \rrbr$ Reed-Muller Code]
\normalfont
Consider a CSS($X, C_2 ; Z, C_1^{\perp}$) code where $C_2 = \text{RM}(1,6) \subset C_1 = \text{RM}(2,6)$ and therefore $C_1^{\perp} = \text{RM}(3,6) \subset C_2^{\perp} = \text{RM}(4,6)$.
The distance of this code is the minimum of the minimum distances of $C_1$ and $C_2^{\perp}$.
It is well-known that the minimum distance of RM($r,m$) is $2^{m-r}$, so the distance of this CSS-T code is $4$.
Therefore, this gives a $\llbr 64,15,4 \rrbr$ code.
Let us quickly check the conditions in Theorem~\ref{thm:transversal_T}.
The code $C_2$ is generated by degree-$1$ monomials in $6$ binary variables $x_1,x_2,\ldots,x_6$, and all of its codewords have even weight.
Using the same strategy that we used for the $\llbr 15,1,3 \rrbr$ RM code, consider the monomial $x_1$ in $C_2$.
Since $C_1^{\perp}$ is generated by all monomials of degree less than or equal to $3$, it contains the monomials $x_1 (1), x_1 (x_i), x_1 (x_i x_j)$ for $i,j \in \{ 2,3,4,5,6 \}$ and $i < j$.
If we project down to $x_1$, then the monomials $1, \tilde{x}_i \tilde{x}_i \tilde{x}_j$ for $i,j \in \{1,2,3,4,5\}$ and $i < j$ exactly generate the code RM($2,5$) which is self-dual.
A similar analysis holds for all the other codewords in $C_2$ which are degree-$1$ polynomials as well.
Finally, the generators of $C_1^{\perp}$ have Hamming weights $8,16,32$ and $64$, so there are no negative signs introduced by the last condition of Theorem~\ref{thm:transversal_T}. 
Thus, this is indeed a CSS-T code.

In order to determine the logical operation realized by transversal $T$, we initially wrote a computer program to generate the vectors in the superposition of all the $2^{15}$ logical computational basis states. 
(Later, in Theorem~\ref{thm:QRM_family}, we derive the exact logical operation analytically.)
Then we calculated the action of $T^{\otimes 64}$ on them by just computing the Hamming weights of the vectors in the superposition.
The effective logical operation was a diagonal unitary with entries $\pm 1$, and there were $13,888$ entries that were $(-1)$ in the diagonal (out of the $2^{15} = 32,768$).
We determined the Boolean function encoding the locations of $1$ and $-1$ in the diagonal, and simplified the naive $13,888$ term sum-of-products (SOP) expression into a $1991$ term SOP expression using the software ``Logic Friday''.
Subsequently, we used ``Mathematica'' to convert this Boolean function into its \emph{algebraic normal form (ANF)} and obtained the following polynomial.
\begin{align}
\label{eq:QRM26_poly}
q(v_1,\ldots,v_{15}) & = v_1 v_{10} v_{15} + v_1 v_{11} v_{14} + v_1 v_{12} v_{13}  \nonumber \\
                     & \quad + v_2 v_7 v_{15} + v_2 v_8 v_{14} + v_2 v_9 v_{13} \nonumber \\
                     & \quad + v_3 v_6 v_{15} + v_3 v_8 v_{12} + v_3 v_9 v_{11}  \nonumber \\ 
                     & \quad + v_4 v_6 v_{14} + v_4 v_7 v_{12} + v_4 v_9 v_{10}  \nonumber \\ 
                     & \quad + v_5 v_6 v_{13} + v_5 v_7 v_{11} + v_5 v_8 v_{10}.
\end{align}
Therefore, the logical diagonal gate can be represented as $U^L \ket{v_1 \cdots v_{15}}_L = (-1)^{q(v_1,\ldots,v_{15})} \ket{v_1 \cdots v_{15}}_L$.
This implies that the gate decomposes into exactly $15$ CC$Z$ gates on the logical qubits, and hence belongs to the $3$rd level of the Clifford hierarchy.
More interestingly, note that the CSS superposition of a given logical computational basis state $\ket{v_1 \cdots v_{15}}_L$ consists of all vectors in the corresponding coset of $C_2$ in $C_1$ generated by $\prod_{i=1}^{15} \bar{X}_i^{v_i}$, i.e., the binary vector representations $x_i$ of the logical operators $\bar{X}_i = E(x_i,0)$.
Therefore, the diagonal of $U^L$ encodes exactly which cosets of RM($1,6$) in RM($2,6$) have all vectors of weight exactly $4$ mod $8$ (diagonal entry $-1$) and which cosets have all vectors of weight $0$ mod $8$ (diagonal entry $1$).
Since the above phase polynomial is a codeword in RM($3,15$) of degree $3$, this suggests a deeper connection to RM codes where the coset weight distribution modulo $8$ is encoded exactly by a codeword in RM($3,15$).
Theorem~\ref{thm:QRM_family} explores this connection more rigorously than the empirical approach described above.
\end{example}

\begin{example}[$\llbr 128,21,4 \rrbr$ Reed-Muller Code]
\normalfont
Similarly, we constructed a $\llbr 128,21,4 \rrbr$ Reed-Muller CSS-T code by setting $C_2 = \text{RM}(1,7) \subset C_1 = \text{RM}(2,7)$, and hence $C_1^{\perp} = \text{RM}(4,7) \subset C_2^{\perp} = \text{RM}(5,7)$.
The $X$-stabilizers are generated by degree-$1$ monomials, and logical $X$ operators are given by the coset representatives for $C_1/C_2$, which are degree-$2$ polynomials.
This implies, for any $a,b \in C_2$ and $x,y \in C_1/C_2$, $a \ast b, a \ast x, x \ast y$ are either degree $2,3$, or $4$ polynomials, all of which belong to $C_1^{\perp}$ by the definition of RM codes.
So $C_1$ is a triorthogonal code that also satisfies the first condition of Theorem~\ref{thm:logical_identity}, and hence transversal $T$ realizes the logical identity.
However, the code supports the application of the $T$ gate on the physical qubits corresponding to the pattern prescribed by any degree-$1$ polynomial, as can be verified from the conditions in Theorem~\ref{thm:transversal_T_Tinv} by setting $t_7 = 0$ and $t_1$ a degree-$1$ polynomial.
%For the pattern given by the monomial $x_1$, i.e., $t = t_1 = 010101 \cdots 0101$, we verified using a computer program that $T^{\otimes t}$ realizes a logical diagonal gate from the $21$st level of the Clifford hierarchy.
Although this example does not fit in Theorem~\ref{thm:QRM_family} that concerns strictly with transversal $Z$-rotations, we verified computationally that the logical gate is non-trivial in this case.
\end{example}

\begin{example}[Reed-Muller Family]
\label{ex:QRM_family}
\normalfont
We can generalize this construction as a Reed-Muller family of $\llbr n = 2^m, \binom{m}{r}, 2^{r} \rrbr$ CSS-T codes defined by $C_2 = \text{RM}(r-1,m), C_1 = \text{RM}(r,m)$, and hence $C_1^{\perp} = \text{RM}(m-r-1,m) \subset C_2^{\perp} = \text{RM}(m-r,m)$.
Using the conditions in Theorem~\ref{thm:transversal_T} and Theorem~\ref{thm:logical_identity}, we see that we need $r \leq \frac{m}{3}$ for transversal $T$ to be supported, and also $r > \frac{m-1}{3}$ for transversal $T$ to not realize the logical identity.
This appears to imply that there is exactly one integer value of $r$ that provides a valid code, but this need not be true since decreasing monomial codes correspond precisely to non-integer values of $r$.
For example, one can take $m = 9, r = 3$ to obtain a valid $\llbr 512,84,8 \rrbr$ code.
Indeed, notice that $C_1^{\perp} = \text{RM}(5,9)$ contains the code $\text{RM}(4,8)$ in the support of any degree-$1$ polynomial, and $\text{RM}(4,8)$ contains the self-dual code ``$\text{RM}(3.5,8)$'', which is generated by all degree at most $3$ monomials as well as the first half of all degree $4$ monomials when they are arranged in lexicographic order.
%However, it is hard to use a naive computer program to determine the logical operator realized by transversal $T$ in this case, so we do not know either the logical gate or its level in the Clifford hierarchy for this example.
Once again, Theorem~\ref{thm:QRM_family} provides the logical gate realized by transversal $T$ on this $\llbr 512,84,8 \rrbr$ code.
\end{example}

The family of quantum Reed-Muller codes in Example~\ref{ex:QRM_family} appears in recent work by Haah and Hastings~\cite{Haah-quantum17b}, and Campbell and Howard~\cite{Campbell-pra17,Campbell-prl17}, where in~\cite{Haah-quantum17b} they focused on distilling CCZ magic states from these codes via physical transversal $T$.
For this reason, they needed the logical CC$Z$s to be on \emph{distinct} triples of (logical) qubits and they provided an analytic construction that guarantees a $\llbr 2^m, 3(2^{m/3}-2), 2^{m/3} \rrbr$ quantum Reed-Muller code satisfying this constraint.
However, using a computational search strategy, they show that in certain cases the number of logical qubits can be increased to produce more disjoint CC$Z$s.
For example, for $m = 9$, they expand the $\llbr 512,18,8 \rrbr$ code from the analytic construction that yields $6$ logical CC$Z$s into a $\llbr 512,30,8 \rrbr$ code that yields $10$ logical CC$Z$s on disjoint triples of logical qubits.

In Theorem~\ref{thm:QRM_family}, we generalize the quantum Reed-Muller family from Example~\ref{ex:QRM_family} for general $\pi/2^{\ell}$ $Z$-rotations, and also prove the exact logical operator realized on these codes.
We believe that, when applied to the transversal $T$ scenario, this result allows one to \emph{analytically} derive the above codes in~\cite{Haah-quantum17b} by using combinatorial arguments to carefully ``peel off'' the additional CC$Z$s that either overlap on qubits involved in existing ones or violate the \emph{generalized triorthogonality} constraints~\cite{Haah-quantum17b}.
This peeling procedure effectively drops all logical qubits that are not involved in the maximum number of disjoint CC$Z$s, $k_{\text{CC}Z}^{\text{max}}$, obtainable on these codes.
Moreover, this approach might lead to an exact characterization of $k_{\text{CC}Z}^{\text{max}}$ for any $m$, without involving the \emph{Lovasz Local Lemma} that appears to provide guarantees only for large $m$.
We leave this investigation for future work.

\subsection{Stabilizer Codes that Support Transversal $Z$-Rotations}
\label{sec:stab_codes_Z_rotations}

We first generalize Lemma~\ref{lem:conj_by_trans_T} to transversal $\pi/2^{\ell}$ $Z$-rotations, that again could be of independent interest. % elsewhere.

\begin{lemma}
\label{lem:conj_by_trans_Z_rot}
Let $E(a,b) \in HW_N$ for $N = 2^n, a,b \in \mathbb{Z}_2^n$.
Then transversal $\tau_{[ 1 ]}^{(\ell)} = \exp\left( \frac{\imath\pi}{2^{\ell}} Z \right), \ell \geq 2$, acts on $E(a,b)$ as
\begin{align}
& \tau_{I_n}^{(\ell)} E(a,b) \left( \tau_{I_n}^{(\ell)} \right)^{\dagger} \nonumber \\
  & = \frac{1}{\left( \sec\frac{2\pi}{2^{\ell}} \right)^{w_H(a)}} \sum_{y \preceq a} \left( \tan\frac{2\pi}{2^{\ell}} \right)^{w_H(y)} (-1)^{b y^T} E(a, b \oplus y),
\end{align}
where $w_H(a) = aa^T$ is the Hamming weight of $a$, and $y \preceq a$ denotes that $y$ is contained in the support of $a$.
\begin{proof}
See Appendix~\ref{sec:proof_conj_by_trans_Z_rot}.
\end{proof}
\end{lemma}

For $\ell = 3$ (transversal $T$), the cosine term produced a $2^{-w_H(a_j)/2}$ factor which we were able to ensure was an integer by enforcing $a_j$ to have even Hamming weight.
Then we produced $2^{w_H(a_j)/2}$ copies of each stabilizer element in order to cancel this factor and thereby reproduced the code projector.
However, for $\ell > 3$, extending this idea requires that $\left( \sec\frac{2\pi}{2^{\ell}} \right)^{w_H(a)}$ cancel the sum of (signed) tangents acquired for each copy of the stabilizer element.
This leads us to an extension of Theorem~\ref{thm:transversal_T}.

\begin{theorem}[Transversal $Z$-rotations]
\label{thm:transversal_Z_rot}
Let $S = \langle \nu_i E(c_i,d_i) ; i = 1,\ldots,r \rangle$ define an $\llbr n,n-r \rrbr$ stabilizer code as in Theorem~\ref{thm:transversal_T}.
Let $Z_S \coloneqq \{ z \in \mathbb{Z}_2^n \colon \epsilon_z E(0,z) \in S \}$ for some $\{ \epsilon_z \}$. 
For any $\epsilon_j E(a_j,b_j) \in S$ with non-zero $a_j$, define the subspace $Z_j \coloneqq \{ v \in Z_S \colon v \preceq a_j \}$ and the set $W_j \coloneqq \{ y \in \mathbb{Z}_2^n \colon y \preceq a_j, y \notin Z_j \}$.
Then the transversal application of the $\exp\left( \frac{\imath\pi}{2^{\ell}} Z \right)$ gate realizes a logical operation on $V(S)$ if and only if the following are true for all such $a_j \neq 0$:
\begin{align}
\sum_{v \in Z_j} \epsilon_v \left( \imath \tan\frac{2\pi}{2^{\ell}} \right)^{w_H(v)} & = \left( \sec\frac{2\pi}{2^{\ell}} \right)^{w_H(a_j)}, \\
\sum_{v \in Z_j} \epsilon_v \left( \imath \tan\frac{2\pi}{2^{\ell}} \right)^{w_H(v \oplus y)} & = 0 \quad \text{for\ all}\ y \in W_j,
\end{align}      
where $\epsilon_v \in \{ \pm 1 \}$ is the sign of $E(0,v)$ in the stabilizer $S$.
%\item For each $z \in \mathbb{Z}_2^n$ such that $z \in A_j$ for some $j \in \{ 1,\ldots,2^r \}$, $\imath^{w_H(z)} E(0,z) \in S$.      
\begin{proof}
See Appendix~\ref{sec:proof_transversal_Z_rot}.
\end{proof}
\end{theorem}

The extension here is only partial in the sense that the conditions on the stabilizer involve trigonometric quantities and we still have to distill finite geometric constraints on the vectors describing the stabilizer elements, similar to Theorems~\ref{thm:transversal_T} and~\ref{thm:transversal_T_Tinv}.
However, under the assumption that $Z_j$ is a self-dual code and $\epsilon_v = 1$ for all $v \in Z_j$, we are able to deduce the following condition on the Hamming weights of $v$ and $a_j$.

\begin{lemma}
\label{lem:divisible}
Let $C$ be an $[m,m/2]$ self-dual code and $\ell \geq 2$.
Then $\sum_{v \in C} \left( \imath \tan\frac{2\pi}{2^{\ell}} \right)^{w_H(v)} = \left( \sec\frac{2\pi}{2^{\ell}} \right)^{m}$ if and only if each $v \in C$ satisfies $w_H(v) = m/2$ or $(m - 2 w_H(v))$ is divisible by $2^{\ell}$.
\begin{proof}
See Appendix~\ref{sec:proof_divisible}.
\end{proof}
\end{lemma}

Finally, although we do not have the full extension of Theorem~\ref{thm:transversal_T} yet, we consider a family of quantum Reed-Muller codes QRM$(r,m)$ that supports $\pi/2^{\ell}$ $Z$-rotations from the Clifford hierarchy, and we also explicitly construct the logical operations induced by transversal $Z$-rotations on these codes.
% This result applies to a general family of \emph{quantum pin codes} as discussed in~\cite[Section V-D]{Vuillot-arxiv19}.
The code QRM$(r,m)$ is a CSS code defined by $C_2 = \text{RM}(r-1,m)$ and $C_1 = \text{RM}(r,m)$.
Hence, we can identify the following relationships:
\begin{align}
\text{$X$-type\ stabilizers}\ & \leftrightarrow\ c \in \text{RM}(r-1,m), \nonumber \\
\text{$Z$-type\ stabilizers}\ & \leftrightarrow\ c \in \text{RM}(m-r-1,m), \nonumber \\
\text{$X$-type\ logical\ operators}\ & \leftrightarrow\ c \in \text{RM}(r,m), \nonumber \\
\text{$Z$-type\ logical\ operators}\ & \leftrightarrow\ c \in \text{RM}(m-r,m).
\end{align}
The parameters for QRM$(r,m)$ are given by $\llbr 2^m, \binom{m}{r}, 2^{\min\{r,m-r\}} \rrbr$.
Recollect that for $v_f \in \mathbb{Z}_2^k$, the CSS basis states are
\begin{align}
\label{eq:css_basis_states}
\ket{v_f}_L & \equiv \frac{1}{|C_2|} \sum_{c \in C_2}  \ket{v_f \cdot G_{C_1/C_2} \oplus c} \nonumber \\
  & = \frac{1}{|C_2|} \sum_{y \in \mathbb{Z}_2^{k_2}}  \ket{v_f \cdot G_{C_1/C_2} \oplus y \cdot G_2},
\end{align}
where $G_{C_1/C_2}$ denotes the generator matrix for the linear subspace of coset representatives for $C_2$ in $C_1$, and $G_2$ denotes the generator matrix for the code $C_2$.
For QRM$(r,m)$, the rows of $G_{C_1/C_2}$ correspond to degree $r$ monomials, each identifying a logical qubit.
Hence, any polynomial $f$ comprised of these monomials corresponds to a distinct logical computational basis state $\ket{v_f}_L$.
So a non-trivial logical $X$ operator is described by a degree $r$ polynomial $f$, but only the degree $r$ terms will determine which logical qubits are acted upon.
Also, this implies that if a particular degree $r$ term is present in $f$, then the corresponding logical qubit is set to $\ket{1}_L$ in $\ket{v_f}_L$.

\addtocounter{example}{-3}

\begin{example}[contd.]
\normalfont
Before we state the general result, let us setup the notation through the $\llbr 64,15,4 \rrbr$ example from Section~\ref{sec:product_of_CCZs}.
Recollect that in this case we have $m = 6$ and $r = 2$, so the logical qubits can be identified with the degree $2$ monomials that define generators for logical $X$ operators.
Hence, the polynomial in~\eqref{eq:QRM26_poly} defining the logical gate realized by physical transversal $T$ can be represented in monomial subscripts as $q(f) \equiv q(v_f)$
\begin{align} 
% & q(f) \equiv q(v_f) \nonumber \\
%
   & = v_{x_1 x_2} v_{x_3 x_4} v_{x_5 x_6} + v_{x_1 x_2} v_{x_3 x_5} v_{x_4 x_6} + v_{x_1 x_2} v_{x_3 x_6} v_{x_4 x_5} \nonumber \\
     & \quad + v_{x_1 x_3} v_{x_2 x_4} v_{x_5 x_6} + v_{x_1 x_3} v_{x_2 x_5} v_{x_4 x_6} + v_{x_1 x_3} v_{x_2 x_6} v_{x_4 x_5}  \nonumber \\
     & \quad + v_{x_1 x_4} v_{x_2 x_3} v_{x_5 x_6} + v_{x_1 x_4} v_{x_2 x_5} v_{x_3 x_6} + v_{x_1 x_4} v_{x_2 x_6} v_{x_3 x_5}  \nonumber \\
     & \quad + v_{x_1 x_5} v_{x_2 x_3} v_{x_4 x_6} + v_{x_1 x_5} v_{x_2 x_4} v_{x_3 x_6} + v_{x_1 x_5} v_{x_2 x_6} v_{x_3 x_4}  \nonumber \\
     & \quad + v_{x_1 x_6} v_{x_2 x_3} v_{x_4 x_5} + v_{x_1 x_6} v_{x_2 x_4} v_{x_3 x_5} + v_{x_1 x_6} v_{x_2 x_5} v_{x_3 x_4},
\end{align}
where each term in the polynomial corresponds to a logical CC$Z$ gate acting on the three logical qubits indexed by the three monomial subscripts, and the sum corresponds to a product of such gates (in the logical unitary space).
In the notation of~\eqref{eq:QRM26_poly}, $v_{x_1 x_2} v_{x_3 x_4} v_{x_5 x_6} \equiv v_1 v_{10} v_{15}$ and so on, which means $v_f = [v_{x_1 x_2}, v_{x_1 x_3}, \ldots, v_{x_5 x_6}]$, i.e., 
\begin{align}
\ket{v_f}_L & = \ket{v_{x_1 x_2}}_L \otimes \ket{v_{x_1 x_3}}_L \otimes \cdots \otimes \ket{v_{x_5 x_6}}_L \nonumber \\
  & = \ket{v_1}_L \otimes \ket{v_2}_L \otimes \cdots \otimes \ket{v_{15}}_L.    
\end{align}
For this code, the rows of $G_{C_1/C_2}$ are evaluations of the $\binom{m}{m/3} = 15$ degree $r=2$ monomials, namely $x_1 x_2, x_1 x_3, x_1 x_4, \ldots, x_5 x_6$.
So, the polynomial $f \in \text{RM}(r,m)$ above is a linear combination of degree $r=2$ monomials, and possibly lower degree monomials that correspond to just $X$-type stabilizers. 
Hence, $v_f \in \mathbb{Z}_2^{15}$ exactly describes which corresponding rows of $G_{C_1/C_2}$ are chosen in this linear combination.
Therefore, if $f = x_1 x_2 + x_3 x_4 + x_5 x_6 + (\text{smaller\ degree\ terms})$, then $\ket{v_{x_1 x_2} v_{x_3 x_4} v_{x_5 x_6}}_L = \ket{111}_L$ and other qubits are set to $\ket{0}_L$, so $q(f) = 1$.
But if $f = x_1 x_2 + x_3 x_4 + x_5 x_6 + x_3 x_5 + x_4 x_6 + (\text{smaller\ degree\ terms})$, then $q(f) = 0$ as this polynomial corresponds to two CC$Z$s applying the phase $-1$.

For stating the general result, it will be convenient to replace the monomial subscripts with binary vectors $p_1, p_2, p_3 = p_{m/r}$. 
So, for example, for the first term $v_{x_1 x_2} v_{x_3 x_4} v_{x_5 x_6}$ these index vectors are given by $p_1 = [1,1,0,0,0,0], p_2 = [0,0,1,1,0,0], p_3 = [0,0,0,0,1,1]$, which each have Hamming weight $r = 2$ and sum up to $\vecnot{1}$.
\end{example}

Define $\mu(x) \coloneqq (-1)^x$, and let $\nu_p(s)$ denote the largest integer $t$ such that $p^t$ divides $s$.

\begin{theorem} 
\label{thm:QRM_family}
Suppose that $1 \leq r \leq m/2$ and $r$ divides $m$.  
Then, the transversal $\exp\left( \frac{\imath\pi}{2^{m/r}} Z \right)$ gate is a logical operator for $\text{QRM}(r,m)$.  
Moreover, up to local corrections, the corresponding logical operator acts on a computational basis state by
\begin{align}
\label{eq:QRM_family_logical_action}
\ket{v_f}_L \mapsto \mu(q(f))\ket{v_f}_L,
\end{align} 
where $f = \sum_{\vecnot{d} \in \mathbb{Z}_2^m} a_{\vecnot{d}} x^{\vecnot{d}} \in \text{RM}(r,m), a_{\vecnot{d}} \in \{0,1\}$ with $a_{\vecnot{d}} = 0$ for all $w_H(\vecnot{d}) > r$, and 
\begin{align}
\label{eq:QRM_thm_poly}
q(f) \equiv q(v_f) = \sum_{(p_1,\ldots,p_{m/r}) \in P} \prod_{j = 1}^{m/r} v_{p_j} \ (\bmod\ 2),
\end{align}
where $P \coloneqq \{ (p_1,\ldots,p_{m/r}) \colon p_j \in \mathbb{Z}_2^m, \sum_{j=1}^{m/r} p_j = \vecnot{1}, w_H(p_j) = r \}$.
In particular, $deg(q) = m/r$ and so by~\cite{Cui-physreva17}, it is a gate from the $(m/r)$-th level of the Clifford hierarchy.
% Also, QRM($r,m$) is $2^{\frac{m}{r}-1} q(v_f)$-(quasi)transversal~\cite{Campbell-pra17}.
\begin{proof}
See Appendix~\ref{sec:proof_QRM_family}.
\end{proof}
\end{theorem}

In general, the above theorem states that the logical gate polynomial $q(f)$ consists of all terms such that the monomials in the subscripts of each term form a unique partition of $m$ variables into $m/r$ groups of $r$ variables each.
Therefore, the number of terms in the polynomial $q(f)$, and hence the number of gates in the induced logical operator is given by $\frac{m!}{(r!)^{m/r} \left( \frac{m}{r} \right)!}$.

\begin{remark}[Quasitransversality]
\normalfont
In~\cite{Campbell-pra17,Campbell-prl17}, Campbell and Howard considered diagonal gates $U_F \in \mathcal{C}^{(3)}$ that can be expressed as $U_F = \sum_{x \in \mathbb{Z}_2^k} \omega^{F(x)} \ketbra{x}$, where $\omega = e^{\imath\pi/4}$ and $F(x) = L(x) + 2Q(x) + 4C(x)$ (mod $8$) is a weighted polynomial with $L(x)$ linear (mod $8$), $Q(x)$ quadratic (mod $4$) and $C(x)$ cubic (mod $2$) polynomials.
So $L$ corresponds to single-qubit $Z$-rotations, $Q$ corresponds to controlled $Z$-rotations, and $C$ corresponds to CC$Z$ gates.
They define a quantum code to be ``$F$-quasitransversal'' if there exists a Clifford $g$ such that $g T^{\otimes n}$ acting on the physical qubits realize the logical gate $U_F$.
In~\cite[Lemma 1]{Campbell-pra17}, they provided the following sufficient condition for a CSS code to be $F$-quasitransversal, which we rewrite in our notation (e.g., see~\eqref{eq:css_basis_states}):
\begin{align}
    w_H(x \cdot G_{C_1/C_2} \oplus y \cdot G_2) \sim_{c} F(x) \ (\bmod\ 8), \ y \in \mathbb{Z}_2^{k_2},
\end{align}
where the subscript ``$c$'' implies that the two sides are Clifford equivalent, i.e., there exists a weighted polynomial $\tilde{F}$ such that by replacing $F(x)$ with $F(x) + 2\tilde{F}(x)$ above, we can replace $\sim_{c}$ with equality.

Now, observe that $u = x \cdot G_{C_1/C_2} \oplus y \cdot G_2 \in C_1$ exactly corresponds to $u = \text{ev}(f)$ for some $f \in \text{RM}(r,m)$ above (with $x = v_f$).
Hence, we note that~\eqref{eq:quasitransversal} exactly matches the (quasi)transversality condition above (with equality and thereby no Clifford correction).
Therefore, QRM($m/3,m$) is $4 q(f)$-(quasi)transversal.
\end{remark}

\begin{remark}[Quantum Pin Codes]
\normalfont
Vuillot and Breuckmann~\cite{Vuillot-arxiv19} recently introduced ``Quantum Pin Codes'' as an abstract framework to synthesize stabilizer codes that support transversal, or partially transversal, physical $Z$-rotations.
These codes are inspired by topological constructions such as color codes~\cite{Kubica-pra15}, but the abstraction extends beyond algebraic topology while retaining transversality properties.
The authors produce several new codes using this formalism.
We note that the above result regarding QRM($r,m$) codes applies to a general family of quantum pin codes as discussed in~\cite[Section V-D]{Vuillot-arxiv19}.
\end{remark}

\section{Conclusion}
\label{sec:conclusion}

In this paper, we used the recent characterization of quadratic form diagonal (QFD) gates~\cite{Rengaswamy-pra19} to derive necessary and sufficient conditions for a stabilizer code to support a physical transversal $T$ gate.
Our Heisenberg approach allowed us to generalize all such existing constructions.
Using this, we showed that for any non-degenerate stabilizer code with this property, there exists an equivalent CSS code that also possesses this property.
So for magic state distillation via transversal $T$ on non-degenerate stabilizer codes, CSS codes are essentially optimal.
We also showed that triorthogonal codes form the most general family of CSS codes that realize logical transversal $T$ via physical transversal $T$.
Among several examples, we constructed a $\llbr 16,3,2 \rrbr$ code using the decreasing monomial formalism, and demonstrated how to check that transversal $T$ realizes logical CC$Z$ with the help of generalized triorthogonality conditions.
This points to a possibly general construction of CSS codes supporting transversal $T$ using this formalism.

We then extended the above results beyond $T$ gates, and derived trigonometric stabilizer conditions for the code to support a transversal $\pi/2^{\ell}$ $Z$-rotation.
However, we were only able to reduce this to finite geometric conditions under some assumptions. 
Finally, we considered a family of quantum Reed-Muller codes and determined the exact logical operation induced by transversal $Z$-rotations on these codes from an alternate viewpoint, using Ax's theorem on residue weights of polynomials.
Although these logical operations involve products of overlapping many-controlled-$Z$ gates, it will be interesting to investigate their utility in magic state distillation and other proposals for universal quantum computation.
In certain systems, finer angle rotations often have better fidelity than coarser angle rotations. 
Hence, these native resources in the lab could be leveraged in combination with these codes to potentially obtain better circuit decompositions.

\section*{Acknowledgment}

This article is dedicated to the memory of David Poulin.

We would like to thank Kenneth Brown and Anirudh Krishna for helpful discussions.
Kenneth Brown suggested us to think about a Bacon-Shor-like code on a $4 \times 4$ lattice, and this led to the first construction of the $\llbr 16,3,2 \rrbr$ code.
Subsequently, during his visit to Duke University, Anirudh Krishna demonstrated that an equivalent $\llbr 16,3,2 \rrbr$ code can be constructed using decreasing monomial codes.
We would also like to thank Andrew Landahl and Earl Campbell for carefully reading earlier drafts of this manuscript and providing helpful feedback.
We are also thankful to Jeongwan Haah for encouraging us to clarify the existence of the self-dual code in the proof of Theorem~\ref{thm:transversal_T}; this motivated us to greatly improve this proof.
N.R. would like to thank Christophe Vuillot, Jean-Pierre Tillich, Earl Campbell, Mark Howard, Michael Beverland, Andrew Landahl and Andrew Cross for fruitful discussions and feedback.
The connections to pin codes, generalized triorthogonality and quasitransversality were made during these conversations.
N.R. would like to thank Christophe Vuillot, Jean-Pierre Tillich, and Earl Campbell also for their hospitality during his visits to their research groups.

This work was supported in part by the National Science Foundation (NSF) under Grant Nos. 1718494 and 1908730. 
The research of M.N. was supported under the grant ODNI/IARPA LogiQ program (W911NF-16-1-0082).
Any opinions, findings, conclusions, and recommendations expressed in this material are those of the authors and do not necessarily reflect the views of these sponsors.

%\bibliographystyle{IEEEtran}
%\bibliography{WCLabrv,WCLnewbib}

\begin{thebibliography}{10}
\providecommand{\url}[1]{#1}
\csname url@samestyle\endcsname
\providecommand{\newblock}{\relax}
\providecommand{\bibinfo}[2]{#2}
\providecommand{\BIBentrySTDinterwordspacing}{\spaceskip=0pt\relax}
\providecommand{\BIBentryALTinterwordstretchfactor}{4}
\providecommand{\BIBentryALTinterwordspacing}{\spaceskip=\fontdimen2\font plus
\BIBentryALTinterwordstretchfactor\fontdimen3\font minus
  \fontdimen4\font\relax}
\providecommand{\BIBforeignlanguage}[2]{{%
\expandafter\ifx\csname l@#1\endcsname\relax
\typeout{** WARNING: IEEEtran.bst: No hyphenation pattern has been}%
\typeout{** loaded for the language `#1'. Using the pattern for}%
\typeout{** the default language instead.}%
\else
\language=\csname l@#1\endcsname
\fi
#2}}
\providecommand{\BIBdecl}{\relax}
\BIBdecl

\bibitem{Rengaswamy-isit20}
\BIBentryALTinterwordspacing
N.~Rengaswamy, R.~Calderbank, M.~Newman, and H.~D. Pfister, ``Classical coding
  problem from transversal {$T$} gates,'' \emph{Accepted\ to\ IEEE Int.\ Symp.\
  Inform.\ Theory}, 2020. [Online]. Available:
  \url{http://arxiv.org/abs/2001.04887}
\BIBentrySTDinterwordspacing

\bibitem{Eastin-prl09}
\BIBentryALTinterwordspacing
B.~Eastin and E.~Knill, ``{Restrictions on Transversal Encoded Quantum Gate
  Sets},'' \emph{Phys. Rev. Lett.}, vol. 102, no.~11, p. 110502, 2009.
  [Online]. Available: \url{http://arxiv.org/abs/0811.4262}
\BIBentrySTDinterwordspacing

\bibitem{Zeng-it07}
\BIBentryALTinterwordspacing
B.~Zeng, A.~Cross, and I.~L. Chuang, ``{Transversality Versus Universality for
  Additive Quantum Codes},'' \emph{IEEE Trans.\ Inform.\ Theory}, vol.~57,
  no.~9, pp. 6272--6284, 2011. [Online]. Available:
  \url{https://arxiv.org/abs/0706.1382}
\BIBentrySTDinterwordspacing

\bibitem{Bravyi-pra12}
\BIBentryALTinterwordspacing
S.~Bravyi and J.~Haah, ``{Magic-state distillation with low overhead},''
  \emph{Phys. Rev. A}, vol.~86, no.~5, p. 052329, 2012. [Online]. Available:
  \url{http://arxiv.org/abs/1209.2426}
\BIBentrySTDinterwordspacing

\bibitem{Kubica-pra15}
\BIBentryALTinterwordspacing
A.~Kubica and M.~E. Beverland, ``{Universal transversal gates with color codes:
  A simplified approach},'' \emph{Phys. Rev. A}, vol.~91, no.~3, p. 032330,
  2015. [Online]. Available: \url{https://arxiv.org/abs/1410.0069}
\BIBentrySTDinterwordspacing

\bibitem{Gottesman-nature99}
\BIBentryALTinterwordspacing
D.~Gottesman and I.~L. Chuang, ``{Demonstrating the viability of universal
  quantum computation using teleportation and single-qubit operations},''
  \emph{Nature}, vol. 402, no. 6760, pp. 390--393, 1999. [Online]. Available:
  \url{http://www.nature.com/articles/46503}
\BIBentrySTDinterwordspacing

\bibitem{Bravyi-pra05}
\BIBentryALTinterwordspacing
S.~Bravyi and A.~Kitaev, ``{Universal quantum computation with ideal Clifford
  gates and noisy ancillas},'' \emph{Phys. Rev. A}, vol.~71, no.~2, p. 022316,
  2005. [Online]. Available: \url{https://arxiv.org/abs/quant-ph/0403025}
\BIBentrySTDinterwordspacing

\bibitem{Landahl-arxiv13}
\BIBentryALTinterwordspacing
A.~J. Landahl and C.~Cesare, ``Complex instruction set computing architecture
  for performing accurate quantum {\$}z{\$} rotations with less magic,''
  \emph{arXiv preprint arXiv:1302.3240}, 2013. [Online]. Available:
  \url{http://arxiv.org/abs/1302.3240}
\BIBentrySTDinterwordspacing

\bibitem{Nam-arxiv19}
\BIBentryALTinterwordspacing
Y.~Nam, J.-S. Chen, N.~C. Pisenti, K.~Wright, C.~Delaney, D.~Maslov, K.~R.
  Brown, S.~Allen, J.~M. Amini, J.~Apisdorf, K.~M. Beck, A.~Blinov, V.~Chaplin,
  M.~Chmielewski, C.~Collins, S.~Debnath, K.~M. Hudek, A.~M. Ducore, M.~Keesan,
  S.~M. Kreikemeier, J.~Mizrahi, P.~Solomon, M.~Williams, J.~D. Wong-Campos,
  D.~Moehring, C.~Monroe, and J.~Kim, ``{Ground-state energy estimation of the
  water molecule on a trapped-ion quantum computer},'' \emph{npj Quantum Inf.},
  vol.~6, no.~1, p.~33, 2020. [Online]. Available:
  \url{http://arxiv.org/abs/1902.10171}
\BIBentrySTDinterwordspacing

\bibitem{Rengaswamy-pra19}
\BIBentryALTinterwordspacing
N.~Rengaswamy, R.~Calderbank, and H.~D. Pfister, ``Unifying the {C}lifford
  hierarchy via symmetric matrices over rings,'' \emph{Phys. Rev. A}, vol. 100,
  no.~2, p. 022304, 2019. [Online]. Available:
  \url{http://arxiv.org/abs/1902.04022}
\BIBentrySTDinterwordspacing

\bibitem{Cui-physreva17}
\BIBentryALTinterwordspacing
S.~X. Cui, D.~Gottesman, and A.~Krishna, ``{Diagonal gates in the Clifford
  hierarchy},'' \emph{Phys. Rev. A}, vol.~95, no.~1, p. 012329, 2017, [Online].
  Available: http://arxiv.org/abs/1608.06596. [Online]. Available:
  \url{https://journals.aps.org/pra/pdf/10.1103/PhysRevA.95.012329}
\BIBentrySTDinterwordspacing

\bibitem{JochymOconnor-prx17}
T.~Jochym-O'Connor, A.~Kubica, and T.~J. Yoder, ``{Disjointness of Stabilizer
  Codes and Limitations on Fault-Tolerant Logical Gates},'' \emph{Phys. Rev.
  X}, vol.~8, no.~2, p. 021047, 2018, [Online]. Available:
  https://arxiv.org/pdf/1710.07256.pdf.

\bibitem{JochymOconnor-prl14}
T.~Jochym-O’Connor and R.~Laflamme, ``Using concatenated quantum codes for
  universal fault-tolerant quantum gates,'' \emph{Physical review letters},
  vol. 112, no.~1, p. 010505, 2014.

\bibitem{Haah-quantum17b}
\BIBentryALTinterwordspacing
J.~Haah and M.~B. Hastings, ``{Codes and Protocols for Distilling {\$}T{\$},
  controlled-{\$}S{\$}, and Toffoli Gates},'' \emph{Quantum}, vol.~2, p.~71,
  2017. [Online]. Available: \url{https://arxiv.org/abs/1709.02832}
\BIBentrySTDinterwordspacing

\bibitem{Campbell-pra17}
\BIBentryALTinterwordspacing
E.~T. Campbell and M.~Howard, ``{Unified framework for magic state distillation
  and multiqubit gate synthesis with reduced resource cost},'' \emph{Phys. Rev.
  A}, vol.~95, no.~2, p. 022316, Feb 2017. [Online]. Available:
  \url{https://journals.aps.org/pra/pdf/10.1103/PhysRevA.95.022316}
\BIBentrySTDinterwordspacing

\bibitem{Campbell-prl17}
\BIBentryALTinterwordspacing
------, ``{Unifying Gate Synthesis and Magic State Distillation},'' \emph{Phys.
  Rev. Lett.}, vol. 118, no.~6, p. 060501, Feb 2017. [Online]. Available:
  \url{https://journals.aps.org/prl/pdf/10.1103/PhysRevLett.118.060501}
\BIBentrySTDinterwordspacing

\bibitem{Vuillot-arxiv19}
\BIBentryALTinterwordspacing
C.~Vuillot and N.~P. Breuckmann, ``{Quantum Pin Codes},'' \emph{arXiv preprint
  arXiv:1906.11394}, 2019. [Online]. Available:
  \url{http://arxiv.org/abs/1906.11394}
\BIBentrySTDinterwordspacing

\bibitem{Rains-arxiv02}
E.~M. Rains and N.~J. Sloane, ``Self-dual codes,'' \emph{arXiv preprint
  arXiv:math/0208001}, 2002.

\bibitem{Nebe-2006}
G.~Nebe, E.~M. Rains, and N.~J.~A. Sloane, \emph{Self-dual codes and invariant
  theory}.\hskip 1em plus 0.5em minus 0.4em\relax Springer, 2006, vol.~17.

\bibitem{Macwilliams-1977}
F.~J. MacWilliams and N.~J.~A. Sloane, \emph{The Theory of Error-Correcting
  Codes}.\hskip 1em plus 0.5em minus 0.4em\relax North-Holland, Amsterdam,
  1977.

\bibitem{McEliece-jpl72}
R.~McEliece, ``Weights modulo 8 in binary cyclic codes,'' \emph{Jet Propulsion
  Lab}, vol.~11, pp. 86--88, 1972.

\bibitem{McEliece-dm72}
R.~J. McEliece, ``Weight congruences for p-ary cyclic codes,'' \emph{Discrete
  Mathematics}, vol.~3, no. 1-3, pp. 177--192, 1972.

\bibitem{Li-prx19}
M.~Li, D.~Miller, M.~Newman, Y.~Wu, and K.~R. Brown, ``2d compass codes,''
  \emph{Physical Review X}, vol.~9, no.~2, p. 021041, 2019.

\bibitem{Campbell-blog16}
\BIBentryALTinterwordspacing
E.~T. Campbell, ``The smallest interesting colour code,'' 2016, blog post.
  [Online]. Available:
  \url{https://earltcampbell.com/2016/09/26/the-smallest-interesting-colour-code/}
\BIBentrySTDinterwordspacing

\bibitem{Bardet-isit16}
\BIBentryALTinterwordspacing
M.~Bardet, V.~Dragoi, A.~Otmani, and J.-P. Tillich, ``Algebraic properties of
  polar codes from a new polynomial formalism,'' in \emph{Proc.\ IEEE Int.\
  Symp.\ Inform.\ Theory}.\hskip 1em plus 0.5em minus 0.4em\relax IEEE, 2016,
  pp. 230--234. [Online]. Available: \url{http://arxiv.org/abs/1601.06215}
\BIBentrySTDinterwordspacing

\bibitem{Bardet-arxiv16}
M.~{Bardet}, V.~{Dragoi}, A.~{Otmani}, and J.-P. {Tillich}, ``{Algebraic
  Properties of Polar Codes From a New Polynomial Formalism},'' [Online].
  Available: http://arxiv.org/pdf/1601.06215.pdf.

\bibitem{Arikan-it09}
E.~Ar{\i}kan, ``Channel polarization: {A} method for constructing
  capacity-achieving codes for symmetric binary-input memoryless channels,''
  \emph{IEEE Trans.\ Inform.\ Theory}, vol.~55, no.~7, pp. 3051--3073, July
  2009.

\bibitem{Krishna-arxiv18}
\BIBentryALTinterwordspacing
A.~Krishna and J.-P. Tillich, ``Magic state distillation with punctured polar
  codes,'' \emph{arXiv preprint arXiv:1811.03112}, 2018. [Online]. Available:
  \url{http://arxiv.org/abs/1811.03112}
\BIBentrySTDinterwordspacing

\bibitem{Tuckett-prl18}
\BIBentryALTinterwordspacing
D.~K. Tuckett, S.~D. Bartlett, and S.~T. Flammia, ``{Ultrahigh Error Threshold
  for Surface Codes with Biased Noise},'' \emph{Phys. Rev. Lett.}, vol. 120,
  no.~5, p. 050505, 2018. [Online]. Available:
  \url{https://link.aps.org/doi/10.1103/PhysRevLett.120.050505}
\BIBentrySTDinterwordspacing

\bibitem{Wilde-2013}
M.~M. Wilde, \emph{Quantum {I}nformation {T}heory}.\hskip 1em plus 0.5em minus
  0.4em\relax Cambridge University Press, 2013.

\bibitem{Rengaswamy-arxiv18}
\BIBentryALTinterwordspacing
N.~Rengaswamy, R.~Calderbank, S.~Kadhe, and H.~D. Pfister, ``{Synthesis of
  Logical Clifford Operators via Symplectic Geometry},'' in \emph{Proc.\ IEEE
  Int.\ Symp.\ Inform.\ Theory}, Jun 2018, pp. 791--795. [Online]. Available:
  \url{http://arxiv.org/abs/1803.06987v1}
\BIBentrySTDinterwordspacing

\bibitem{Boykin-arxiv99}
\BIBentryALTinterwordspacing
P.~Boykin, T.~Mor, M.~Pulver, V.~Roychowdhury, and F.~Vatan, ``{On universal
  and fault-tolerant quantum computing: a novel basis and a new constructive
  proof of universality for Shor's basis},'' in \emph{40th Annu. Symp. Found.
  Comput. Sci. (Cat. No.99CB37039)}, 1999, pp. 486--494. [Online]. Available:
  \url{https://arxiv.org/abs/quant-ph/9906054}
\BIBentrySTDinterwordspacing

\bibitem{Nielsen-2010}
M.~A. Nielsen and I.~L. Chuang, \emph{Quantum {C}omputation and {Q}uantum
  {I}nformation}.\hskip 1em plus 0.5em minus 0.4em\relax Cambridge University
  Press, 2010.

\bibitem{Linke-nas17}
\BIBentryALTinterwordspacing
N.~M. Linke, D.~Maslov, M.~Roetteler, S.~Debnath, C.~Figgatt, K.~A. Landsman,
  K.~Wright, and C.~Monroe, ``Experimental comparison of two quantum computing
  architectures,'' \emph{Proceedings of the National Academy of Sciences}, vol.
  114, no.~13, pp. 3305--3310, 2017. [Online]. Available:
  \url{https://www.pnas.org/content/114/13/3305/}
\BIBentrySTDinterwordspacing

\bibitem{Randriambololona-aagct15}
H.~Randriambololona, ``On products and powers of linear codes under
  componentwise multiplication,'' \emph{Algorithmic arithmetic, geometry, and
  coding theory}, vol. 637, pp. 3--78, 2015.

\bibitem{Gottesman-arxiv98}
\BIBentryALTinterwordspacing
D.~Gottesman, ``The heisenberg representation of quantum computers,'' in
  \emph{Intl. Conf. on Group Theor. Meth. Phys.}\hskip 1em plus 0.5em minus
  0.4em\relax International Press, Cambridge, MA, 1998, pp. 32--43. [Online].
  Available: \url{https://arxiv.org/abs/quant-ph/9807006}
\BIBentrySTDinterwordspacing

\bibitem{Anderson-prl14}
J.~T. Anderson, G.~Duclos-Cianci, and D.~Poulin, ``{Fault-Tolerant Conversion
  between the Steane and Reed-Muller Quantum Codes},'' \emph{Phys. Rev. Lett.},
  vol. 113, no.~8, p. 080501, 2014, [Online]. Available:
  http://arxiv.org/abs/1403.2734.

\bibitem{Quan-jpmt18}
\BIBentryALTinterwordspacing
D.~Quan, L.~Zhu, C.-X. Pei, and B.~C. Sanders, ``Fault-tolerant conversion
  between adjacent {R}eed--{M}uller quantum codes based on gauge fixing,''
  \emph{J. Phys. A Math. Theor.}, vol.~51, no.~11, p. 115305, 2018. [Online].
  Available: \url{http://arxiv.org/abs/1703.03860}
\BIBentrySTDinterwordspacing

\bibitem{Haah-pra18}
\BIBentryALTinterwordspacing
J.~Haah, ``Towers of generalized divisible quantum codes,'' \emph{Phys. Rev.
  A}, vol.~97, no.~4, p. 042327, 2018. [Online]. Available:
  \url{https://arxiv.org/abs/1709.08658}
\BIBentrySTDinterwordspacing

\bibitem{Watson-pra15}
\BIBentryALTinterwordspacing
F.~H.~E. Watson, E.~T. Campbell, H.~Anwar, and D.~E. Browne, ``Qudit colour
  codes and gauge colour codes in all spatial dimensions,'' \emph{Phys. Rev.
  A}, vol.~92, no.~2, p. 022312, 2015. [Online]. Available:
  \url{http://arxiv.org/abs/1503.08800}
\BIBentrySTDinterwordspacing

\bibitem{Ax-ajm64}
\BIBentryALTinterwordspacing
J.~Ax, ``{Zeroes of Polynomials Over Finite Fields},'' \emph{Am. J. Math.},
  vol.~86, no.~2, p. 255, Apr 1964. [Online]. Available:
  \url{https://www.jstor.org/stable/2373163?origin=crossref}
\BIBentrySTDinterwordspacing

\bibitem{Hou-actaa16}
X.-d. Hou, ``Polynomials meeting {A}x’s bound,'' \emph{Acta Arithmetica},
  vol. 176, no.~1, pp. 65--80, 2016.

\end{thebibliography}


\newpage

\appendices

\onecolumn

%\clearpage
%
%\title{Supplemental Material for the Paper\\ ``On Optimality of CSS Codes for Transversal $T$''}
%
%
% \author{%
%   \IEEEauthorblockN{Narayanan Rengaswamy,~\IEEEmembership{Student Member,~IEEE},
%                     Robert Calderbank,~\IEEEmembership{Fellow,~IEEE},
%                     Michael Newman, and\\
%                     Henry D. Pfister~\IEEEmembership{Senior Member,~IEEE}}%
%   \thanks{N. Rengaswamy, R. Calderbank, and H.D. Pfister are with the
%           Department of Electrical and Computer Engineering,
%           Duke University,
%           Durham, North Carolina 27708, USA.
%           M. Newman is with the Departments of Electrical and Computer Engineering, Chemistry and Physics,
%           Duke University,
%           Durham, North Carolina 27708, USA.
%           Email: \{narayanan.rengaswamy, robert.calderbank, henry.pfister\}@duke.edu, mgnewman@umich.edu}%
%    \thanks{Part of this work was presented at the 2020 IEEE International Symposium on Information Theory~\cite{Rengaswamy-isit20}.}
%%    \thanks{This paper has supplementary downloadable material available at http://ieeexplore.ieee.org, provided by the author. The material includes appendices containing proofs for many results. Contact narayanan.rengaswamy@duke.edu for further questions about this work.}
%  }
%
%
%{\maketitle}

\section{Classical Reed-Muller Codes}
\label{sec:rm_codes}

Given an integer $m \geq 1$, let $x_1, x_2, \ldots, x_m$ be binary variables and we adopt the convention that $x_1$ represents the least significant bit (LSB) and $x_m$ represents the most significant bit (MSB).
These variables can also be interpreted as \emph{monomials} of degree $1$, and we can construct degree $t$ monomials $x_{i_1} x_{i_2} \cdots x_{i_t}$ where $i_j \in \{ 1,\ldots,m \}$.
The set of all monomials in $m$ variables is denoted by $\mathcal{M}_m$.
A degree $t$ polynomial $f$ on $m$ variables is a binary linear combination of monomials such that the maximum degree term(s) has (have) degree $t$.
Any polynomial $f \in \mathbb{F}_2[x_1,\ldots,x_m]$ can be associated one-to-one to its evaluation vector $\text{ev}(f) \coloneqq [\, f(x_m,\ldots,x_1)\, ]_{(x_m,\ldots,x_1) \in \mathbb{F}_2^m} \in \mathbb{F}_2^{2^m}$.
Note that the unique degree $0$ monomial is taken to be $1$ whose evaluation vector is the all-$1$s vector. 

For $0 \leq r \leq m$, the binary Reed-Muller code RM($r,m$) is generated by evaluation vectors of all monomials on $m$ binary variables with degree at most $r$, i.e.,
\begin{align}
\text{RM}(r,m) & \coloneqq \{ \text{ev}(f) \in \mathbb{F}_2^{2^m} \colon f \in \mathbb{F}_2[x_1,\ldots,x_m], \text{deg}(f) \leq r \} \nonumber \\
  & = \langle \text{ev}(f) \in \mathbb{F}_2^{2^m} \colon f \in \mathcal{M}_m, \text{deg}(f) \leq r \rangle.
\end{align}
Hence, the dimension of RM($r,m$) is given by $k = \sum_{t=0}^{r} \binom{m}{t}$.
It is well-known that the minimum distance of RM($r,m$) is $2^{m-r}$ and that the dual of RM($r,m$) is RM($m-r-1,m$)~\cite{Macwilliams-1977}.
If $\text{ev}(f) \in \text{RM}(r,m)$, then we also write $f \in \text{RM}(r,m)$.

\section{Proofs for All Results}
\label{sec:all_proofs}

In all the proofs below we will use a few observations or identities repeatedly, so we mention them here.
\begin{enumerate}

\item[(O1)] As discussed in Section~\ref{sec:pauli_gp}, for $a,b,x \in \mathbb{Z}^n$ we have $E(a, b + 2x) = (-1)^{ax^T}$ and $E(a + 2x, b) = (-1)^{bx^T} E(a,b)$.
When multiplying two Pauli matrices we have the identities
\begin{align}
E(a,b) E(c,d) & = (-1)^{\syminn{[a,b]}{[c,d]}} E(c,d) E(a,b) \\
   & = \imath^{bc^T - ad^T \bmod 4} E(a+c, b+d) \\
   & = \imath^{bc^T - ad^T \bmod 4} E(a+c, (b \oplus d) + 2 (b \ast d) ) \\
   & = \imath^{bc^T - ad^T \bmod 4} (-1)^{(a + c) (b \ast d)^T} E( (a \oplus c) + 2 (a \ast c), b \oplus d) \\
   & =  \imath^{bc^T - ad^T \bmod 4} (-1)^{(a \oplus c) (b \ast d)^T + (b \oplus d) (a \ast c)^T} E(a \oplus c, b \oplus d).
\end{align}

\item[(O2)] For any binary subspace $A \subseteq \mathbb{Z}_2^n$, due to symmetry in binary subspaces we have $\sum_{x \in A} (-1)^{xv^T} = |A| \cdot \mathbbm{I}(v \in A^{\perp})$.
In other words, we have $|x \in A \colon xv^T \equiv 0| = |x \in A \colon xv^T \equiv 1|$ if and only if $v \notin A^{\perp}$.

\item[(O3)] For $a \in \mathbb{Z}_2^n$, a set such as $A \coloneqq \{ x \in \mathbb{Z}_2^n \colon x \preceq a \}$ is a subspace of $\mathbb{Z}_2^n$ since $x_1, x_2 \preceq a \Rightarrow (x_1 \oplus x_2) \preceq a$.
Hence, using (O2) we observe that $\sum_{x \in A} (-1)^{xv^T} = |A| \cdot \mathbbm{I}(v \in A^{\perp}) = 2^{w_H(a)} \cdot \mathbbm{I}(v \preceq \bar{a})$, where $\bar{a}$ is the ones' complement of $a$.

\item[(O4)] Let $a,b \in \mathbb{Z}_2^n$ have disjoint supports so that $a \ast b = 0$. Then we observe that
\begin{align}
\sum_{v \in \mathbb{Z}_2^n} \imath^{va^T} (-\imath)^{vb^T} & \overset{(i)}{=} \sum_{w \preceq \overline{(a \oplus b)}} \left( \sum_{v_a \preceq a} \imath^{v_a a^T} \right) \left( \sum_{v_b \preceq b} (-\imath)^{v_b b^T} \right) \\
  & = \sum_{w \preceq \overline{(a \oplus b)}} \left( \sum_{v_a \preceq a} \imath^{w_H(v_a)} \right) \left( \sum_{v_b \preceq b} (-\imath)^{w_H(v_b)} \right) \\
  & \overset{(ii)}{=} 2^{n - w_H(a) - w_H(b)} (1 + \imath)^{w_H(a)} (1 - \imath)^{w_H(b)}.
\end{align}
In step (i) we split any $v \in \mathbb{Z}_2^n$ uniquely as $v = v_a \oplus v_b \oplus w$, where $v_a$ is supported only on $a$, $v_b$ is supported only on $b$, and $w$ is supported outside the support of $a$ and $b$.
In step (ii), for each of the two sums over $v_a$ and $v_b$, we notice that the inner products $v_a a^T$ and $v_b b^T$ take values $q \in \{ 0,1,\ldots,w_H(v_a) \}$ and $q \in \{ 0,1,\ldots,w_H(v_b) \}$ according to the Hamming weights $w_H(v_a)$ and $w_H(v_b)$, respectively.
Hence, there are $\binom{n}{q}$ vectors $v_a \preceq a$ (resp. $v_b \preceq b$) that produce the inner product $q$, and this is captured in the binomial expansion of $(1 + \imath)^{w_H(a)}$ (resp. $(1 - \imath)^{w_H(b)}$).

\end{enumerate}

\subsection{Proof of Theorem~\ref{thm:transversal_T}}
\label{sec:proof_transversal_T}

We will make use of the above observations in the following proof.
%As mentioned earlier in the proof of Lemma~\ref{lem:conj_by_trans_T}, 
We proved in~\cite{Rengaswamy-pra19} that given a tensor product $\tau_{R_1}^{(\ell)} \otimes \tau_{R_2}^{(\ell)} \otimes \cdots \otimes \tau_{R_n}^{(\ell)}$, the result is also of the form $\tau_R^{(\ell)}$ with 
\begin{align}
R = 
\begin{bmatrix}
R_1 & 0 & \cdots & 0 \\
0 & R_2 & \cdots & 0 \\
\vdots & \vdots & \ddots & \vdots \\
0 & 0 & \cdots & R_n 
\end{bmatrix}.
\end{align}
Hence, for an $n$-qubit system, the transversal application of $T$ corresponds to $R = I_n$ and $\ell = 3$.
Based on the discussion in Section~\ref{sec:general_QFD}, for the case of transversal $T$, we need
\begin{align}
T^{\otimes n} \Pi_S (T^{\otimes n})^{\dagger} & = \frac{1}{2^r} \sum_{j = 1}^{2^r} \epsilon_j T^{\otimes n} E(a_j,b_j) (T^{\otimes n})^{\dagger} \\
  & = \frac{1}{2^r} \sum_{j = 1}^{2^r} \epsilon_j \cdot \frac{1}{2^{w_H(a_j)/2}} \sum_{y \preceq a_j} (-1)^{b_j y^T} E(a_j, b_j \oplus y) \\
  & = \frac{1}{2^r} \sum_{j = 1}^{2^r} \frac{\epsilon_j}{2^{\frac{w_H(a_j)}{2}}} \left[ E(a_j, b_j) + \sum_{\substack{y \preceq a_j\\y \neq 0}} (-1)^{b_j y^T} E(a_j, b_j \oplus y) \right] \\
\label{eq:transT_equality}
  & = \frac{1}{2^r} \sum_{j = 1}^{2^r} \epsilon_j E(a_j,b_j),
\end{align}
where only for the last step we have assumed that transversal $T$ preserves the code subspace.
Note that whenever $a_j = 0$, the denominator is $1$ and the inner summation is trivial since only $0 \preceq 0$.
Therefore, each such stabilizer $E(0,b_j)$ is retained unchanged (as we would expect since $T^{\otimes n}$ is diagonal and commutes with diagonal Paulis), and we only need to analyze the case $a_j \neq 0$.

For any index $j = h \in \{ 1,2,\ldots,2^r \}$, first we observe that we need $w_H(a_h)$ to be even in order to make the denominator an integer, which can be canceled by producing $2^{w_H(a_h)/2}$ copies of the stabilizer element $E(a_h,b_h)$ (with the appropriate sign $\epsilon_h \in \{ \pm 1 \}$) in the summation over all $2^r$ stabilizer elements.
This clearly shows that the first condition in the theorem is necessary for transversal $T$ to preserve the code space.
(We will call the sum over $j \in \{1,\ldots,2^r\}$ as the ``outer summation'' and the sum over $y \preceq a_j$ as the ``inner summation''; also, for a given $j = h$ we refer to the corresponding $E(a_h,b_h)$ as the ``outer summation term''.)
The only way to produce copies of $E(a_h,b_h)$ is through stabilizers $E(a_h,b_h')$ such that $b_h' \oplus y = b_h$ for some $y \preceq a_h$.
(Note: These two stabilizers correspond to two different outer summation terms for some indices $j$ and $j'$, where $a_j = a_{j'} = a_h$ but $b_j = b_h$ is distinct from $b_{j'} = b_h \oplus y$.)
If two such Paulis $E(a_h, b_h \oplus y), E(a_h, b_h \oplus z)$ must belong to $S$ then we need $E(a_h,b_h \oplus y)$ and $E(a_h,b_h \oplus z)$ to commute, which means we need $a_h y^T \equiv a_h z^T$ (mod $2$).
Since $y = 0$ must be included, we need $a_h z^T = zz^T = w_H(z) \equiv 0$ (mod $2$) for all such choices $z \preceq a_h$.
% There are $n_j \coloneqq 2^{w_H(a_j)-1}$ such even weight vectors $z \preceq a_j$ but we need exactly $2^{w_H(a_j)/2}$ of them, so we can pick $p_j \coloneqq w_H(a_j)/2$ linearly independent ones. %in $\prod_{i = 0}^{p_j - 1} (n_j - 2^i)$ ways.
% Let us denote the span of these vectors $z \preceq a_j$ to be $A_j$, and we have $|A_j| = 2^{w_H(a_j)/2}$.

There are $n_h \coloneqq 2^{w_H(a_h)-1}$ such even weight vectors $z \preceq a_h$ but only a subset of them might correspond to stabilizers in the inner summation.
Hence, let us define
\begin{align}
Z_h \coloneqq \{ z \preceq a_h \colon \epsilon_z E(0,z) \in S\ \text{for\ some}\ \epsilon_z \in \{ \pm 1 \} \}
% = \{ z \in Z_S \colon z \preceq a_h \},
\end{align}
% as per the definition of $Z_S$ 
as in the statement of the theorem.
It is clear that $Z_h$ is a binary subspace of even weight vectors, so let its dimension be $t_h \leq w_H(a_h) - 1$.
As mentioned above, every $z \in Z_h$ provides an outer summation term $E(a_h, b_h \oplus z)$ that produces a copy of $E(a_h,b_h)$ in its corresponding inner summation.
So we need at least $2^{p_h}$ such $z$'s in order to cancel the factor in the denominator, where $p_h \coloneqq w_H(a_h)/2$, i.e., $t_h \geq p_h$.

In order to prove that the second and third conditions in the theorem are necessary for transversal $T$ to preserve the code space, we need to ensure two things based on the equality imposed in~\eqref{eq:transT_equality}:
\begin{itemize}
    \item[(a)] For any $h \in \{ 1,2,\ldots,2^r \}$, over all outer summation indices $j$ there are a net of $2^{w_H(a_h)/2}$ copies of $E(a_h, b_h)$ (with the right sign) so that this stabilizer element is preserved under conjugation of the code projector by transversal $T$.
    
    \item[(b)] For any $h \in \{ 1,2,\ldots,2^r \}$, for each $y \notin Z_h$ and $y \preceq a_h$, over all outer summation indices $j$ there are exactly the same number of positive and negative copies of the non-stabilizer element $E(a_h, b_h \oplus y)$, which cancel out.
\end{itemize}
Let us express these two conditions mathematically by collecting signs of the respective Pauli elements.
For notational clarity, if the outer summation term is $E(a_h, b_h)$ then we write its sign as $\epsilon_{(a_h, b_h)}$ and similarly if the term is $E(a_h, b_h \oplus z)$ then we write its sign as $\epsilon_{(a_h, b_h \oplus z)}$.
For conditions (a) and (b), we respectively require
\begin{align}
\sum_{z \in Z_h} \epsilon_{(a_h, b_h \oplus z)} (-1)^{(b_h \oplus z) z^T} & = \sum_{z \in Z_h} \epsilon_{(a_h, b_h \oplus z)} (-1)^{b_h z^T} = \epsilon_{(a_h, b_h)} 2^{w_H(a_h)/2}, \\
\sum_{z \in Z_h} \epsilon_{(a_h, b_h \oplus z)} (-1)^{(b_h \oplus z) (z \oplus y)^T} & = (-1)^{b_h y^T} \sum_{z \in Z_h} \epsilon_{(a_h, b_h \oplus z)} (-1)^{b_h z^T} (-1)^{zy^T} = 0.
\end{align}
If we ignore the overall sign $(-1)^{b_h y^T}$ in the second equation, then we see that it is impossible to have $(-1)^{z y^T} = 1$ for all $z \in Z_h$, because otherwise we have a contradiction between the two conditions.
Therefore, since $y \notin Z_h$, by the symmetry of binary vectors spaces (observation (O2)), we must have $|\{ z \in Z_h \colon zy^T \equiv 0 \}| = |\{ z \in Z_h \colon zy^T \equiv 1 \}|$.
To proceed further we will find the following lemma about binary subspaces to be useful.

\begin{lemma}
\label{lem:C_self_dual}
For even $n$, let $C$ be an $[n,k \geq \frac{n}{2}]$ binary linear code comprising only of even weight codewords.
Then the following are equivalent:
\begin{enumerate}
    \item $C$ contains its dual $C^{\perp}$.
    \item $C$ contains an $[n,\frac{n}{2}]$ self-dual subcode $C_s$.
    \item Any vector $y \in \mathbb{Z}_2^n \setminus C$ satisfies $|\{ x \in C \colon xy^T \equiv 0 \}| = |\{ x \in C \colon xy^T \equiv 1 \}|$.
\end{enumerate}
\end{lemma}
\begin{IEEEproof}
First we show that 2) implies 3).
Let $G \in \mathbb{Z}_2^{k \times n}$ be a generator matrix for $C$ with rows $g_1,g_2,\ldots,g_k$ such that $g_1,g_2,\ldots,g_{n/2}$ form a submatrix $G_s$ that generates the self-dual subcode $C_s$.
Then any $y \in \mathbb{Z}_2^n \setminus C$ cannot be orthogonal to all rows of $G_s$, since otherwise $y \in C_s$.
Let $yg_1^T = 1$ (mod $2$) without loss of generality.
Then for any other $g_i$ such that $yg_i^T = 1$ (mod $2$), one can replace $g_i$ with $g_i \oplus g_1$ to form a new matrix $G'$ with rows $g_1,g_2',g_3',\ldots,g_k'$ that still spans $C$ (i.e., some $g_i'$ can be equal to $g_i$).
Hence, now $y(g_i')^T = 0$ (mod $2$) for all $i=2,3,\ldots,k$ and $yx^T = 1$ (mod $2$) for all $x$ in the coset $g_1 \oplus \text{span}(g_2',g_3',\ldots,g_k')$.
This shows that $|\{ x \in C \colon xy^T \equiv 0 \}| = |\{ x \in C \colon xy^T \equiv 1 \}|$.

% We start by assuming that $C$ does not contain a self-dual code.
% Hence, if $G \in \mathbb{Z}_2^{k \times n}$ is a generator matrix for $C$ with rows $g_1,g_2,\ldots,g_k$, then at most a subset of $(\frac{n}{2} - 1)$ rows form a self-orthogonal code.
% In that case one can always find a vector $y \in \mathbb{Z}_2^n \setminus C$ that satisfies $Gy^T = 0$ (mod $2$).
% This would imply that we have found a vector $y \in \mathbb{Z}_2^n \setminus C$ that violates $|\{ x \in C \colon xy^T \equiv 0 \}| = |\{ x \in C \colon xy^T \equiv 1 \}|$, which is a contradiction.
% Hence, $C$ must contain a self-dual code.
% Finally, note that if $C$ indeed contained a self-dual code, then solving for $Gy^T = 0$ will produce a $y$ that belonged to $C$, since here $y$ is orthogonal to the self-dual code.

Next we show that 3) implies 1) and 1) implies 2).
Now we start by assuming that every $y \in \mathbb{Z}_2^n \setminus C$ satisfies $|\{ x \in C \colon xy^T \equiv 0 \}| = |\{ x \in C \colon xy^T \equiv 1 \}|$.
This means that all vectors orthogonal to $C$ are contained in $C$, i.e., $C^{\perp} \subseteq C$.
Without loss of generality, assume that we have a generator matrix $G$ for $C$ with rows $g_1,g_2,\ldots,g_k$ such that $g_1,g_2,\ldots,g_{n-k}$ form a submatrix $G^{\perp}$ that generates the dual code $C^{\perp}$.
Then we can enlarge $C^{\perp}$ into a dimension $(n-k+1)$ self-orthogonal code by adding $g_{n-k+1}$ to the rows of $G^{\perp}$, since $g_{n-k+1} \in C$ is dual to $C^{\perp}$ and to itself (as it has even weight).
Now notice that all vectors orthogonal to this new $C^{\perp}$ are still in the rowspace of $G$ since the dual of the span of the first $(n-k)$ rows are contained in the rowspace of $G$.
Hence, it is possible to find a row between $g_{n-k+2}$ and $g_k$ that can be used to enlarge $C^{\perp}$ into a dimension $(n-k+2)$ self-orthogonal code.
Intuitively, as $C^{\perp}$ is enlarged, its dual keeps shrinking starting from $C$ at the beginning.
We can continue this process and, since we assumed $k \geq n/2$, it stops only when $C^{\perp}$ is a dimension $n/2$ self-dual code, as required.
\end{IEEEproof}

Let $\tilde{Z}_h$ denote the subspace $Z_h$ where all the indices outside the support of $a_h$ are punctured, i.e., $\tilde{Z}_h \subseteq \{ 0,1 \}^{w_H(a_h)}$.
Setting $C = \tilde{Z}_h$ in Lemma~\ref{lem:C_self_dual}, we see that $\tilde{Z}_h$ must contain its dual $(\tilde{Z}_h)^{\perp}$ and also a dimension $w_H(a_h)/2$ self-dual code $\tilde{Z}_h^s$.
In other words, $Z_h$ must contain a dimension $w_H(a_h)/2$ self-dual code in the support of $a_h$.
This proves the necessity of the second condition in the theorem.

Let us compute the sign $\epsilon_{(a_h, b_h \oplus z)}$ of $E(a_h, b_h \oplus z)$ assuming that $\epsilon_h E(a_h, b_h), \epsilon_z E(0,z) \in S$.
Using~\eqref{eq:Eab_multiply},~\eqref{eq:Eab_2x} we calculate
\begin{align}
\epsilon_h E(a_h, b_h) \cdot \epsilon_z E(0,z) & = \epsilon_h \epsilon_z \imath^{-a_h z^T} E(a_h, b_h + z) \\
   & = \epsilon_h \epsilon_z \imath^{zz^T} E(a_h, (b_h \oplus z) + 2 (b_h \ast z)) \\
   & = \epsilon_h \epsilon_z \imath^{zz^T} (-1)^{a_h (b_h \ast z)^T} E(a_h, b_h \oplus z) \\
   & = \epsilon_h \epsilon_z \imath^{zz^T} (-1)^{b_h z^T} E(a_h, b_h \oplus z) \\
\Rightarrow \epsilon_{(a_h, b_h \oplus z)} & = \epsilon_{(a_h, b_h)} \epsilon_{(0,z)} \imath^{zz^T} (-1)^{b_h z^T},
\end{align}
where we have used observation (O1) and for the fourth equality we have assumed that $z \preceq a_h$.
Now we start analyzing the sum for condition (a) above by observing that $w_H(z \oplus v) = zz^T + vv^T - 2zv^T$.
\begin{align}
\Gamma & \coloneqq \sum_{z \in Z_h} \epsilon_{(a_h, b_h \oplus z)} (-1)^{b_h z^T} \\
   & = \sum_{z \in Z_h} \epsilon_{(a_h, b_h)} \epsilon_{(0,z)} \imath^{zz^T} (-1)^{b_h z^T} (-1)^{b_h z^T} \\
   & = \epsilon_{(a_h, b_h)} \sum_{z \in \tilde{Z}_h} \epsilon_{(0,z)} \imath^{zz^T} \\
\Rightarrow \Gamma^2 & = \sum_{z,v \in \tilde{Z}_h} \epsilon_{(0,z)} \epsilon_{(0,v)} \imath^{zz^T + vv^T} \\
   & = \sum_{z,v \in \tilde{Z}_h} \epsilon_{(0,z \oplus v)} \imath^{w_H(z \oplus v)} (-1)^{(z \oplus v) v^T} \\
   & = \sum_{u \in \tilde{Z}_h} \epsilon_{(0,u)} \imath^{uu^T} \left( \sum_{v \in \tilde{Z}_h} (-1)^{uv^T} \right)\ (u \coloneqq z \oplus v) \\
   & = |\tilde{Z}_h| \sum_{u \in (\tilde{Z}_h)^{\perp}} \epsilon_{(0,u)} \imath^{uu^T}\ (\text{by\ observation\ (O3)}).
\end{align}
We need this sum over $(\tilde{Z}_h)^{\perp}$ to be $|(\tilde{Z}_h)^{\perp}|$ so that $|\Gamma| = 2^{w_H(a_h)/2}$ as required.
Therefore, it is clear that we need $\epsilon_{(0,u)} \coloneqq \imath^{uu^T}$ for all $u \in (\tilde{Z}_h)^{\perp}$.
Again, note that these vectors $u$ essentially represent $n$-qubit pure $Z$-type stabilizers even though $(\tilde{Z}_h)^{\perp}$ is a space where all indices outside the support of $a_h$ have been punctured.
The subtlety is that $(\tilde{Z}_h)^{\perp} \neq \tilde{(Z_h^{\perp})}$ since the latter computes the dual over $\{ 0,1 \}^n$, which is not what we want here.
Let us also verify that condition (b) above is satisfied. % for all $y \notin \tilde{Z}_h$.
\begin{align}
\Delta & \coloneqq \sum_{z \in Z_h} \epsilon_{(a_h, b_h \oplus z)} (-1)^{b_h z^T} (-1)^{zy^T} \\
  & = \sum_{z \in Z_h} \epsilon_{(a_h, b_h)} \epsilon_{(0,z)} \imath^{zz^T} (-1)^{b_h z^T} (-1)^{b_h z^T} (-1)^{zy^T} \\
  & = \epsilon_{(a_h, b_h)} \sum_{z \in \tilde{Z}_h} \epsilon_{(0,z)} \imath^{zz^T} (-1)^{zy^T} \\
  & = \epsilon_{(a_h, b_h)} \sum_{\substack{w \in \tilde{Z}_h/(\tilde{Z}_h)^{\perp}\\ u \in (\tilde{Z}_h)^{\perp}}} \epsilon_{(0, w \oplus u)} \imath^{w_H(w \oplus u)} (-1)^{(w \oplus u)y^T} \\
  & = \epsilon_{(a_h, b_h)} \sum_{w \in \tilde{Z}_h/(\tilde{Z}_h)^{\perp}} \epsilon_{(0,w)} \imath^{ww^T} (-1)^{wy^T} \nonumber \\
  & \hspace*{1.5cm} \left( \sum_{u \in (\tilde{Z}_h)^{\perp}} \epsilon_{(0,u)} \imath^{uu^T} (-1)^{(w \oplus y) u^T} \right) \\
  & = 0,
\end{align}
since $uw^T = 0$ for all $u \in (\tilde{Z}_h)^{\perp}$ and, again, by symmetry of binary vector spaces and because $y \notin \tilde{Z}_h$, we have $|u \in (\tilde{Z}_h)^{\perp} \colon uy^T \equiv 0| = |u \in (\tilde{Z}_h)^{\perp} \colon uy^T \equiv 1|$ (observation (O2)).
Thus, this establishes the necessity of the third condition in the theorem.

Now for the converse we assume that the first two conditions in the theorem hold true and that $\epsilon_{(0,u)} \coloneqq \imath^{uu^T}$ for all $u \in \tilde{Z}_h^{s}$, where $\tilde{Z}_h^{s}$ is the self-dual code present inside $\tilde{Z}_h$.
Once again, we just have to show that conditions (a) and (b) above are satisfied under these assumptions.
For (b), the above calculation for $\Delta$ itself suffices since $(\tilde{Z}_h)^{\perp} \subseteq \tilde{Z}_h^s$, so we are only left to show that $\Gamma = \epsilon_{(a_h, b_h)} 2^{w_H(a_h)/2}$.
We observe that
\begin{align}
\Gamma & = \epsilon_{(a_h, b_h)} \sum_{z \in \tilde{Z}_h} \epsilon_{(0,z)} \imath^{zz^T} \\
  & = \epsilon_{(a_h, b_h)} \sum_{w \in \tilde{Z}_h/\tilde{Z}_h^s} \sum_{u \in \tilde{Z}_h^s} \epsilon_{(0,w \oplus u)} \imath^{ww^T + uu^T - 2wu^T} \\
  & = \epsilon_{(a_h, b_h)} \sum_{w \in \tilde{Z}_h/\tilde{Z}_h^s} \epsilon_{(0,w)} \imath^{ww^T} 
  \left( \sum_{u \in \tilde{Z}_h^s} \epsilon_{(0,u)} \imath^{uu^T} (-1)^{wu^T} \right) \\
  & = \epsilon_{(a_h, b_h)} \sum_{w \in \tilde{Z}_h/\tilde{Z}_h^s} \epsilon_{(0,w)} \imath^{ww^T} 
  \left( \sum_{u \in \tilde{Z}_h^s} (-1)^{wu^T} \right) \\
  & = \epsilon_{(a_h, b_h)} |\tilde{Z}_h^s| \\
  & = \epsilon_{(a_h, b_h)} 2^{w_H(a_h)/2},
\end{align}
since only $w = 0 \in \tilde{Z}_h/\tilde{Z}_h^s$ is orthogonal to all $u \in \tilde{Z}_h^s$ (observation (O2)).
This completes the proof of the theorem. \hfill \IEEEQEDhere

\subsection{Proof of Lemma~\ref{lem:conj_by_trans_T_Tinv}}
\label{sec:proof_conj_by_trans_T_Tinv}

We will use the observations listed at the beginning of Appendix~\ref{sec:all_proofs} to complete this proof.
As mentioned earlier in the proof of Theorem~\ref{thm:transversal_T}, we proved in~\cite{Rengaswamy-pra19} that given a tensor product of diagonal unitaries $\tau_{R_1}^{(\ell)} \otimes \tau_{R_2}^{(\ell)} \otimes \cdots \otimes \tau_{R_n}^{(\ell)}$, the result is also of the form $\tau_R^{(\ell)}$ with 
\begin{align}
R = 
\begin{bmatrix}
R_1 & 0 & \cdots & 0 \\
0 & R_2 & \cdots & 0 \\
\vdots & \vdots & \ddots & \vdots \\
0 & 0 & \cdots & R_n 
\end{bmatrix}.
\end{align}
Hence, for an $n$-qubit system, the transversal application of $T$ gate corresponds to $R = I_n$ and $\ell = 3$.
%Based on similar arguments in the beginning of Lemma~\ref{sec:proof_conj_by_trans_T}, it is easy to see that the symmetric matrix corresponding to $T^{\otimes t}$ is $R = D_t$, where $t = t_1 + 7 t_7$.
Similarly, it is easy to see that the symmetric matrix corresponding to $T^{\otimes t}$ is $R = D_t$, where $t = t_1 + 7 t_7$ and $D_t$ is a diagonal $n \times n$ matrix with the diagonal set to $t$.
Then, according to~\eqref{eq:tau_Eab}, we see that
\begin{align}
\tilde{R}(D_t,a,3) & = 3 D_{a \ast t} - (D_{1-a} D_t D_a + D_a D_t D_{1-a} + 2 D_{a \ast t}) = D_{a \ast t} = D_{a \ast t_1} + 7 D_{a \ast t_7}, \\
\phi(D_t,a,b,3) & = - a D_t a^T = - at^T = -at_1^T - 7 at_7^T = - w_H(a \ast t_1) + w_H(a \ast t_7)\ (\text{mod}\ 8).
\end{align}
We need to calculate $c_{D_{a \ast t},x}^{(2)} = \frac{1}{\sqrt{2^n}} \sum_{v \in \mathbb{Z}_2^n} (-1)^{vx^T} \imath^{v D_{a \ast t_1} v^T} (-\imath)^{v D_{a \ast t_7} v^T}$ for all $x \in \mathbb{Z}_2^n$.
For $x = 0$, we observe
\begin{align}
c_{D_{a \ast t},0}^{(2)} & = \frac{1}{\sqrt{2^n}} \sum_{v \in \mathbb{Z}_2^n} \imath^{v D_{a \ast t_1} v^T} (-\imath)^{v D_{a \ast t_7} v^T} \\
  & = \frac{1}{\sqrt{2^n}} \sum_{v \in \mathbb{Z}_2^n} \imath^{v (a \ast t_1)^T} (-\imath)^{v (a \ast t_7)^T} \\
  & = \frac{1}{\sqrt{2^n}} 2^{n - w_H(a \ast t_1) - w_H(a \ast t_7)} (1 + \imath)^{w_H(a \ast t_1)} (1 - \imath)^{w_H(a \ast t_7)} \ (\text{observation\ (O4)}) \\
  & = e^{\frac{\imath \pi}{4} \left[ w_H(a \ast t_1) - w_H(a \ast t_7) \right]} 2^{\left( n - w_H(a \ast t') \right)/2} \ (\text{since}\ e^{\pm \frac{\imath \pi}{4}} = (1 \pm \imath)/\sqrt{2}).
\end{align}
For the case $x \neq 0$, we %simplify the steps in Section~\ref{sec:proof_conj_by_trans_T} 
calculate as follows.
Recollect that we have defined $t' = t_1 + t_7 \in \mathbb{Z}_2^n$.
\begin{align}
c_{D_{a \ast t}, x}^{(2)} & = \frac{1}{\sqrt{2^n}} \sum_{v \in \mathbb{Z}_2^n} \imath^{v \left[ (a \ast t_1) + 3(a \ast t_7) + 2x \right]^T} \\
  & = \frac{1}{\sqrt{2^n}} \sum_{v \preceq (a \ast t')} \imath^{v \left[ (a \ast t_1) + 3(a \ast t_7) + 2x \right]^T} \sum_{w \preceq \overline{(a \ast t')}} (-1)^{w x^T} \ (\text{observation\ (O4)}) \\
  & = 2^{\frac{n}{2} - w_H(a \ast t')} \left( \sum_{v \preceq (a \ast t') \oplus x} \imath^{v \left[ (a \ast t_1) + 3(a \ast t_7) + 2x \right]^T} \right) \left( \sum_{w \preceq x} \imath^{w \left[ (a \ast t_1) + 3(a \ast t_7) + 2x \right]^T} \right),\ x \preceq (a \ast t').
\end{align}
In the last step, unless $x \preceq (a \ast t')$ it is easy to see that the inner summation in the second step vanishes (observation (O3)).
Furthermore, we have used this property of $x$ to split the sum into indices on the support of $x$ and the others (observation (O4)).
For convenience, let us denote by $A$ the support of $(a \ast t') \oplus x$, by $A_1$ the support of $(a \ast t_1) \oplus (x \ast a \ast t_1)$, and by $A_7$ the support of $(a \ast t_7) \oplus (x \ast a \ast t_7)$.
Note that $A = \text{supp}(a \ast t') \setminus \text{supp}(x)$ and hence $vx^T = 0$.
For simplicity, we will write $\{ v \in \mathbb{Z}_2^n \colon \text{supp}(v) \subseteq A \}$ as $v \preceq A$.
Then, using observation (O4), we can write the first sum above as
\begin{align}
\sum_{v \preceq A} \imath^{v (a \ast t_1)^T} (-\imath)^{v (a \ast t_7)^T} & = \sum_{z_1 \preceq A_1} \imath^{z_1 (a \ast t_1)^T} \sum_{z_2 \preceq A_7} (-\imath)^{z_2 (a \ast t_7)^T} \\
  & = (1 + \imath)^{w_H(a \ast t_1) - w_H(x \ast a \ast t_1)} (1 - \imath)^{w_H(a \ast t_7) - w_H(x \ast a \ast t_7)} \\
  & = e^{\frac{\imath\pi}{4} \left[ \left( w_H(a \ast t_1) - w_H(a \ast t_7) \right) - \left( w_H(x \ast a \ast t_1) - w_H(x \ast a \ast t_7) \right)  \right]} 2^{\left( w_H(a \ast t') - w_H(x) \right)/2}.
\end{align}
Now, again using observation (O4), we can calculate the second sum similarly as follows.
\begin{align}
\sum_{w \preceq x} \imath^{w \left[ (a \ast t_1) + 3(a \ast t_7) + 2x \right]^T} & = \sum_{w \preceq x \ast (a \ast t_1)} (-\imath)^{ww^T} \sum_{z \preceq x \ast (a \ast t_7)} \imath^{zz^T} \\
  & = (1 - \imath)^{w_H(x \ast a \ast t_1)} (1 + \imath)^{w_H(x \ast a \ast t_7)} \\
  & = e^{\frac{-\imath\pi}{4} \left[ w_H(x \ast a \ast t_1) - w_H(x \ast a \ast t_7) \right]} 2^{w_H(x)/2}.
\end{align}
Combing these two results and substituting back we get,
\begin{align}
c_{D_{a \ast t},x}^{(2)} = 
\begin{cases}
2^{\left( n - w_H(a \ast t') \right)/2} e^{\frac{\imath\pi}{4} \left[ \left( w_H(a \ast t_1) - w_H(a \ast t_7) \right) - 2 \left( w_H(x \ast a \ast t_1) - w_H(x \ast a \ast t_7) \right) \right]}  & \text{if}\ x \preceq (a \ast t'), \\
0 & \text{otherwise}.
\end{cases}
\end{align}
Recollect from~\eqref{eq:tau_Eab_expand} that the action of $\tau_R^{(\ell)}$ on a Pauli matrix $E(a,b)$ is given by
\begin{align}
\tau_R^{(\ell)} E(a,b) ( \tau_R^{(\ell)} )^{\dagger} & = \frac{1}{\sqrt{2^n}} \xi^{\phi(R,a,b,\ell)} \sum_{x \in \mathbb{Z}_2^n} c_{\tilde{R}(R,a,\ell),x}^{(\ell-1)} \imath^{-ax^T} E(a, b + aR + x).
\end{align}
Hence, using observation (O1), we can calculate the action of $T^{\otimes t}$ on $E(a,b)$ under conjugation to be
\begin{align}
& T^{\otimes t} E(a,b) \left( T^{\otimes t} \right)^{\dagger} \nonumber \\
 & = \frac{1}{\sqrt{2^n}} e^{\frac{-\imath\pi}{4} \left[ w_H(a \ast t_1) - w_H(a \ast t_7) \right]} \sum_{x \preceq (a \ast t')} c_{D_{a \ast t},x}^{(2)} \imath^{-a x^T} E(a, b + aD_t + x) \\
  & = \frac{2^{\left( n - w_H(a \ast t') \right)/2}}{\sqrt{2^n}} \sum_{x \preceq (a \ast t')} \imath^{-x \left[ (a \ast t_1) - (a \ast t_7) \right]^T - x \left[ (a \ast t_1) + (a \ast t_7) \right]^T} E(a, b + (a \ast t_1) + 7 (a \ast t_7) + x) \\
  & = \frac{1}{2^{w_H(a \ast t')/2}} \sum_{x \preceq (a \ast t')}  (-1)^{x (a \ast t_1)^T} E(a, b + ((a \ast t') + x) + 6 (a \ast t_7) ) \\
  & = \frac{1}{2^{w_H(a \ast t')/2}} \sum_{x \preceq (a \ast t')}  (-1)^{x (a \ast t_1)^T + a (a \ast t_7)^T} E\left( a, b + \left( (a \ast t') \oplus x + 2 (x \ast (a \ast t')) \right) \right) \\
  & = \frac{1}{2^{w_H(a \ast t')/2}} \sum_{x \preceq (a \ast t')}  (-1)^{x (a \ast t_1)^T + w_H(a \ast t_7)} E\left( a, b + \left( (a \ast t') \oplus x + 2 x \right) \right) \\
  & = \frac{1}{2^{w_H(a \ast t')/2}} \sum_{x \preceq (a \ast t')}  (-1)^{x (a \ast t_1)^T + w_H(a \ast t_7) + xa^T} E\left( a, b \oplus \left( (a \ast t') \oplus x \right) + 2 b \ast ((a \ast t') \oplus x) \right) \\
  & = \frac{1}{2^{w_H(a \ast t')/2}} \sum_{x \preceq (a \ast t')}  (-1)^{x (a \ast t_7)^T + (a \ast t_7) (a \ast t_7)^T + a \left[ b \ast ((a \ast t') \oplus x) \right]^T} E\left( a, b \oplus \left( (a \ast t') \oplus x \right) \right) \\
  & = \frac{1}{2^{w_H(a \ast t')/2}} \sum_{x \preceq (a \ast t')}  (-1)^{(a \ast t_7) \left[(a \ast t_7) \oplus x \right]^T + a \left[ b \ast ((a \ast t') \oplus x) \right]^T} E\left( a, b \oplus \left( (a \ast t') \oplus x \right) \right) \\
  & = \frac{1}{2^{w_H(a \ast t')/2}} \sum_{y \preceq (a \ast t')}  (-1)^{(a \ast t_7) \left[y \oplus (a \ast t_1) \right]^T + a ( b \ast y )^T} E\left( a, b \oplus y \right) \ (y \coloneqq (a \ast t') \oplus x) \\
  & = \frac{1}{2^{w_H(a \ast t')/2}} \sum_{y \preceq (a \ast t')}  (-1)^{y(a \ast t_7)^T + a ( b \ast y )^T} E\left( a, b \oplus y \right) \\
  & = \frac{1}{2^{w_H(a \ast t')/2}} \sum_{y \preceq (a \ast t')}  (-1)^{(b + t_7) y^T} E\left( a, b \oplus y \right).
\end{align}
The last step follows from $y \preceq (a \ast t') \Rightarrow y \preceq a \Rightarrow (b \ast y) \preceq a$ and $y(a \ast t_7)^T = w_H((y \ast a) \ast t_7) = w_H(y \ast t_7) = y t_7^T$.  \hfill \IEEEQEDhere

\subsection{Proof of Corollary~\ref{cor:conj_by_trans_T_Tinv}}
\label{sec:proof_cor_conj_by_trans_T_Tinv}

We have $t = \sum_{j=1}^{7} j t_j$ with $t_j \ast t_{j'} = 0$ for all $j \neq j'$.
Then we can write 
\begin{align*}
T^{\otimes t} = T^{t_1 + t_3 + t_5 + t_7} P^{t_2 + t_3 + t_6 + t_7} Z^{t_4 + t_5 + t_6 + t_7},
\end{align*}
where $Z^{t_4 + t_5 + t_6 + t_7} = E(0, t_4 + t_5 + t_6 + t_7)$.
Now we can compute $T^{\otimes t} E(a,b) (T^{\otimes t})^{\dagger}$ as follows.
Firstly we observe
\begin{align}
E(0, t_4 + t_5 + t_6 + t_7) \cdot E(a,b) \cdot E(0, t_4 + t_5 + t_6 + t_7) = (-1)^{a (t_4 + t_5 + t_6 + t_7)^T} E(a,b).
\end{align}
Next, using observation (O1), and observing that $P^q E(a,b) (P^{\dagger})^q = E(a,b+ (a \ast q))$ for $q \in \mathbb{Z}_2^n$, we can calculate
\begin{align}
P^{t_2 + t_3 + t_6 + t_7} \cdot & \, (-1)^{a (t_4 + t_5 + t_6 + t_7)^T} E(a,b) \cdot (P^{\dagger})^{t_2 + t_3 + t_6 + t_7} \nonumber \\
  & = (-1)^{a (t_4 + t_5 + t_6 + t_7)^T} E(a, b + (a \ast (t_2 + t_3 + t_6 + t_7)) ) \\
  & = (-1)^{a (t_4 + t_5 + t_6 + t_7)^T + a (b \ast (t_2 + t_3 + t_6 + t_7))^T} E(a, b \oplus (a \ast (t_2 + t_3 + t_6 + t_7)) )  \\
  & = (-1)^{a (t_4 + t_5 + t_6 + t_7)^T + a (b \ast (\tilde{t}_2 + \tilde{t}_3))^T} E(a, b \oplus (a \ast (\tilde{t}_2 + \tilde{t}_3)) ) \\
  & = (-1)^{a (t_4 + t_5 + t_6 + t_7)^T + w_H(a \ast b \ast \tilde{t}_2) + w_H(a \ast b \ast \tilde{t}_3)} E(a, c ),
\end{align}
where we have defined $c \coloneqq b \oplus (a \ast (\tilde{t}_2 + \tilde{t}_3))$ for convenience.
Finally, we can invoke Lemma~\ref{lem:conj_by_trans_T_Tinv} with the $t_1, t_7$ in that lemma taken respectively to be $t_1 + t_3 + t_5 + t_7$ and $0$ here.
Then we have
\begin{align}
& T^{t_1 + t_3 + t_5 + t_7} \cdot (-1)^{a (t_4 + t_5 + t_6 + t_7)^T + w_H(a \ast b \ast \tilde{t}_2) + w_H(a \ast b \ast \tilde{t}_3)} E(a,c) \cdot (T^{\dagger})^{t_1 + t_3 + t_5 + t_7} \nonumber \\    
  & = \frac{(-1)^{a (t_4 + t_5 + t_6 + t_7)^T + w_H(a \ast b \ast \tilde{t}_2) + w_H(a \ast b \ast \tilde{t}_3)}}{2^{w_H(a \ast (\tilde{t}_1 + \tilde{t}_3))/2}} \sum_{y \preceq a \ast (\tilde{t}_1 + \tilde{t}_3)} (-1)^{cy^T} E(a, c \oplus y) \\
  & = \frac{(-1)^{a (t_4 + t_5 + t_6 + t_7)^T + w_H(a \ast b \ast \tilde{t}_2) + w_H(a \ast b \ast \tilde{t}_3)}}{2^{w_H(a \ast (\tilde{t}_1 + \tilde{t}_3))/2}} \sum_{y \preceq a \ast (\tilde{t}_1 + \tilde{t}_3)} (-1)^{by^T + y (a \ast (\tilde{t}_2 + \tilde{t}_3))^T} E(a, b \oplus [(a \ast (\tilde{t}_2 + \tilde{t}_3)) \oplus y]) \\
  & = \frac{(-1)^{a (t_4 + t_5 + t_6 + t_7)^T + w_H(a \ast (\tilde{t}_2 + \tilde{t}_3)) }}{2^{w_H(a \ast (\tilde{t}_1 + \tilde{t}_3))/2}} \sum_{(a \ast \tilde{t}_2) \preceq z \preceq a \ast (\tilde{t}_1 + \tilde{t}_2 + \tilde{t}_3)} (-1)^{bz^T + z (a \ast (\tilde{t}_2 + \tilde{t}_3))^T} E(a, b \oplus z),
\end{align}
where we have defined $z \coloneqq (a \ast (\tilde{t}_2 + \tilde{t}_3)) \oplus y$.
In the last step, first we observe that $(a \ast \tilde{t}_2)$ is always present in $z$ and is unaffected by the range of $y$ above since $\tilde{t}_1 \ast \tilde{t}_2 = 0 = \tilde{t}_3 \ast \tilde{t}_2$.
Hence, we write that $z$ contains and is contained in $(a \ast \tilde{t}_2)$.
Next, since the change of variables does not change the support of $y$ inside $(a \ast \tilde{t}_1)$, the new variable $z$ also loops over all vectors contained in $(a \ast \tilde{t}_1)$.
Finally, since $z$ adds $(a \ast \tilde{t}_3)$ to $y$ (modulo 2), we see that $y$ undergoes a ones' complement in the support of $(a \ast \tilde{t}_3)$.
However, since $y$ takes all possible values in the support of $(a \ast \tilde{t}_3)$, we conclude that $z$ still retains that property.
Furthermore, under the change of variables, we have cancelled the new factors $(-1)^{b (a \ast \tilde{t}_2)^T}$ and $(-1)^{b (a \ast \tilde{t}_3)^T}$ correspondingly with the existing factors $(-1)^{w_H(b \ast a \ast \tilde{t}_2)}$ and $(-1)^{w_H(b \ast a \ast \tilde{t}_3)}$.
Now we observe that 
\begin{align}
z (a \ast (\tilde{t}_2 + \tilde{t}_3))^T = w_H(z \ast (a \ast \tilde{t}_2)) + w_H((z \ast a) \ast \tilde{t}_3) = w_H(a \ast \tilde{t}_2) + w_H(z \ast \tilde{t}_3) = w_H(a \ast \tilde{t}_2) + \tilde{t}_3 z^T,
\end{align}
since clearly $z \preceq a$.
Substituting this back, and noting that $\tilde{t}_3 = t_3 + t_7$, we obtain 
\begin{align}
T^{\otimes t} E(a,b) \left( T^{\otimes t} \right)^{\dagger} = \frac{ (-1)^{a (t_3 + t_4 + t_5 + t_6)^T} }{2^{w_H(a \ast (\tilde{t}_1 + \tilde{t}_3))/2}} \sum_{(a \ast \tilde{t}_2) \preceq z \preceq (a \ast (\tilde{t}_1 + \tilde{t}_2 + \tilde{t}_3))}  (-1)^{(b + \tilde{t}_3) z^T} E\left( a, b \oplus z \right),
\end{align}
which is the final expression given in the statement of the lemma. \hfill \IEEEQEDhere

\subsection{Proof of Corollary~\ref{cor:csst_sufficient}}
\label{sec:proof_csst_sufficient}

For convenience, we will use the notation $A,B,C,D$ to also refer to the subspaces $\langle A \rangle, \langle B \rangle, \langle C \rangle, \langle D \rangle$ generated by the matrices $A,B,C,D$, respectively.
We assume that $A$ and $C$ are disjoint, that $B$ and $D$ are disjoint, and that $C$ and $D$ have full rank, all without loss of generality.
Note that if some of these are not satisfied, then we can perform suitable row operations to subsume rows $\begin{bmatrix} a & 0 \end{bmatrix}$ into $\begin{bmatrix} C & 0 \end{bmatrix}$ and rows $\begin{bmatrix} 0 & b \end{bmatrix}$ into $\begin{bmatrix} 0 & D \end{bmatrix}$.
We will prove the result for $t = [1,1,\ldots,1]$, i.e., transversal $T$, but the extension to any $t \in \{0,1,7\}^n$ is straightforward as we comment at the end of the proof.
Firstly, it is clear that the CSS code has the transversal $T$ property because this depends on the subspace $\langle A,C \rangle$ being even and the existence of a self-dual code in $D$ within the support of each vector in the subspace $\langle A,C \rangle$. 
Dropping $B$ does not affect these properties. 
It is also clear that the CSS code still encodes $k$ qubits if $A$ has full rank, but if this is violated then some rows of the stabilizer matrix are removed to provide room for more than $k$ logical qubits (without affecting the transversal $T$ property).

Secondly, the distance of the CSS code is lower bounded by the minimum of the minimum weights of $\langle A,C \rangle^{\perp}$ and $D^{\perp}$. 
Since $\begin{bmatrix} 0 & \langle A,C \rangle^{\perp} \end{bmatrix}$ belongs to the normalizer of the given stabilizer code, and $D$ has minimum weight at least $d$ by non-degeneracy, we know that the minimum weight of $\langle A,C \rangle^{\perp}$ is at least $d$. 
Originally, $\begin{bmatrix} \langle B,D \rangle^{\perp} & 0 \end{bmatrix}$ is in the normalizer of the given stabilizer, so the minimum weight of $\langle B,D \rangle^{\perp}$ is at least $d$ as well. 
Since $\begin{bmatrix} C & 0 \end{bmatrix}$ was initially in the stabilizer, we know that $C \subset \langle B,D \rangle^{\perp}$ and minimum weight of $C$ must be at least $d$ by non-degeneracy. 
However, $A \subset D^{\perp}$ and $\begin{bmatrix} A & 0 \end{bmatrix}$ was not originally part of the stabilizer, so it appears that $A$ might have vectors of weight less than $d$. 
But by non-degeneracy, minimum weight of $D$ is $d$. 
So the minimum weight of any self-dual code $C_x \subset D$ is $d$, for any $x \in \langle A,C \rangle$. 
Hence, this means that $w_H(x) \geq d$ since $x \in C_x$
\footnote{Strictly speaking, we are only concerned with the minimum weight of $D^{\perp} \setminus \langle A, C \rangle$ and not of $D^{\perp}$, i.e., we do not require that the new CSS code is also non-degenerate. 
However, the above shows that the minimum weight of $\langle A,C \rangle$ is already at least $d$ due to Theorem~\ref{thm:transversal_T}.}.

Now consider a $z \in D^{\perp}$. 
We want to show that a minimal weight $z$ has weight at least $d$. 
Suppose, for the sake of contradiction, that $w_H(z)$ is minimal but is strictly less than $d$.
%Note that we are not going to assume that $z$ is outside $D$ explicitly, since by the assumption for contradiction where we take $w_H(z) < d$, this is implicit. 
Now consider any $x \in \langle A,C \rangle$ and look at the projection $(x \ast z)$. 
By assumption, $z$ is orthogonal to $D$ and hence is orthogonal to $C_x$. 
But the inner product of $z$ with any vector in $C_x$ only depends on the projection $(x \ast z)$. 
Therefore, $(x \ast z)$ is orthogonal to $C_x$, and hence belongs to $C_x$ as $C_x$ is self-dual.
Observe that $w_H(x \ast z) \leq w_H(z) < d$, which implies that we have found an element $(x \ast z) \in C_x$ that has weight less than $d$. 
This is a contradiction since minimum weight of $C_x$ is $d$, and this completes the proof for transversal $T$.
Note that for any other $t = t_1 + 7t_7$, as Theorem~\ref{thm:transversal_T_Tinv} suggests, we simply replace $x \in \langle A,C \rangle$ in the above argument with $(x \ast t')$, where $t' = t_1 + t_7$.  \hfill \IEEEQEDhere

\subsection{Proof of Theorem~\ref{thm:logical_identity}}
\label{sec:proof_logical_identity}

Let $G_1 = 
\begin{bmatrix}
G_{C_1/C_2} \\
G_2
\end{bmatrix}$ be a generator matrix for the code $C_1$.
Then by the CSS construction, the vectors $x = \bigoplus_{i = 1}^{k} c_i x_i$, where $c_i \in \{0,1\}$ and $x_i$ form the $k$ rows of the coset generator matrix $G_{C_1/C_2}$, determine all logical $X$ operators $E(x,0)$ for the code CSS($X, C_2; Z, C_1^{\perp}$).
Similarly, vectors $a \in C_2$ determine the $X$-type stabilizers $E(a,0)$ for the code. 
Therefore, $(x \oplus a) \in C_1$ represents all possible $X$-type representatives of all logical $X$ operators for the CSS-T code.
Recollect that by the CSS-T conditions, $C_2 \subset C_1^{\perp} \Rightarrow a \in C_1^{\perp}$ as well, and in fact $\tilde{a}$ belongs to the dual $\tilde{Z}_a^{\perp}$ of the (punctured) subspace $Z_a$ in $C_1^{\perp}$ that is supported on $a$ (see Theorem~\ref{thm:transversal_T} for notation), so that $\imath^{w_H(a)} E(0,a) \in S$.
By assumption, we have transversal $T$ acting trivially on the logical qubits, so transversal $P = T^2$ must also act trivially on the logical qubits.
Using this fact, and the identity $PXP^{\dagger} = Y$, let us observe the action of transversal $P$ on a logical $X$ representative $E(x \oplus a, 0)$.
We have
\begin{align}
P^{\otimes n} E(x \oplus a, 0) \left( P^{\otimes n} \right)^{\dagger} = E(x \oplus a, x \oplus a) = E(x \oplus a, 0) \cdot \imath^{w_H(x \oplus a)} E(0, x \oplus a),
\end{align}
where the second equality follows from the identity~\eqref{eq:Eab_multiply}.
Note that $w_H(x \oplus a) = w_H(x) + w_H(a) - 2 xa^T \equiv w_H(x) + w_H(a)$ (mod $4$), since $xa^T \equiv 0$ (mod $2$) due to the fact that logical $X$ operators must commute with $Z$-type stabilizers, and $\imath^{w_H(a)} E(0,a) \in S$.
(Recollect that $E(x,0)$ and $E(0,a)$ commute if and only if their symplectic inner product $\syminn{[x,0]}{[0,a]} = xa^T = 0$.)
Hence, $\imath^{w_H(x \oplus a)} E(0, x \oplus a) = \imath^{w_H(x)} E(0,x) \cdot \imath^{w_H(a)} E(0,a)$. 
Since $P^{\otimes n}$ must act trivially on logical qubits, we require that $P^{\otimes n} E(x \oplus a, 0) \left( P^{\otimes n} \right)^{\dagger} \equiv E(x \oplus a, 0)$, where the equivalence class is defined by multiplication with stabilizer elements.
For this equivalence to be true, we need $\imath^{w_H(x)} E(0,x) \in S$. 
This proves the first condition stated in the theorem.

The above ensures that $P^{\otimes n}$ acts like the logical identity.
Let us now examine $T^{\otimes n}$ along similar lines by using the identity $TXT^{\dagger} = e^{-\imath\pi/4} Y\, P$.
We require
\begin{align}
T^{\otimes n} E(x \oplus a, 0) \left( T^{\otimes n} \right)^{\dagger} & = e^{-\frac{\imath\pi}{4} w_H(x \oplus a)} E(x \oplus a, x \oplus a)\, P^{x \oplus a} \\ 
  & = e^{-\frac{\imath\pi}{4} w_H(x \oplus a)} E(x \oplus a, 0) \cdot \imath^{w_H(x \oplus a)} E(0, x \oplus a)\, P^{x \oplus a} \\
  & = e^{-\frac{\imath\pi}{4} w_H(x \oplus a)} E(x \oplus a, 0)\, P^{x \oplus a} \cdot \imath^{w_H(x \oplus a)} E(0, x \oplus a)\\
  & \equiv E(x \oplus a, 0).
\end{align}
Here the notation $P^{x \oplus a}$ means that the phase gate is applied to the qubits in the support of $(x \oplus a)$.
From the above calculation for $P^{\otimes n}$, we know that $\imath^{w_H(x \oplus a)} E(0, x \oplus a) \in S$, so for the last equivalence to be true, we need to ensure that $P^{x \oplus a}$ acts like the logical identity.
%Since we can write $x \oplus a = x + a - 2(x \ast a)$, we can split this term into three separate operations $P^x \cdot P^a \cdot P^{-2 (x \ast a)} = P^x \cdot P^a \cdot Z^{(x \ast a)} = P^x \cdot P^a \cdot E(0, x \ast a)$.
%Since $w_H(x \oplus a) = w_H(x) + w_H(a) - 2 w_H(x \ast a)$, we combine the phase $e^{\frac{2\pi\imath}{4} w_H(x \ast a)}$ with $E(0, x \ast a)$ to arrive at the constraint that $\imath^{w_H(x \ast a)} E(0, x \ast a) \in S$.
%This proves the second condition stated in the theorem.
%We are now left to ensure that $P^x$ and $P^a$ act like the logical identity.
Let us examine its action on an arbitrary logical $X$ representative $E(y \oplus b, 0)$ for $y \in C_1/C_2, b \in C_2$.
We require
\begin{align}
P^{x \oplus a} E(y \oplus b, 0) \left( P^{x \oplus a} \right)^{\dagger} & = E(y \oplus b, (y \oplus b) \ast (x \oplus a)) \\
  & = E(y \oplus b, 0) \cdot \imath^{w_H((y \oplus b) \ast (x \oplus a))} E(0, (y \oplus b) \ast (x \oplus a)) \\
  & \equiv E(y \oplus b, 0).
\end{align}
Observe that this is satisfied for the case $y = x, b = a$ by the arguments above for $P^{\otimes n}$.
Clearly, the constraint we need is that $\imath^{w_H((y \oplus b) \ast (x \oplus a))} E(0, (y \oplus b) \ast (x \oplus a)) \in S$.
Since this must hold for all valid $x,y,a,b$, by setting $a = b = 0$ we obtain the third condition of the theorem.
Similarly, by setting $a = y = 0$ and $x = y = 0$ respectively, we obtain the second and fourth conditions of the theorem.
It can be verified that these alone ensure that $\imath^{w_H((y \oplus b) \ast (x \oplus a))} E(0, (y \oplus b) \ast (x \oplus a)) \in S$ for all combinations of $x,y,a,b$, since we can split $(y \oplus b) \ast (x \oplus a) = (y \ast x) \oplus (y \ast a) \oplus (b \ast x) \oplus (b \ast a)$ and using the Hamming weight identity we used above.
Finally, we will show that the last three conditions amount to triorthogonality.

Note that since we need $\imath^{w_H(y \ast x)} E(0, y \ast x) \in S$, it must be true that $(y \ast x) \in C_1^{\perp}$ since by the CSS construction pure $Z$-type stabilizers arise from the code $C_1^{\perp}$.
As logical $X$ operators and $X$-type stabilizers must each commute with $Z$-type stabilizers, by the symplectic inner product constraint we can see that this implies $z (y \ast x)^T = w_H(z \ast y \ast x) \equiv 0$ (mod $2$) for any $z \in C_1/C_2$ or $z \in C_2$.
Similarly, since we need $\imath^{w_H(b \ast a)} E(0, b \ast a) \in S$, we also have $w_H(z \ast b \ast a) \equiv 0$ for any $z \in C_1/C_2$ or $z \in C_2$.
These are exactly the triorthogonality conditions in Definition~\ref{def:triorthogonality}.
The first condition of Definition~\ref{def:triorthogonality} follows from the facts that $C_2 \subset C_1^{\perp}$, and since $\imath^{w_H(y \ast x)} E(0, y \ast x), \imath^{w_H(y \ast a)} E(0, y \ast a) \in S$ must be Hermitian, the phase has to be $\pm 1$, which implies $w_H(y \ast x) = xy^T \equiv 0, w_H(y \ast a) = ay^T \equiv 0$ (mod $2$) for any $x,y \in C_1/C_2$ and $a \in C_2$.
Hence, triorthogonality of $G_1$ is a necessary condition for transversal $T$ to realize the logical identity on a CSS-T code. \hfill \IEEEQEDhere

\subsection{Proof of Theorem~\ref{thm:logical_trans_T}}
\label{sec:proof_logical_trans_T}

The proof uses a very similar strategy as for Theorem~\ref{thm:logical_identity}.
%As we did there, let $G_1 = 
%\begin{bmatrix}
%G_{C_1/C_2} \\
%G_2
%\end{bmatrix}$ be a generator matrix for the code $C_1$.
%Then by the CSS construction, the vectors $x = \sum_{i = 1}^{k} c_i x_i$, where $c_i \in \{0,1\}$ and $x_i$ form the $k$ rows of the coset generator matrix $G_{C_1/C_2}$, determine all logical $X$ operators $E(x,0)$ for the code CSS($X, C_2; Z, C_1^{\perp}$).
%Similarly, vectors $a \in C_2$ determine the $X$-type stabilizers $E(a,0)$ for the code. 
As we observed there, $(x \oplus a) \in C_1$ for $x \in C_1, a \in C_2$ represents all possible $X$-type representatives of all logical $X$ operators for the CSS-T code.
% Recollect that by the CSS-T conditions, $C_2 \subset C_1^{\perp} \Rightarrow a \in C_1^{\perp}$ as well, so that $\imath^{w_H(a)} E(0,a) \in S$.
Recollect that by the CSS-T conditions, $C_2 \subset C_1^{\perp} \Rightarrow a \in C_1^{\perp}$ as well, and in fact $\tilde{a}$ belongs to the dual $\tilde{Z}_a^{\perp}$ of the (punctured) subspace $Z_a$ in $C_1^{\perp}$ that is supported on $a$ (see Theorem~\ref{thm:transversal_T} for notation), so that $\imath^{w_H(a)} E(0,a) \in S$.
By assumption, we have physical transversal $T$ acting as transversal $T$ on the logical qubits, so physical transversal $P = T^2$ must also act as transversal $P$ on the logical qubits.
Using this fact, and the identity $PXP^{\dagger} = Y$, let us observe the action of transversal $P$ on a logical $X$ representative $E(x \oplus a, 0)$.
We require
\begin{align}
P^{\otimes n} E(x \oplus a, 0) \left( P^{\otimes n} \right)^{\dagger} = E(x \oplus a, x \oplus a) & = E(x \oplus a, 0) \cdot \imath^{w_H(x \oplus a)} E(0, x \oplus a) \\
  & \equiv \imath^{w_H(c)} E(x \oplus a, 0) E(0,z),
\end{align}
where the second equality follows from~\eqref{eq:Eab_multiply}, and $z \in C_2^{\perp}/C_1^{\perp}$ is the corresponding logical $Z$ string for the logical $X$ string $x$%
\footnote{Using the notation $x = \bigoplus_{i=1}^{k} c_i x_i$ in the theorem statement, $E(x \oplus a, 0) = E(a,0) \prod_{i=1}^{k} \bar{X}_i^{c_i}$. Since $(P \otimes P) (X \otimes X) (P^{\dagger} \otimes P^{\dagger}) = Y \otimes Y = \imath^2 (X \otimes X) (Z \otimes Z)$ physically, the calculation above translates this to the logical level to produce $w_H(c)$, where $\bar{X}_i = E(x_i,0)$ and $\bar{Z}_i = E(0,z_i)$.}.
This also means $xz^T \equiv w_H(c)$ (mod $2$) since the respective pairs of logical $X$ and $Z$ must anti-commute.
Note that $w_H(x \oplus a) = w_H(x) + w_H(a) - 2 xa^T \equiv w_H(x) + w_H(a)$ (mod $4$) because $xa^T \equiv 0$ (mod $2$) due to the fact that logical $X$ operators must commute with $Z$-type stabilizers (and $\imath^{w_H(a)} E(0,a) \in S$).
(Recollect that $E(x,0)$ and $E(0,a)$ commute if and only if their symplectic inner product $\syminn{[x,0]}{[0,a]} = xa^T = 0$.)
Hence, $\imath^{w_H(x \oplus a)} E(0, x \oplus a) = \imath^{w_H(x)} E(0,x) \cdot \imath^{w_H(a)} E(0,a)$, and for the last equivalence to be true above, we need $\imath^{w_H(x) - w_H(c)} E(0,x \oplus z) \in S$. 
Therefore, $w_H(x) - w_H(c) \equiv 0$ (mod $2$) for this matrix to be Hermitian. 
Observing the case when $w_H(c) = 1$ and hence $x = x_i$ for some $i \in \{1,\ldots,k\}$, this implies that $w_H(x_i)$ must be odd. 
Moreover, since $a \in C_2 \subset C_1^{\perp}$, we see that $x_i a^T = 0$, and $z_i a^T = 0$ because $z_i \in C_2^{\perp}$.
Also, for any $x_j \in C_1/C_2$ that is distinct from $x_i$, we have $z_i x_j^T = 0$ by definition, and since the above condition means $x_i \oplus z_i \in C_1^{\perp}$, we also have $x_i x_j^T = 0$.
So we can assume $z_i = x_i$ to be the corresponding logical $Z$ string for $\bar{X}_i$ as well, in which case the above condition becomes trivial and the required equivalence is satisfied.
In this scenario, since $x_i \oplus z_i = 0 \Rightarrow E(0,x_i \oplus z_i) = I_N$, we need $w_H(x_i) \equiv 1$ (mod $4$) exactly as for a non-trivial stabilizer group we need to make sure that $-I_N \notin S$.
%This proves the first condition stated in the theorem.

The above ensures that $P^{\otimes n}$ acts desirably and so $E(x \oplus a, x \oplus a)$ indeed corresponds to the logical $Y$ operator $\bar{Y}_c$ corresponding to the given logical $X$ string $x = \bigoplus_{i=1}^{k} c_i x_i \in C_1/C_2$.
Let us now examine $T^{\otimes n}$ along similar lines by using the identity $TXT^{\dagger} = e^{-\imath\pi/4} Y\, P$.
We require
\begin{align}
T^{\otimes n} \bar{X}_c \left( T^{\otimes n} \right)^{\dagger} = T^{\otimes n} E(x \oplus a, 0) \left( T^{\otimes n} \right)^{\dagger} = e^{-\frac{\imath\pi}{4} w_H(x \oplus a)} E(x \oplus a, x \oplus a)\, P^{x \oplus a} 
%
%  & = e^{-\frac{\imath\pi}{4} w_H(x \oplus a)} E(x \oplus a, 0) \cdot \imath^{w_H(x \oplus a)} E(0, x \oplus a)\, P^{x \oplus a} \\
%  & = e^{-\frac{\imath\pi}{4} w_H(x \oplus a)} E(x \oplus a, 0)\, P^{x \oplus a} \cdot \imath^{w_H(x \oplus a)} E(0, x \oplus a)\\
  \equiv e^{-\frac{\imath\pi}{4} w_H(c)} \bar{Y}_c\, \bar{P}_c,
\end{align}
where $\bar{P}_c$ denotes the logical phase gate corresponding to the given logical $X$ string $x = \bigoplus_{i=1}^{k} c_i x_i$, and the notation $P^{x \oplus a}$ means that the phase gate is applied to the qubits in the support of $(x \oplus a)$.
We have verified above that the $\bar{Y}_c$ condition is satisfied, and when $P^{x \oplus a}$ acts on $\bar{X}_c = E(x \oplus a, 0)$, it indeed acts desirably.
Therefore, to verify $P^{x \oplus a} \equiv \bar{P}_c$, we just need to ensure that $P^{x \oplus a}$ acts like the logical identity on all $E(y \oplus b, 0)$ for $y \neq x \in C_1/C_2$ and any $b \in C_2$.
This part of the proof is identical to the corresponding arguments in the proof of Theorem~\ref{thm:logical_identity}, and so this proves that triorthogonality of $G_1$ is a necessary condition.
Finally, we observe that we need the condition $w_H(x \oplus a) \equiv w_H(c)$ (mod $8$) for the above equivalence to hold, and this completes the proof of the theorem. \hfill \IEEEQEDhere

\subsection{Proof of Corollary~\ref{cor:triortho_general}}
\label{sec:proof_triortho_general}

The Bravyi-Haah construction allows for a Clifford correction after the transversal $T$ gate in order to exactly realize logical transversal $T$.
In order to prove the equivalence of their construction to Theorem~\ref{thm:logical_trans_T}, we need to show that, by requiring their Clifford correction to be trivial, we arrive at the same conditions as listed above.
Since triorthogonality of $G_1$ is a common constraint in both Theorem~\ref{thm:logical_trans_T} and the Bravyi-Haah construction, we are left to verify that the Hamming weight condition above coincides with the condition for the Clifford correction to be trivial.

Let $C_2$ be an $[n,k_2]$ code so that the number of rows in $G_2$ is $k_2$.
Let $y = \bigoplus_{i=1}^{k + k_2} d_i y_i$ with $d_i = c_i, y_i = x_i$ for $i = 1,\ldots,k$ and $y_i = a_i$ for $i > k$, where $a_i$ are the rows of $G_2$.
The Clifford correction depends on the phase $\imath^{Q(d)}$ and is trivial when
\begin{align}
Q(d) \coloneqq \sum_{i=1}^{k + k_2} \Gamma_i d_i - 2 \sum_{i < j} \Gamma_{ij} d_i d_j \equiv 0\ (\text{mod}\ 4),
\end{align}
where $w_H(y_i) = 
\begin{cases}
2 \Gamma_i + 1 & \text{for}\ 1 \leq i \leq k, \\
2 \Gamma_i & \text{for}\ i > k,
\end{cases}$
and $y_i y_j^T = 2 \Gamma_{ij}$.
This is because, by their construction $w_H(x_i)$ is odd.
Substituting for $\Gamma_i$ and $2 \Gamma_{ij}$, we get
\begin{align}
\sum_{i=1}^{k} c_i \frac{(w_H(x_i) - 1)}{2} + \sum_{j=1}^{k_2} d_{k+j} \frac{w_H(a_j)}{2} - \sum_{\substack{i < j\\i,j = 1}}^{k+k_2} d_i d_j (y_i y_j^T) & \equiv 0\ (\text{mod}\ 4) \\
\Leftrightarrow \sum_{i=1}^{k} c_i w_H(x_i) - w_H(c) + \sum_{j = 1}^{k_2} d_{k+j} w_H(a_j) - 2 \sum_{\substack{i < j\\i,j = 1}}^{k+k_2} d_i d_j (y_i y_j^T) & \equiv 0\ (\text{mod}\ 8) \\
%
%\Rightarrow w_H(x \oplus a) - 4 \sum_{i < j} d_i d_j (y_i y_j^T) & \equiv w_H(c)\ (\text{mod}\ 8) \\
%
\Leftrightarrow w_H(x \oplus a) & \equiv w_H(c)\ (\text{mod}\ 8).
\end{align}
%We can ignore the $4 y_i y_j^T$ terms because $x_i x_j^T \equiv 0, x_i a_j^T \equiv 0, a_i a_j^T \equiv 0$ (mod $2$) implies $y_i y_j^T \equiv 0$ (mod $2$).
For the last step, using the fact that $w_H(x \oplus a) = w_H(x) + w_H(a) - 2 xa^T$ recursively for $x = \bigoplus_{i=1}^{k} c_i x_i, a = \bigoplus_{j=1}^{k_2} d_{k+j} a_j$, it can be verified that 
% $w_H(x \oplus a) = \sum_{i=1}^{k} c_i w_H(x_i) + \sum_{j = 1}^{k_2} d_{k+j} w_H(a_j) - 2 \sum_{i < j} d_i d_j (y_i y_j^T)$.
\begin{align}
w_H(x \oplus a) & = w_H\left( \bigoplus_{i=1}^{k} c_i x_i \oplus \bigoplus_{j=1}^{k_2} d_{k+j} a_j \right) \\
  & = w_H\left( \bigoplus_{i=1}^{k} c_i x_i \right) + w_H\left( \bigoplus_{j=1}^{k_2} d_{k+j} a_j \right) - 2 \left( \bigoplus_{i=1}^{k} c_i x_i \right) \left( \bigoplus_{j=1}^{k_2} d_{k+j} a_j \right)^T \\
  & = \left[ c_1 w_H(x_1) + w_H\left( \bigoplus_{i=2}^{k} c_i x_i \right) - 2 c_1 x_1 \left( \bigoplus_{i=2}^{k} c_i x_i \right)^T \right] \nonumber \\
  & \qquad + \left[ d_{k+1} w_H(a_1) + w_H\left( \bigoplus_{j=2}^{k_2} d_{k+j} a_j \right) - 2 d_{k+1} a_1 \left( \bigoplus_{j=2}^{k_2} d_{k+j} a_j \right)^T \right] \nonumber \\
  & \qquad - 2 \left( \bigoplus_{i=1}^{k} d_i y_i \right) \left( \bigoplus_{j=k+1}^{k+k_2} d_j y_j \right)^T \\
  & = \left[ c_1 w_H(x_1) + w_H\left( \bigoplus_{i=2}^{k} c_i x_i \right) + d_{k+1} w_H(a_1) + w_H\left( \bigoplus_{j=2}^{k_2} d_{k+j} a_j \right) \right] \\
  & \qquad - 2 \left[ d_1 y_1 \left( \bigoplus_{i=2}^{k} d_i y_i \right)^T + d_{k+1} y_{k+1} \left( \bigoplus_{j=k+2}^{k+k_2} d_j y_j \right)^T + \left( \bigoplus_{i=1}^{k} d_i y_i \right) \left( \bigoplus_{j=k+1}^{k+k_2} d_j y_j \right)^T \right] \\
  & \ \ {\large \vdots} \ (\text{continue\ recursion\ for}\ i=2,\ldots,k-1\ \text{and}\ j=k+2,\ldots,k+k_2-1) \nonumber \\
  & = \sum_{i=1}^{k} c_i w_H(x_i) + \sum_{j = 1}^{k_2} d_{k+j} w_H(a_j) - 2 \sum_{\substack{i < j\\i,j = 1}}^{k+k_2} d_i d_j (y_i y_j^T).
\end{align}
This completes the proof.  \hfill \IEEEQEDhere

\subsection{Proof of Lemma~\ref{lem:conj_by_trans_Z_rot}}
\label{sec:proof_conj_by_trans_Z_rot}

In this proof, we will see that the ideas used to calculate the coefficients for the transversal $T$ gate generalize to the transversal application of power-of-$2$ roots of $T$, namely $\tau_{R}^{(\ell)} \coloneqq 
\begin{bmatrix}
1 & 0 \\
0 & e^{2\pi\imath/2^\ell}
\end{bmatrix} = 
\begin{bmatrix}
1 & 0 \\
0 & \xi
\end{bmatrix}
$ with $R = [\, 1\, ]$.
When this $Z$-rotation is applied transversally, we have $R = I_n$ as before for the $T$ gate.
Then from~\eqref{eq:tau_Eab} we get $\phi(R,a,b,\ell) = (1 - 2^{\ell-2}) a I_n a^T = (1 - 2^{\ell-2}) w_H(a)$ and $\tilde{R} \coloneqq \tilde{R}(R,a,\ell) = (1 + 2^{\ell-2}) D_{a I_n} - (D_{1-a} I_n D_a + D_a I_n D_{1-a} + 2 D_{a I_n D_a}) = (1 + 2^{\ell-2}) D_a - (0 + 0 + 2 D_a) = (2^{\ell-2} - 1) D_a$.

First consider the case $a \neq [1,1,\ldots,1]$ and let $x \in \Fn$ such that $\text{supp}(x) \nsubseteq \text{supp}(a)$, so that there exists some $j \in \{1,\ldots,n\}$ satisfying $a_j = 0, x_j = 1$.
Once again, let $\tilde{x} = [x_1,x_2,\ldots,x_{j-1},x_{j+1},\ldots,x_n]$ and similarly for $\tilde{a}, \tilde{v}$.
Then we observe that
\begin{align}
c_{(2^{\ell-2} - 1) D_a,x}^{(\ell-1)} & = \frac{1}{\sqrt{2^n}} \sum_{v \in \Fn} (-1)^{vx^T} (\xi^2)^{v \tilde{R} v^T} \\
  & = \frac{1}{\sqrt{2^n}} \sum_{v \in \Fn} (-1)^{vx^T} \xi^{2(2^{\ell-2} - 1) va^T} \\
  & = \frac{1}{\sqrt{2^n}} \sum_{\substack{\tilde{v} \in \mathbb{F}_2^{n-1}\\(v_j = 0)}} (-1)^{\tilde{v} \tilde{x}^T} \xi^{(2^{\ell-1} - 2) \tilde{v} \tilde{a}^T} - \frac{1}{\sqrt{2^n}} \sum_{\substack{\tilde{v} \in \mathbb{F}_2^{n-1}\\(v_j = 1)}} (-1)^{\tilde{v} \tilde{x}^T} \xi^{(2^{\ell-1} - 2) \tilde{v} \tilde{a}^T} \\
  & = 0.
\end{align}
So we only need to consider coefficients corresponding to $x \preceq a$.
For the remaining calculations, it is convenient to note that $\xi^{2^{\ell-1} - 2} = - \xi^{-2}$.
Once again, for the case $x = [0,0,\ldots,0]$ and arbitrary $a \in \Fn$, we observe that
\begin{align}
\label{eq:trans_root_T_coeff_x_0}
c_{(2^{\ell-2} - 1) D_a,0}^{(\ell-1)} = \frac{1}{\sqrt{2^n}} \sum_{v \in \Fn} (-\xi^{-2})^{v a^T} = \frac{1}{\sqrt{2^n}} 2^{n - w_H(a)} (1 - \xi^{-2})^{w_H(a)} = \sqrt{2^n} \left( \frac{1 - \xi^{-2}}{2} \right)^{w_H(a)}.
\end{align}
For $x \neq [0,0,\ldots,0]$, and $x \preceq a$, let $j \in \text{supp}(x) \subseteq \text{supp}(a)$ so that $x_j = a_j = 1$ and $e_j x^T = 1$.
Then for each $v$ such that $vx^T \equiv 0$, we have $(v \oplus e_j) x^T \equiv 1$.
So we calculate
\begin{align}
& c_{(2^{\ell-2} - 1) D_a,x}^{(\ell-1)} \nonumber \\
  & = \frac{1}{\sqrt{2^n}} \sum_{v \in \Fn} (-1)^{vx^T} (-\xi^{-2})^{v a^T} \\
  & = \frac{1}{\sqrt{2^n}} \sum_{\substack{v \in \Fn\\vx^T \equiv 0}} (-\xi^{-2})^{v a^T} - \frac{1}{\sqrt{2^n}} \sum_{\substack{v \in \Fn\\vx^T \equiv 1}} (-\xi^{-2})^{v a^T} \\
  & = \frac{1}{\sqrt{2^n}} \sum_{\substack{v \in \Fn\\vx^T \equiv 0}} \left[ (-\xi^{-2})^{v a^T} - (-\xi^{-2})^{(v \oplus e_j) a^T} \right] \\
  & = \frac{1}{\sqrt{2^n}} \sum_{\substack{v \in \Fn\\vx^T \equiv 0}} \left[ (-\xi^{-2})^{v a^T} - (-\xi^{-2})^{(v + e_j - 2 (v \ast e_j)) a^T} \right] \\
  & = \frac{1}{\sqrt{2^n}} \sum_{\substack{v \in \Fn\\vx^T \equiv 0}} (-\xi^{-2})^{v a^T} \left[ 1 - (-\xi^{-2})^{a_j - 2 v_j a_j}  \right] \\
  & = \frac{(1 + \xi^{-2})}{\sqrt{2^n}} \sum_{\substack{\tilde{v} \colon \tilde{v} \tilde{x}^T \equiv 0\\(v_j = 0)}} (-\xi^{-2})^{\tilde{v} \tilde{a}^T} + \frac{(1 + \xi^2)}{\sqrt{2^n}} \sum_{\substack{\tilde{v} \colon \tilde{v} \tilde{x}^T \equiv 1\\(v_j = 1)}} (-\xi^{-2}) \cdot (-\xi^{-2})^{\tilde{v} \tilde{a}^T} \\
  & = 
\begin{dcases}
\left( \frac{1 + \xi^{-2}}{\sqrt{2}} \right) \frac{1}{\sqrt{2^{n-1}}} \sum_{\tilde{v} \in \mathbb{F}_2^{n-1}} (-\xi^{-2})^{\tilde{v} \tilde{a}^T} & \text{if}\ x = e_j, j \in \text{supp}(a), \\
\left( \frac{1 + \xi^{-2}}{\sqrt{2}} \right) \left[ \frac{1}{\sqrt{2^{n-1}}} \sum_{\substack{\tilde{v} \in \mathbb{F}_2^{n-1}\\\tilde{v} \tilde{x}^T \equiv 0}} (-\xi^{-2})^{\tilde{v} \tilde{a}^T} - \frac{1}{\sqrt{2^{n-1}}} \sum_{\substack{\tilde{v} \in \mathbb{F}_2^{n-1}\\\tilde{v} \tilde{x}^T \equiv 1}} (-\xi^{-2})^{\tilde{v} \tilde{a}^T} \right] & \text{if}\ w_H(x) \geq 2, x \preceq a
\end{dcases} \\
  & = \left( \frac{1 + \xi^{-2}}{\sqrt{2}} \right) \times 
\begin{dcases}
c_{(2^{\ell-2} - 1) D_{\tilde{a}},0}^{(\ell-1)} & \text{if}\ x = e_j, j \in \text{supp}(a), \\
c_{(2^{\ell-2} - 1) D_{\tilde{a}}, \tilde{x}}^{(\ell-1)} & \text{if}\ w_H(x) \geq 2, x \preceq a.
\end{dcases}
\end{align}
When $w_H(x) = 1$, i.e., $x = e_j$, we calculate 
\begin{align}
c_{(2^{\ell-2} - 1) D_a, e_j}^{(\ell-1)} = \frac{(1 + \xi^{-2})}{\sqrt{2}} \sqrt{2^{n-1}} \left( \frac{1 - \xi^{-2}}{2} \right)^{w_H(a) - 1} = \left( \frac{1 + \xi^{-2}}{1 - \xi^{-2}} \right) \sqrt{2^n} \left( \frac{1 - \xi^{-2}}{2} \right)^{w_H(a)}.
\end{align}
We can simplify the first factor as
\begin{align}
\frac{1 + \xi^{-2}}{1 - \xi^{-2}} \cdot \frac{1 - \xi^2}{1 - \xi^2} = \frac{1 + \xi^{-2} - \xi^2 - 1}{1 - \xi^{-2} - \xi^2 + 1} = \frac{-2\imath \sin\frac{2\pi}{2^{\ell-1}}}{2 \left( 1 - \cos\frac{2\pi}{2^{\ell-1}} \right)} = \frac{-2\imath \sin\frac{2\pi}{2^\ell} \cos\frac{2\pi}{2^\ell}}{2 \sin^2\frac{2\pi}{2^\ell}} = -\imath \cot\frac{2\pi}{2^\ell}.
\end{align}
Therefore, in general we have
\begin{align}
c_{(2^{\ell-2} - 1) D_a, x}^{(\ell-1)} & = 
\begin{dcases}
\left( \frac{1 + \xi^{-2}}{\sqrt{2}} \right)^{w_H(x)} \sqrt{2^{n-w_H(x)}} \left( \frac{1 - \xi^{-2}}{2} \right)^{w_H(a) - w_H(x)} & \text{if}\ x \preceq a, \\
0 & \text{otherwise}
\end{dcases} \\
\label{eq:D_ak_coeff_value}
  & = 
\begin{dcases}
\left( -\imath \cot\frac{2\pi}{2^\ell} \right)^{w_H(x)} \sqrt{2^n} \left( \frac{1 - \xi^{-2}}{2} \right)^{w_H(a)} & \text{if}\ x \preceq a, \\
0 & \text{otherwise}.
\end{dcases}
\end{align}
Using this expression, we proceed to calculate the action of $\tau_{I_n}^{(\ell)}$ on an arbitrary Pauli operator $E(a,b)$. 
\begin{align}
& \tau_{I_n}^{(\ell)} E(a,b) (\tau_{I_n}^{(\ell)})^{\dagger} \nonumber \\
  & = \frac{1}{\sqrt{2^n}} \xi^{\phi(I_n,a,b,\ell)} \sum_{x \in \Fn} c_{(2^{\ell-2} - 1) D_a, x}^{(\ell-1)} \imath^{-a x^T} E(a, b + a I_n + x) \\
  & = \frac{1}{\sqrt{2^n}} \xi^{(1 - 2^{\ell-2}) w_H(a)} \sum_{x \preceq a} \left( -\imath \cot\frac{2\pi}{2^\ell} \right)^{w_H(x)} \sqrt{2^n} \left( \frac{1 - \xi^{-2}}{2} \right)^{w_H(a)} \imath^{-w_H(x)} E(a, b + a + x) \\
  & = \left( \frac{\xi(1 - \xi^{-2})}{2 \imath} \right)^{w_H(a)} \sum_{x \preceq a} \left( - \cot\frac{2\pi}{2^\ell} \right)^{xx^T} E(a, b + a + x) \\
  & = \left( \frac{\xi - \xi^{-1}}{2 \imath} \right)^{w_H(a)} \sum_{x \preceq a} \left( - \cot\frac{2\pi}{2^\ell} \right)^{xx^T} (-1)^{a(b \ast (a \oplus x))^T + a x^T} E(a, b \oplus (a \oplus x)) \\
  & = \left( \frac{2\imath \sin\frac{2\pi}{2^\ell}}{2 \imath} \right)^{w_H(a)} \sum_{y \preceq a} \left( \cot\frac{2\pi}{2^\ell} \right)^{(a - y) (a - y)^T} (-1)^{a(b \ast y)^T} E(a, b \oplus y) \\
  & = \left( \sin\frac{2\pi}{2^\ell} \right)^{w_H(a)} \left( \cot\frac{2\pi}{2^\ell} \right)^{w_H(a)} \sum_{y \preceq a} \left( \tan\frac{2\pi}{2^\ell} \right)^{w_H(y)} (-1)^{a(b \ast y)^T} E(a, b \oplus y) \\
  & = \left( \cos\frac{2\pi}{2^\ell} \right)^{w_H(a)} \sum_{y \preceq a} \left( \tan\frac{2\pi}{2^\ell} \right)^{w_H(y)} (-1)^{a(b \ast y)^T} E(a, b \oplus y) \\
  & = \frac{1}{\left( \sec\frac{2\pi}{2^\ell} \right)^{w_H(a)}} \sum_{y \preceq a} \left( \tan\frac{2\pi}{2^\ell} \right)^{w_H(y)} (-1)^{by^T} E(a, b \oplus y).
\end{align}
Note that we have used the fact that $y \coloneqq a \oplus x = a - x$ since $x \preceq a$, and also that $a x^T = xx^T, a y^T = yy^T$.
\hfill\IEEEQEDhere

\subsection{Proof of Theorem~\ref{thm:transversal_Z_rot}}
\label{sec:proof_transversal_Z_rot}

As in the proof of Theorem~\ref{thm:transversal_T}, we need to determine necessary and sufficient conditions for the equality
\begin{align}
\tau_{I_n}^{(\ell)} \Pi_S \left( \tau_{I_n}^{(\ell)} \right)^{\dagger} & = \frac{1}{2^r} \sum_{j = 1}^{2^r} \epsilon_j \tau_{I_n}^{(\ell)} E(a_j,b_j) \left( \tau_{I_n}^{(\ell)} \right)^{\dagger} \\
  & = \frac{1}{2^r} \sum_{j = 1}^{2^r} \frac{\epsilon_j}{\left( \sec\frac{2\pi}{2^\ell} \right)^{w_H(a_j)}} \sum_{y \preceq a_j} \left( \tan\frac{2\pi}{2^\ell} \right)^{w_H(y)} (-1)^{b_j y^T} E(a_j, b_j \oplus y) \\
  & = \frac{1}{2^r} \sum_{j = 1}^{2^r} \epsilon_j E(a_j,b_j) \\
  & = \Pi_S.
\end{align}
Note that for $j \neq j'$, either $a_j \neq a_{j'}$ or at least $b_j \neq b_{j'}$ if $a_j = a_{j'}$.
By commutativity of stabilizers, for $v_1, v_2 \in Z_j$, we need $\syminn{[a_j, b_j \oplus v_1]}{[a_j, b_j \oplus v_2]} = 0$ which implies $a_j(v_1 \oplus v_2) = 0$ (mod $2$), or equivalently, $w_H(v_1 \oplus v_2) = 0$ (mod $2$).
Hence, all vectors in $Z_j$ must have even weight for any $a_j \neq 0$.

Let $\epsilon_v^{(j)} E(a_j, b_j \oplus v) \in S$ for $v \in Z_j$, for some $\epsilon_v^{(j)} \in \{ \pm 1 \}$.
Note that it is unclear if we need $\epsilon_v^{(j)} = \epsilon_j (-1)^{b_j v^T}$ always, as the expression above might suggest, and from the calculation in Theorem~\ref{thm:transversal_T}.
In general, if we collect all the coefficients for $E(a_j,b_j)$, then dividing it by $\left( \sec\frac{2\pi}{2^\ell} \right)^{w_H(a_j)}$ must leave $\epsilon_j$ alone in the numerator, i.e.,
\begin{align}
\epsilon_j + \sum_{v \in Z_j \setminus \{0\}} \epsilon_v^{(j)} (-1)^{(b_j \oplus v) v^T} \left( \tan\frac{2\pi}{2^\ell} \right)^{w_H(v)} & = \epsilon_j \left( \sec\frac{2\pi}{2^\ell} \right)^{w_H(a_j)} \\
\Rightarrow \sum_{v \in Z_j} \epsilon_v^{(j)} (-1)^{b_j v^T} \left( \tan\frac{2\pi}{2^\ell} \right)^{w_H(v)} & = \epsilon_j \left( \sec\frac{2\pi}{2^\ell} \right)^{w_H(a_j)} \ (\text{since}\ w_H(v)\ \text{even}).
\end{align}
Let $\epsilon_v E(0,v) \in S$ for $v \in Z_j$, for some $a_j$, and $\epsilon_v \in \{ \pm 1 \}$.
Then we calculate
\begin{align}
\epsilon_j E(a_j,b_j) \cdot \epsilon_v E(0,v) & = \epsilon_j \epsilon_v \imath^{-a_j v^T} E(a_j, b_j + v) \\
   & = \epsilon_j \epsilon_v \imath^{vv^T} E(a_j, b_j \oplus v + 2(b_j \ast v)) \\
   & = \epsilon_j \epsilon_v \imath^{vv^T} (-1)^{b_j v^T} E(a_j, b_j \oplus v) \\
   & \eqqcolon \epsilon_v^{(j)} E(a_j, b_j \oplus v).
\end{align}
Hence, $\epsilon_v^{(j)} = \epsilon_j \epsilon_v \imath^{vv^T} (-1)^{b_j v^T}$, where $w_H(v) = vv^T$ is even.
(Note that by the property of any non-trivial stabilizer, $E(0,0) = I_N$ has sign $+1$ always, so $\epsilon_0 = 1$.)
Substituting this value back, we get the condition
\begin{align}
\sum_{v \in Z_j} \epsilon_v \imath^{vv^T} \left( \tan\frac{2\pi}{2^\ell} \right)^{w_H(v)} & = \left( \sec\frac{2\pi}{2^\ell} \right)^{w_H(a_j)} \\
\Rightarrow \sum_{v \in Z_j} \epsilon_v \left( \imath \tan\frac{2\pi}{2^\ell} \right)^{w_H(v)} & = \left( \sec\frac{2\pi}{2^\ell} \right)^{w_H(a_j)},
\end{align}
for some choices of $\epsilon_v \in \{ \pm 1 \}$, except when $v = 0$ as commented above.
This proves the first condition in the theorem.

Now we need to ensure that all terms $E(a_j, b_j \oplus y)$ for $y \in W_j$ get cancelled properly, for all $a_j$, so that the code projector $\Pi_S$ is indeed preserved.
Observe that terms corresponding to the coset $\{ b_j \oplus y \colon y \preceq a_j \}$ can only appear in the inner summations corresponding to those $i \in \{ 1,2,\ldots,2^r \}$ where $\epsilon_i E(a_i, b_i) = \epsilon_v^{(j)} E(a_j, b_j \oplus v)$ with $v \in Z_j$.
Moreover, for every $y \in B_j$, the Pauli term $E(a_j, b_j \oplus y)$ appears exactly once in the inner summation corresponding to $\epsilon_v^{(j)} E(a_j, b_j \oplus v)$.
We just have to collect these coefficients and set their sum to zero.
Hence, for each $y \in B_j$ we need
\begin{align}
\sum_{v \in Z_j} \epsilon_v^{(j)} (-1)^{(b_j \oplus v) (v \oplus y)^T} \left( \tan\frac{2\pi}{2^\ell} \right)^{w_H(v \oplus y)} & = 0 \\
\Rightarrow \sum_{v \in Z_j} \epsilon_j \epsilon_v \imath^{vv^T} (-1)^{b_j v^T + (b_j \oplus v) (v \oplus y)^T} \left( \tan\frac{2\pi}{2^\ell} \right)^{w_H(v \oplus y)} & = 0 \\
\Rightarrow \sum_{v \in Z_j} \epsilon_v \left( \imath \tan\frac{2\pi}{2^\ell} \right)^{w_H(v \oplus y)} & = 0.
\end{align}
In the last equation, we can reduce the exponent $w_H(v \oplus y)$ to just $w_H(v) - 2vy^T$ since the additional term $w_H(y)$ does not affect the equality to zero.
This proves the second condition in the theorem. \hfill \IEEEQEDhere

\subsection{Proof of Lemma~\ref{lem:divisible}}
\label{sec:proof_divisible}

The weight enumerator of the code $C$ is $W_C(x,y) = \sum_{i=0}^{m} A_i x^{m-i} y^i = \sum_{v \in C} x^{m - w_H(v)} y^{w_H(v)}$, where $A_i$ is the number of codewords in $C$ of Hamming weight $i$.
The MacWilliams identities for a self-dual code are $W_C(x,y) = \frac{1}{|C|} W_C(x+y, x-y)$, where $|C|$ is the number of codewords in $C$.
Then we observe that
\begin{align}
\sum_{v \in C} \left( \imath \tan\frac{2\pi}{2^{\ell}} \right)^{w_H(v)} & = \left( \sec\frac{2\pi}{2^{\ell}} \right)^{m} \\
\Rightarrow \sum_{v \in C} \left( \imath \tan\frac{2\pi}{2^{\ell}} \right)^{w_H(v)} \left( \cos\frac{2\pi}{2^{\ell}} \right)^{m} & = 1 \\
\Rightarrow \sum_{v \in C} \left( \imath \sin\frac{2\pi}{2^{\ell}} \right)^{w_H(v)} \left( \cos\frac{2\pi}{2^{\ell}} \right)^{m - w_H(v)} & = 1 \\
\Rightarrow W_C\left( \cos\frac{2\pi}{2^{\ell}}, \imath\sin\frac{2\pi}{2^{\ell}} \right) & = 1 \\
\Rightarrow \frac{1}{|C|} \sum_{v \in C} \left( \cos\frac{2\pi}{2^{\ell}} + \imath \sin\frac{2\pi}{2^{\ell}} \right)^{m - w_H(v)} \left( \cos\frac{2\pi}{2^{\ell}} - \imath \sin\frac{2\pi}{2^{\ell}} \right)^{w_H(v)} & = 1 \\
\Rightarrow \frac{1}{|C|} \sum_{v \in C} \left( \cos\frac{2\pi}{2^{\ell}} + \imath \sin\frac{2\pi}{2^{\ell}} \right)^{m - 2w_H(v)} & = 1,
\end{align}
where the last step follows from the fact that $\exp\left( \frac{-2\pi\imath}{2^{\ell}} \right) = \cos\frac{2\pi}{2^{\ell}} - \imath \sin\frac{2\pi}{2^{\ell}}$.
We note that $w_H(v)$ is even for all $v \in C$ and observe three cases.
\begin{enumerate}

\item If $w_H(v) = m/2$, then the term $v$ contributes $1$ to the sum.

\item If $w_H(v) < m/2$, then the term $v$ contributes $\left( \cos\frac{2 (m - 2 w_H(v)) \pi}{2^{\ell}} + \imath \sin\frac{2 (m - 2 w_H(v)) \pi}{2^{\ell}} \right)$.

\item If $w_H(v) > m/2$, then the term $v$ contributes $\left( \cos\frac{2 (2 w_H(v) - m) \pi}{2^{\ell}} - \imath \sin\frac{2 (2 w_H(v) - m) \pi}{2^{\ell}} \right)$.

\end{enumerate}
Since the all-$1$s vector is always present in a self-dual code, we pair the terms $v$ and $w = \vecnot{1} \oplus v$ such that $w_H(v) < m/2$ and $w_H(w) = m - w_H(v)$.
Hence, the term $w$ contributes $\left( \cos\frac{2 (m - 2 w_H(v)) \pi}{2^{\ell}} - \imath \sin\frac{2 (m - 2 w_H(v)) \pi}{2^{\ell}} \right)$.
Therefore, we have the condition $\frac{1}{|C|} \sum_{v \in C} \cos\frac{2 (m - 2 w_H(v)) \pi}{2^{\ell}} = 1$ that is satisfied if and only if each term in the sum equals $1$. 
Indeed, this happens if and only if each $v \in C$ satisfies either $w_H(v) = m/2$ or $2^{\ell}$ divides $(m - 2 w_H(v))$.   \hfill \IEEEQEDhere

\subsection{Proof of Theorem~\ref{thm:QRM_family}}
\label{sec:proof_QRM_family}

From the example above and~\eqref{eq:css_basis_states}, we realize that any $f \in \text{RM}(r,m)$ corresponds to a vector $u = v_f \cdot G_{C_1/C_2} \oplus y \cdot G_2 \in C_1$ in the CSS superposition for $\ket{v_f}_L$.
Define $\zeta_t \coloneqq e^{\frac{2\pi \imath}{t}}$ and note that on any state $\ket{u}$, transversal $\exp\left( \frac{\imath\pi}{2^{m/r}} Z \right)$ maps 
\begin{align}
\label{eq:u_zeta}
\ket{u} \mapsto \zeta_{2^{\frac{m}{r}}}^{w_H(u)} \ket{u}.
\end{align}
By fixing $v_f$ we fix the degree $r$ terms in $f$, and by sweeping over all $y \in \mathbb{Z}_2^{k_2}$ we exhaust all choices of degree at most $(r-1)$ terms in $f$, thereby examining all states $\ket{u}$ in the CSS superposition corresponding to $\ket{v_f}_L$.
For the logical operator to be well-defined as a diagonal gate acting as per~\eqref{eq:QRM_family_logical_action}, we need to show that $w_H(u)$ (mod $2^{m/r}$) depends only on the degree $r$ terms in $f$.
Thus, we are interested in $\nu_2(w_H(u))$ for different $u$ in a single coset of $\text{RM}(r-1,m)$ in $\text{RM}(r,m)$.

First, let us consider $\text{QRM}(1,m)$ separately for simplicity.  
Here, if $\ket{u} = \ket{00\cdots0}$, then for any $w \in u + \text{RM}(0,m)$, $\nu_2(w_H(w)) = m$ and so $\ket{w} \mapsto \ket{w}$.  
However, if $u = \text{ev}(f)$ with $\text{deg}(f) = 1$, then for any $w \in u + \text{RM}(0,m) = \{ u, u \oplus \vecnot{1} \}$, $w$ corresponds to a codimension-$1$ affine plane so that $\nu_2(w_H(w)) = m-1$, and so $\ket{w} \mapsto -\ket{w}$.
Hence, the logical diagonal unitary has diagonal entries $(1,-1,-1,\ldots,-1)$, which is equivalent to $(-1,1,1,\ldots,1)$ up to a global phase of $(-1)$.
Thus, up to local corrections (i.e., a logical transversal $X$ gate correction), transversal application of physical $Z$-rotation $\exp\left( \frac{\imath\pi}{2^{m/r}} Z \right)$ implements a logical $\text{C}^{m-1}Z$ gate.
This captures the $\llbr 8,3,2 \rrbr$ code we discussed previously in Section~\ref{sec:logical_ccz}.

Now consider the more general case where $r \geq 2$.  
We are interested in calculating $w_H(u) = 2^m - N(f)$ (mod $2^{m/r}$), where $N(f)$ denotes the number of zeros of $f$ over $\mathbb{F}_2$. 
Then, following the proofs of Ax's theorem~\cite{Ax-ajm64, McEliece-dm72, Hou-actaa16}, note that
\begin{align} 
N(f) & = \frac{1}{2}\sum_{\vecnot{x} = (x_0,x_1,\ldots,x_m) \in \mathbb{F}_2^{m+1}} \mu(x_0 f(x_1,\ldots,x_m)) \\
     & = \frac{1}{2}\sum_{\vecnot{x} \in \mathbb{F}_2^{m+1}} \mu\left( x_0 \sum_{\vecnot{d} \in \mathbb{F}_2^m} a_{\vecnot{d}} x^{\vecnot{d}} \right) \\ 
     & = \frac{1}{2}\sum_{\vecnot{x} \in \mathbb{F}_2^{m+1}} \prod_{\vecnot{d} \in \mathbb{F}_2^m} \mu\left( x_0 a_{\vecnot{d}} x^{\vecnot{d}} \right) \\
     & = \frac{1}{2}\sum_{\vecnot{x} \in \mathbb{F}_2^{m+1}} \prod_{\vecnot{d} \in \mathbb{F}_2^m} \left( 1 - 2 x_0 a_{\vecnot{d}} x^{\vecnot{d}} \right).
\end{align}
Define the function $t$ on $\mathbb{F}_2$ by $t(0) = 1$, $t(1) = -2$.  
Then distributing the product, we can express $N(f)$ as 
\begin{align}
N(f) = \frac{1}{2}\sum_i \sum_{\vecnot{x} \in \mathbb{F}_2^{m+1}} \prod_{\vecnot{d} \in \mathbb{F}_2^m} \left( t(i(\vecnot{d})) a_{\vecnot{d}}^{i(\vecnot{d})}(x_0x^{\vecnot{d}})^{i(\vecnot{d})} \right),
\end{align}
where the summation over indicators $i$ runs over all Boolean functions $i \colon \mathbb{F}_2^m \rightarrow \mathbb{F}_2$. % with $i(\vecnot{d}) \in \{0,1\}$.  
We want to calculate $\nu_2(w_H(u))$, where $u = \text{ev}(f)$, and we observe that $\nu_2(w_H(u)) = \nu_2(2^m - N(f)) = \nu_2(N(f))$.
Hence, we are interested in the $2$-adic valuation of $N(f)$, so we group terms from this sum into products and rewrite this as 
\begin{align}
N(f) = \frac{1}{2}\sum_i \left(\prod_{\vecnot{d} \in \mathbb{F}_2^m} a_{\vecnot{d}}^{i(\vecnot{d})} \right) \left(\prod_{\vecnot{d} \in \mathbb{F}_2^m} t(i(\vecnot{d})) \right) \left(\sum_{\vecnot{x} \in \mathbb{F}_2^{m+1}} \prod_{\vecnot{d} \in \mathbb{F}_2^m} \left( x_0 x^{\vecnot{d}} \right)^{i(\vecnot{d})}\right).
\end{align}
Observe that for each function $i$, the first product is binary, and the remaining two terms are each powers of $2$, so the whole term is a power of $2$ and has a $2$-adic valuation.
The last term is a power of $2$ because it is precisely the Hamming weight of the monomial $\prod_{\vecnot{d} \in \mathbb{F}_2^m} \left( x_0 x^{\vecnot{d}} \right)^{i(\vecnot{d})}$.

Now we are interested in the quantity $\nu_2\left( \left[ \prod_{\vecnot{d} \in \mathbb{F}_2^m} t(i(\vecnot{d})) \right] \left[ \sum_{\vecnot{x} \in \mathbb{F}_2^{m+1}} \prod_{\vecnot{d}} (x_0x^{\vecnot{d}})^{i(\vecnot{d})} \right] \right)$, since the smallest value among all indicating functions $i$ will determine $\nu_2(N(f))$.
Observe that when $i$ is the zero function, this quantity takes the maximal value of $2^{m+1}$, and hence does not affect $\nu_2(N(f))$.
When $i$ is not the zero function, we can calculate
\begin{align}
\nu_2\left(\sum_{\vecnot{x} \in \mathbb{F}_2^{m+1}} \prod_{\vecnot{d} \in \mathbb{F}_2^m} \left( x_0 x^{\vecnot{d}} \right)^{i(\vecnot{d})}\right) & = m - w_H\left( \sum_{\vecnot{d} \in \mathbb{F}_2^m} i(\vecnot{d})\vecnot{d} \right) \text{ and} \\
\nu_2\left(\prod_{\vecnot{d} \in \mathbb{F}_2^m} t(i(\vecnot{d}))\right) & = \sum_{\vecnot{d} \in \mathbb{F}_2^m} i(\vecnot{d}).
\end{align}
So we conclude that $\nu_2\left( \left[ \prod_{\vecnot{d} \in \mathbb{F}_2^m} t(i(\vecnot{d})) \right] \left[ \sum_{\vecnot{x} \in \mathbb{F}_2^{m+1}} \prod_{\vecnot{d} \in \mathbb{F}_2^m} \left( x_0 x^{\vecnot{d}} \right)^{i(\vecnot{d})} \right] \right) = m - w_H\left( \sum_{\vecnot{d} \in \mathbb{F}_2^m} i(\vecnot{d}) \vecnot{d} \right) + \sum_{\vecnot{d} \in \mathbb{F}_2^m} i(\vecnot{d})$.  
Now, because $f \in \text{RM}(r,m)$, $a_{\vecnot{d}}$ in the first term of each $i$ in $N(f)$ ensures that only terms with $\text{deg}(\vecnot{d}) \leq r$ survive. 
Hence, we have $w_H\left( \sum_{\vecnot{d} \in \mathbb{F}_2^m} i(\vecnot{d}) \vecnot{d} \right) \leq r \sum_{\vecnot{d} \in \mathbb{F}_2^m} i(\vecnot{d})$, with equality occurring only when all $\vecnot{d}$ with $i(\vecnot{d}) = 1$ are disjoint degree $r$ terms, i.e., weight $r$ vectors. % must be at least $\left\lceil\left|\sum_{\vecnot{d}} i(\vecnot{d})\vecnot{d}\right|/r\right\rceil$.  
From this, we can conclude
\begin{align}
\nu_2\left( \left[ \prod_{\vecnot{d} \in \mathbb{F}_2^m} t(i(\vecnot{d})) \right] \left[ \sum_{\vecnot{x} \in \mathbb{F}_2^{m+1}} \prod_{\vecnot{d} \in \mathbb{F}_2^m} \left( x_0 x^{\vecnot{d}} \right)^{i(\vecnot{d})} \right] \right) & \geq m - w_H\left( \sum_{\vecnot{d} \in \mathbb{F}_2^m} i(\vecnot{d}) \vecnot{d} \right) + \frac{1}{r} w_H\left( \sum_{\vecnot{d} \in \mathbb{F}_2^m} i(\vecnot{d})\vecnot{d} \right) \\
  & \geq \frac{m}{r}.
\end{align}
The second inequality holds because $m - t + \frac{t}{r} - \frac{m}{r} = (m-t) (1 - \frac{1}{r}) \geq 0$, since $t = w_H\left( \sum_{\vecnot{d} \in \mathbb{F}_2^m} i(\vecnot{d}) \vecnot{d} \right) \leq m$ and $r \geq 2$.

Furthermore, because $r|m$, we have equality if and only if $w_H\left( \sum_{\vecnot{d} \in \mathbb{F}_2^m} i(\vecnot{d}) \vecnot{d}\right) = m$ and $\sum_{\vecnot{d} \in \mathbb{F}_2^m} i(\vecnot{d}) = m/r$.  
In other words, the $2$-adic valuation of $N(f)$ is solely determined by those functions $i$ for which exactly $m/r$ \emph{disjoint} terms $\vecnot{d}$, each of weight $r$, have $i(\vecnot{d}) = 1$.
Put together, these conditions exactly define the coefficient products appearing in $q(f)$ in~\eqref{eq:QRM_thm_poly}.  
Let $P'$ denote the set of all such $i$ satisfying these conditions, so that this set has a bijective mapping to the set $P$ defined in the theorem statement.  
%Note also that $\nu_2(w_H(v_f)) = \nu_2(2^m - N(f)) = \nu_2(N(f))$.  
Then returning to $N(f)$, and noting that only those terms $i$ which contribute $2^{m/r}$ matter, we see that 
\begin{align}
w_H(u) & = 2^m - N(f) \\
  & \equiv - N(f) \ (\bmod\ 2^{\frac{m}{r}}) \\
  & \equiv 2^{\frac{m}{r}-1} \sum_{i \in P'} \prod_{\vecnot{d} \in \mathbb{F}_2^m} a_{\vecnot{d}}^{i(\vecnot{d})}\ (\bmod\ 2^{\frac{m}{r}}) \\
  & = 2^{\frac{m}{r}-1} \sum_{(p_1,\ldots,p_{m/r}) \in P} \prod_{j = 1}^{m/r} v_{p_j} \\
\label{eq:quasitransversal}
  & = 2^{\frac{m}{r}-1} q(f)\ (\bmod\ 2^{\frac{m}{r}}),
\end{align}
by construction of $P'$ and $P$.  
Here, $u$ determines which $a_{\vecnot{d}} = 1$, or equivalently which $v_{p_j} = 1$ ($u \leftrightarrow v_f$), since $u = \text{ev}(f)$ points to a specific coset of RM($r-1,m$) in RM($r,m$).
As $q(f)$ is oblivious to lower-order terms in $f$ (that correspond to $X$-type stabilizers), each coset indeed has a well-defined weight residue $(\text{mod}\ 2^{\frac{m}{r}})$, and thus the induced logical operation is also well-defined. 
Accordingly, by~\eqref{eq:u_zeta}, the logical action on (logical) computational basis vectors is defined by 
\begin{align}
\ket{v_f}_L \mapsto \zeta_{2^{m/r}}^{2^{m/r - 1} q(f)} \ket{v_f}_L = \mu(q(f)) \ket{v_f}_L.
\end{align}
% Finally, note that~\eqref{eq:quasitransversal} exactly matches the (quasi)transversality condition in~\cite[Lemma 1]{Campbell-pra17}.
This completes the proof.   \hfill \IEEEQEDhere

\end{document}